\documentclass[12pt,a4paper]{article}

\usepackage{jheppub}

\usepackage{etoolbox}
    \makeatletter
    \patchcmd{\maketitle}{\@fpheader}{}{}{}
    \makeatother

\usepackage{amssymb}
\usepackage{mathtools}

\usepackage{relsize,exscale}
\usepackage{tensor}

\usepackage[utf8]{inputenc}
\usepackage[T1]{fontenc}
\usepackage{lmodern}

\usepackage{mathrsfs}

\usepackage[]{subdepth}

\renewcommand{\ss}{\scriptscriptstyle}

\newcommand{\Lint}{\mathop{\mathlarger{\int}} \! \! }

\newcommand{\ch}{ \, {\rm ch} }

\usepackage{array}

\usepackage{pgfplots}
\usetikzlibrary{decorations.pathmorphing}
\usetikzlibrary{arrows}
\usetikzlibrary{patterns}
\usetikzlibrary{snakes}

\newcommand{\dl}{ \bigl( \hspace{-0.8mm}\bigr| }
\newcommand{\dr}{ \bigl| \hspace{-0.8mm}\bigr) }

\usepackage[numbers]{natbib}

\title{Abelian Higgs model in power-law inflation: \\ 
the propagators in the unitary gauge}

\author[1,2]{Dra\v{z}en Glavan,}
\emailAdd{drazen.glavan@uclouvain.be}
\author[3]{Anja Marunovi\'c,}
\emailAdd{anja.marun@gmail.com}
\author[4]{Tomislav Prokopec,}
\emailAdd{t.prokopec@uu.nl}
\author[5]{Zahra Zahraee}
\emailAdd{zr.zahraee@physics.mcgill.ca}

\affiliation[1]{Centre for Cosmology, Particle Physics and Phenomenology (CP3),
	\\
	Universit\'{e} catholique de Louvain, 
	\\
	Chemin du Cyclotron 2, 1348 Louvain-la-Neuve, Belgium}
\affiliation[2]{Division of Theoretical Physics, Rudjer Bo{\v s}kovi{\' c} Institute,
	\\
	Bijeni{\v c}ka cesta 54, HR-10\,000 Zagreb, Croatia }
\affiliation[3]{Amsterdam, The Netherlands}
\affiliation[4]{Institute for Theoretical Physics (ITF), Spinoza Institute
	\& $\rm EMME\Phi$,
	\\
	Faculty of Science, Utrecht University, 
	\\
	Postbus 80195, 3508 TD Utrecht, The Netherlands }
\affiliation[5]{Department of Physics, McGill University, 
	\\
	3600 Rue University, 
	Montr\'eal, QC, H3A 2T8 Canada}

\abstract{
We consider the Abelian Higgs model in the broken phase as a spectator
in cosmological spaces of
general $D$ space-time dimensions, and allow for the condensate to be time-dependent.
We fix the unitary gauge using Dirac's formalism for constrained systems, and then quantize
the gauge-fixed system. Vector and scalar perturbations develop time-dependent masses.
We work out their propagators assuming the cosmological background 
is that of power-law inflation, characterized by a constant 
principal slow-roll parameter, and that the scalar condensate is in the 
attractor regime, scaling as the Hubble rate.
Our propagators correctly reduce to known results in 
the Minkowski and de Sitter space limits.
We use the vector propagator to compute the equal-time correlators
of electric and magnetic fields and find that at super-Hubble
separations the former is enhanced, while the latter is suppressed
compared to the vacuum fluctuations of the massless vector field.
These correlators satisfy the hierarchy governed by Faraday's law.
}

\preprint{{\tt CP3-20-03}}

\begin{document}

\maketitle

\vskip+2cm

\section{Introduction}
\label{sec: Introduction}

Quantum effects in primordial inflation are particularly strong for light
fields non-conformally coupled to the expanding space-time. There are two such fields
 -- the graviton and the light/massless scalar -- that experience
abundant gravitational particle production 
even at tree level~\cite{Parker:1968mv,Parker:1969au,Parker:1971pt}. 
This can lead to sizable quantum loop corrections to their evolution, 
or the evolution other fields they couple to.
There exists a considerable body of literature on one-loop quantum effects in de Sitter space, 
see {\it e.g.} 
Ref.~\cite{Woodard:2014jba} and references 
therein,
that ascertains that, whenever loop corrections are important,
they grow in time as the logarithm of the scale factor,~$\sim\!\ln(a)$,
as generally predicted by Weinberg's theorem~\cite{Weinberg:2005vy,Weinberg:2006ac}.

The primordial inflating Universe is often approximated by de Sitter space
(more precisely the expanding Poincar\'{e} patch of 
de Sitter space)
 with a constant Hubble rate,~$\dot{H}\!=\!0$. Even though this is true in many cases,
 the actual slow-roll inflation space-time deviated from the exact de Sitter. 
 This deviation is conveniently 
 parametrized by slow roll parameters, of which 
the most important is the principal slow-roll 
parameter~$\epsilon\!=\!-\dot{H}/H^2\!\lesssim\!0.01$. Taking account of these deviations is particularly 
important for the correct modeling of the scalar curvature perturbation.
Furthermore, these deviations can be important when growing secular corrections 
induced by quantum loops are concerned. 
The logarithmic growth of de Sitter might turn into a power-law growth,
schematically,
\begin{equation}
\ln(a) \qquad \quad
\longrightarrow \qquad \quad
\frac{a^{ \# \epsilon}-1}{\# \epsilon} \, ,
\label{first eq}
\end{equation}
where~$\#$ is a numerical factor.
For~$\ln(a)\lesssim1/\epsilon$ the two are virtually indistinguishable, as assumed by Weinberg's
theorem. However, for longer inflation
when $\ln(a)\!\gg\!1/\epsilon\!\sim\!10^2$ the scaling
becomes drastically different.
It is unclear for how long inflation lasted, and 
it is imperative to investigate not only if there are large secular quantum effects 
in the primordial universe, but also how do they behave in more realistic inflationary spaces.
This is particularly important in cases where~$\#$ in Eq.~(\ref{first eq}) is a positive number, which can happen
when light fields are considered~\cite{Janssen:2008px,Glavan:2019uni}, 
and depends on the interplay of gravitational particle 
production of field quanta, and the redshifting of created quanta.

For this reason we are motivated to study quantum effects in 
{\it power-law inflation}~\cite{Lucchin:1984yf,La:1989za}, 
in which~$0\leq \epsilon\!=\!{\rm const.}\!<\!1$, 
as well as in more general slow roll inflationary
models and beyond. 
Power-law inflation is excluded as a model of primordial
inflation by the cosmic microwave background 
observations~\cite{Akrami:2018vks}, and it does not capture the
effect of an evolving principal slow-roll parameter. Nevertheless, it should be a better background model than de Sitter for
studying quantum loop effects in the primordial universe, as it does capture the 
effect of a non-vanishing
principal slow-roll parameter, and notably, it is still tractable enough to allow
for analytic computations. 
Thus far there have been several reports on loop computations in power-law inflation,
and general slow-roll inflation, some utilizing perturbative 
methods~\cite{Abramo:1998hj,Janssen:2008dw,Janssen:2009pb,Herranen:2013raa,Glavan:2019uni},
and some employing resummation/non-perturbative techniques~\cite{Prokopec:2015owa,Cho:2015pwa,Markkanen:2017edu}.
Noteworthy is also the recent
construction~\cite{Kyriazis:2019xgj}  of the one-loop effective potential 
in general inflationary models in which the principal slow roll parameter 
$\epsilon(t)$ is a general function of time.

\medskip

Gauge fields couple conformally to gravity, and by themselves they are not sensitive to
the expansion of the Universe.~\footnote{
The classical local conformal (Weyl) symmetry 
is broken by quantum loop effects via the conformal 
anomaly~\cite{Dolgov:1993vg,Drummond:1979pp}, and thus the conformally coupled 
fields can feel the expansion of the Universe without coupling to the light scalar
or the graviton. However, the effect of 
the anomaly on the vector field dynamics is typically 
quite small~\cite{Prokopec:2001nc,Benevides:2018mwx}.
}
However, the effects of the expansion can be mediated to them by coupling to
either gravitons or to light scalars. 
This happens already at tree level if the scalar develops a condensate which
breaks confomal invariance of the gauge fields, or at the loop level where 
conformal symmetry is broken due to infrared (IR) fluctuations of the graviton or the light scalar.
The former case corresponds to the mechanism of 
{\it spontaneous symmetry breaking}.
In the latter case inflation generates a medium of long wavelength 
quantum fluctuations in which a quantum field can lose its classical conformal symmetry,
which corresponds to the mechanism of {\it dynamical symmetry breaking}. One of 
the simplest, and arguably most interesting, examples of such a phenomenon is
the photon dynamics in presence of a light, charged scalar in inflation. The simplest,
and most striking, manifestation of the local conformal symmetry breaking is 
the photon mass generation in scalar electrodynamics 
in de Sitter space~\cite{Prokopec:2002jn,Prokopec:2002uw,Prokopec:2003tm}.
Even though it originates from quantum loops, this mass  
contributes to the effective action as a local, tree-level, mass term,
akin to the mass in the Brout-Englert-Higgs
(BEH) mechanism that operates in the standard model.
However, contrary to the mass generation in the standard model, which
is mediated by the classical Higgs condensate, 
the inflationary mass generation is predominantly due to the long wavelength (super-Hubble)
fluctuations of a light (or massless) charged scalar. Nevertheless, at the level of the 
effective action the description of the two phenomena is identical,
when treated in the adiabatic approximation.

Motivated by these insights, in this work we study 
the {\it non-minimally coupled Abelian Higgs model} in power-law inflation, 
a particular instance of {\it scalar electrodynamics} 
in which the mass 
term of the complex scalar 
$\Phi\!=\!\frac{1}{\sqrt{2}}(\phi \!+\! i\chi)$ is substituted by the 
non-minimal coupling $\xi$ to the Ricci curvature scalar $R$,
\begin{align}
&
S\bigl[ A_\mu, \Phi, \Phi^* \bigr] \!=\!\! \Lint d^{D\!}x \, \sqrt{-g} \,
	\biggl[
	- \frac{1}{4} g^{\mu\rho} g^{\nu\sigma} F_{\mu\nu} F_{\rho\sigma}
\nonumber \\
&	\hspace{5cm}
	- g^{\mu\nu} \bigl( D_\mu^* \Phi^* \bigr) \bigl( D_\nu \Phi \bigr)
	- \lambda \bigl( \Phi^* \Phi \bigr)^{\!4}
	- \xi R \Phi^* \Phi
	\biggr] \, .
\label{intro action}
\end{align}
Here~$D_{\mu} \!=\! \partial_\mu + i q A_\mu$ is the~$U(1)$ 
covariant derivative, $q$ the $U(1)$ charge, and
$F_{\mu\nu} \!=\! \partial_\mu A_\nu - \partial_\nu A_\mu$ 
is the field strength tensor associated with the vector potential $A_\mu$.
The action is written in~$D$ space-time dimensions 
to facilitate dimensionally regulated 
quantum loop computations.
There are multiple reasons for studying this particular model.
Firstly, it allows one to study the effects of the time-dependent scalar 
condensate in the attractor regime of power-law inflation, 
where both the expansion of the universe
and the evolution of the condensate play a significant role.
Secondly, the model in~(\ref{intro action}) can be used to describe 
the effects that the infrared quantum fluctuations
of a light scalar impart on the vector field they couple 
to~\cite{Prokopec:2007ak}. Thirdly,  the action~(\ref{intro action}) 
can be used as a toy model for the electroweak sector of the standard model both in 
inflationary as well as in post-inflationary epochs.
In what follows we elaborate on the former two.

\medskip

{\bf Complex scalar condensate.}
The evolution of spectator scalar fields~\footnote{ Spectator fields in inflation 
are the ones whose backreaction on the expansion rate can be neglected.}
in inflation often exhibits attractor regimes,
where information about initial conditions is lost.
In power-law inflation,
apart from the symmetric solution with a vanishing 
condensate~$\Phi\!=\!0$, the complex scalar
field of the model in~(\ref{intro action}) can develop a 
symmetry-breaking condensate 
that traces the evolution of the Hubble rate~$\Phi \!\propto\! H$,
which in power-law inflation evolves according to~$-\dot{H}/H^2\!=\!\epsilon\!=\!{\rm const.}$
This attractor solution has been considered before in the 
literature~\cite{Janssen:2008px,Koksma:2009tc,Janssen:2009pb}. 
The vector field, which is assumed not to have a condensate, is conformally 
coupled to gravity at tree-level in the absence of a scalar condensate.
The non-vanishing scalar condensate
is responsible for the spontaneous symmetry breaking mechanism,
as it induces a mass-like term for the vector field,
which is more properly seen as an effective  non-minimal coupling of the
vector field to the Ricci curvature scalar, on the account of its time-dependence 
tracing the evolution of the Hubble rate,
\begin{equation}
\bigl( \Phi^*\Phi \bigr) g^{\mu\nu} \!A_\mu A_\nu
	\propto H^2 g^{\mu\nu} \! A_\mu A_\nu
	\propto R \, g^{\mu\nu} \!A_\mu A_\nu \, .
\label{condensate scaling}
\end{equation}
This is how the conformal coupling of the vector to gravity is broken
in  the symmetry-breaking attractor solution in power-law inflation, and how 
the vector becomes sensitive to the expansion of space-time at tree-level.

The fluctuations of the complex scalar around the condensate
already are non-minimally coupled to gravity 
in~(\ref{intro action}), but the presence of a condensate effectively renormalizes 
the coupling constant (see Sec.~\ref{sec: Scalar perturbations} for more details).
Since the non-minimal coupling can be seen as a time-dependent
mass term, one expects similarities with the Abelian 
Higgs model in flat space that develops time-dependent masses.
This work is mostly devoted to quantifying the effects that the 
time-dependent condensate~(\ref{condensate scaling}) has on the 
linear quantum fluctuations of the Abelian Higgs model in~(\ref{intro action}).
This is accomplished by working out the propagators of both scalar 
and vector linear fluctuations,
thus setting up the stage for studying the loop corrections to the 
inflationary dynamics of the model.

\medskip

{\bf Infrared scalar fluctuations.} 
 In scalar electrodynamics models where the complex scalar does 
not develop a condensate the vector field remains massless at tree-level in cosmological
spaces, and is not sensitive to the expansion.
However, a light complex scalar can induce a mass of the vector purely due to large
infrared (super-Hubble) quantum fluctuations.
In Refs.~\cite{Prokopec:2002jn,Prokopec:2002uw,Prokopec:2003bx} a perturbative one-loop
analysis was carried out in de Sitter
assuming a massless scalar. 
The one-loop vacuum 
polarization was computed in Ref.~\cite{Prokopec:2002jn}, and it was found that the
correction terms it provides to the effective equation of motion for the vector potential 
mode function at late times takes the form of a momentum-dependent mass term,
\begin{equation}
m_A^2 = \frac{(qH)^2}{2\pi^2} \times \ln \Bigl(\frac{k}{H} \Bigr) 
\, ,
\label{photon mass: perturbative}
\end{equation}
where~$k$ is the modulus of the mode wave vector. The effective 
mode function equation was then solved in 
Ref.~\cite{Prokopec:2003iu}, 
where it was found that the electric field gets secularly enhanced compared to 
the vacuum case,~$E^i\!\sim\!a^{-3/2}$, 
while the magnetic field receives a correction to its amplitude,
but the scaling remains that of the conformal vacuum,~$B^i\!\sim\!a^{-2}$. 
The effect on the electric field can be
effectively captured by a vector field with a secularly growing mass~$m_A\!\sim\!\ln(a)$,
but this 
local approximation to the vacuum polarization
gives too little magnetic field.
The subsequent work~\cite{Prokopec:2003tm} adapted the analysis to the 
case of a light non-minimally coupled scalar, reporting an effective vector mass generation,
\begin{equation}
m_A^2 = 
	\frac{(q H)^2}{2\pi^2} \times \frac{3 H^2}{2 \bigl( m_\phi^2 + 12 \xi H^2 \bigr)}
\, ,
\label{photon mass: perturbative massive}
\end{equation}
where $m_\phi \! \ll \! H$ is the scalar mass, and~$\xi$ is its non-minimal coupling constant.

Perturbative studies~\cite{Prokopec:2002jn,Prokopec:2002uw,Prokopec:2003tm} 
of vector mass generation by a massless/light complex scalar
all pointed to the need for non-perturbative studies.
By making a suitable generalization of Starobinsky's stochastic formalism~\cite{Starobinsky:1986fx},
Ref.~\cite{Prokopec:2007ak} showed that, after a sufficiently long time, 
massless scalar quantum electrodynamics in de Sitter space 
settles to a nonperturbative, 
interacting vacuum state, 
in which the scalar remains perturbatively light, but the vector develops a 
non-perturbatively large masss, respectively,
\begin{equation}
m_\phi^2 \approx 0.45 \times  \frac{(q H)^2}{4\pi^2} \, ,
\qquad \qquad
m_A^2 \approx 3.3 \times H^2 \, ,
 \label{nonperturbative photon and scalar masses in de Sitter}
\end{equation}
where $m_A$ is the mass which characterizes the coincident vector propagator.
This interacting vacuum state has lower energy than any free Gaussian state.
Curiously, inserting the non-perturbative scalar 
mass~(\ref{nonperturbative photon and scalar masses in de Sitter})
 into~(\ref{photon mass: perturbative massive}) 
gives for the vector mass $m_A^2 \!\simeq\! 2.2 \!\times\! H^2$, which is not far 
from its true value given in~(\ref{nonperturbative photon and scalar masses in de Sitter}). 
This $\sim \!30\%$ difference can be attributed to 
the non-Gaussian nature of the interacting vacuum state. 
The masses in~(\ref{nonperturbative photon and scalar masses in de Sitter})
are generated by the large infrared quantum fluctuations, {\it i.e.} by the composite
operators developing a condensate, and thus are a result of a {\it dynamical symmetry-breaking}
mechanism in de Sitter space.
Vector field perturbations in de Sitter inflationary setting can affect inflationary 
observables~\cite{Chua:2018dqh,Lu:2019tjj} and
the dynamics of the standard model during inflation~\cite{Chen:2016nrs}, 
which may have observable ramifications, thus warranting a closer investigation.

It would be very interesting to understand how the dynamical symmetry breaking picture 
from de Sitter space generalizes to more general cosmological spaces, such as
power-law inflation considered here.
To this end, we draw inspiration from the similarities between the spontaneous and dynamical
symmetry-breaking mechanisms, in particular from the scaling of the scalar 
condensate~$\langle \hat{\Phi} \rangle \! \propto\! H$
in the first case, and the presumed scaling of the scalar 
fluctuations~$\langle \hat{\Phi}^\dag \hat{\Phi} \rangle\!\propto\!H^2$
in the second case. The latter scaling is supported by the non-perturbative results pertaining
to power-law inflationary dynamics of a real scalar 
field~$\phi$ with a quartic self-interaction term, 
$\lambda\phi^4/4$. 
In Refs.~\cite{Prokopec:2015owa,Cho:2015pwa} the stochastic 
formalism~\cite{Starobinsky:1986fx}
was considered in power-law inflation, where it was found that, 
at late times, the scalar variance indeed scales as~$H^2$,
\begin{equation}
\langle \hat{\phi}^2 \rangle
	= H^2 \int_{-\infty}^{\infty} \! d\sigma \, \sigma^2 \, \rho(\sigma) \, ,
	\qquad
	\rho(\sigma) = \rho_0 \, \exp\biggl[ \frac{-2}{ (3\!-\!\epsilon) A } 
		\biggl( \frac{\lambda}{4} \sigma^4 - \frac{(3\!-\!\epsilon) \epsilon}{2} \sigma^2 \biggr) \biggr] \, ,
\label{phi2 scaling}
\end{equation}
where~$\sigma$ is a dimensionless integration variable, 
$\rho_0$ and~$A$ are dimensionless constants 
that depend on~$\epsilon$ and $\lambda$, where $\lambda$ 
is the quartic self-coupling constant, and~$\sigma$ is a dimensionless integration variable. 
It is interesting that~$\epsilon\!>\!0$ generates a 
negative correction to the effective mass term 
$\propto\epsilon$
appearing in the probability 
distribution function~$\rho(\sigma)$.
The scaling in~(\ref{phi2 scaling})
was further corroborated for the same model in Ref.~\cite{Markkanen:2017edu} by 
solving the gap equation resulting from the 2PI Dyson-Schwinger equation in power-law inflation.

The scaling of scalar fluctuations in the scalar electrodynamics 
model~$\langle \hat{\Phi} \hat{\Phi}^\dag\rangle \!\propto\! H^2$ is qualitatively 
different from the de Sitter case, where this quantity is just a constant, and if correct
might account for interesting effects imparted on the vector.
We cannot confirm that this is the case without a detailed computation,
which indeed would be worthwhile pursuing. 
Furthermore, the vector propagators in cosmological spaces, both
the massive one constructed here and the massless one
left for future work, 
constitute fundamental building blocks for studying physical effects 
from the interactions of gauge fields with matter and gravity in inflation. 
Examples illustrating their importance are
such phenomena as large scale cosmological
magnetic and electric fields~\cite{Prokopec:2001nc,Prokopec:2003iu,Prokopec:2004au,Ratra:1991bn,Durrer:2013pga,Fujita:2019pmi,Brandenburg:2018ptt,Glavan:2019uni,Maleknejad:2012fw,Subramanian:2015lua}, 
generation of gravitational waves~\cite{Bartolo:2016ami,Caprini:2015zlo,Caprini:2006jb,Caprini:2001nb},
{\it etc.}  

Perturbative studies may, however, not be enough to quantify interesting effects,
and in order to fully capture the late time dynamics of 
interacting quantum fields in inflation, one may have 
to resort to non-perturbative methods. At present two such
reliable methods are known,
Starobinsky's stochastic inflation~\cite{Starobinsky:1986fx}
and functional renormalization group methods~\cite{Wetterich:1992yh}.
It is known that the non-perturbative infrared effects are strong enough 
to restore symmetries during inflation. For example, in both the 
scalar field theory with a global $O(N)$ symmetry and in the real scalar self-interacting theory,
an initially broken symmetry gets 
restored~\cite{Lazzari:2013boa,Serreau:2013eoa,Guilleux:2015pma} and the scalars acquire 
a calculable mass
 (see Ref.~\cite{Janssen:2009pb} for an earlier, perturbative analysis of the same problem).
There are two works known to us in which 
symmetry restoration in theories involving gauge fields is investigated in de Sitter 
by means of non-perturbative methods:  
Ref.~\cite{Prokopec:2007ak}~\footnote{  Even
though Ref.~\cite{Prokopec:2007ak} never mentions that the results presented can be 
used in studies of symmetry restoration, the universality of the asymptotically 
late-time scalar and vector field masses
implies that.}
uses stochastic methods and Ref.~\cite{Gonzalez:2016jrn}
uses functional renormalization group methods to study the late time dynamics 
of the vector and scalar fields of scalar electrodynamics in de Sitter.
These works found that, at late times, the symmetry gets restored and the vector
and the scalar acquire a calculable mass~(\ref{nonperturbative photon and scalar masses in de Sitter}).
Both of the non-perturbative approaches in these studies utilized the massive 
vector field propagator in de Sitter space to capture the interactions between the two fields.
For analogous studies in power-law inflation it is important to have the massive vector field
propagator that we construct here.

\medskip

{\bf Gauge choice.} 
The Abelian Higgs model in~(\ref{intro action}) that we consider is a gauge 
theory, and its quantization necessarily requires fixing a gauge.
In this work we impose
the {\it unitary gauge}, which we implement using Dirac's formalism for systems with first-class 
constraints~\cite{Dirac:2011,Gitman:1990qh,Rothe:2010dzf}.
This amounts to enforcing the condition~${\rm Im}(\Phi)\!=\!0$ that fully fixes the gauge.
Here we opt for the unitary gauge for two reasons: (i) it is physically transparent, as it
allows for explicit reduction of phase space leaving two fields -- a massive/non-minimally coupled
scalar and a massive/non-minimally coupled vector, and (ii) it leads to analytically tractable
equations, and allows us to work out the two-point functions analytically.
The form that the propagators take is simple enough to allow for explicit loop computations.

The propagators in the unitary gauge are constructed here with the intent of 
eventually employing them in loop computations. Having that in mind, we should point
out that in flat space loop computations the unitary gauge is usually disfavoured compared
to the~$U(1)$ variant of `t Hooft background field gauge, also known as the~$R_\xi$ 
gauge~\cite{tHooft:1971qjg,Fujikawa:1972fe}
 which is an average gauge, determined at tree level by the 
gauge-fixing functional,
\begin{equation}
S_{\rm gf}\bigl[A_\mu,\chi \bigr]
	= \Lint d^{D\!}x \, \sqrt{-g} \, \biggl[ - \frac{1}{2\alpha} 
		\Bigl( g^{\mu\nu} \nabla_\mu A_\nu - \alpha q \overline{\phi} \chi \Bigr)^{\!2} \biggr] \, ,
\label{gf functional}
\end{equation}
where~$\alpha$ is an arbitrary gauge-fixing parameter,~\footnote{
We denote the gauge-fixing parameter of the~$R_\xi$ gauge by~$\alpha$ instead of the
more common~$\xi$, in order not to confuse it with the non-minimal coupling 
we denote by~$\xi$.}
and where the complex scalar is split into its real and imaginary 
parts,~$\Phi\!=\! \bigl[ ( \overline{\phi}+\varphi )+i\chi \bigr]/\sqrt{2}$,
with~$\overline{\phi}$ being the scalar condensate. The reasons why the~$R_\xi$ gauge is favoured
are: (i) The unitary gauge cannot be applied to the symmetric phase where the condensate
vanishes, which can lead to some-non-perturbative issues~\cite{Anderberg:2008gs};
(ii) The vector propagator in the unitary gauge is more singular than the propagator
in the~$R_\xi$ gauge, which makes the perturbation theory {\it not} power-counting renormalizable,
and in fact requires additional counterterms be added to the effective action
at each perturbative order to ensure
renormalizability. The additional counterterms in flat space can be resummed into a single 
term,~$\delta \mathscr{L} \!=\!  i \delta^4(0)\ln\bigl( | \Phi | /\overline{\phi} \bigr)$,
which was first derived using canonical quantization 
methods~\cite{Weinberg:1973ew,Weinberg:1973ua,Appelquist:1973ms}, and 
subsequently by path-integral 
methods~\cite{Lee:1972fj,Lee:1972yfa,Lee:1974zg,Becchi:1974xu,Becchi:1974md}, thus establishing
the connection with the BPHZ (Bogolyubov-Parasiuk-Hepp-Zimmermann) renormalizability
\cite{Bogoliubov:1957gp,Hepp:1966eg,Zimmermann:1968mu} 
which is known to hold in $R_\xi$ gauges~\cite{tHooft:1971qjg}.
Furthermore, it was shown in Ref.~\cite{Lee:1973fw} that the unitary gauge yields 
correct results for the on-shell ($S$-matrix) calculations. 
Recently, a simple proof of renormalizability of the unitary gauge was provided
in Ref.~\cite{Tupper:2014uva}, by making use of the background field method
(in which the gauge fixing term is gauge invariant under the background field gauge transformations).

Of course, in flat space one has the luxury of choosing between different propagators 
at disposal, as is also the case in de Sitter space. However, in power-law inflation no
massive vector propagators have been reported thus far, and one either needs to
construct the vector propagator in the covariantized~$R_\xi$ gauge, or deal with
the three issues mentioned above 
when using the propagator reported here. We think that these 
issues will not represent an obstacle if the unitary gauge vector propagator is employed with due care.
The issue in (i) is not worrying since non-perturbative issues do not arise in 
perturbative computations in the broken phase. The renormalization issue in (ii) does appear
in perturbative computations, and needs to be addressed, even more so having in mind
that the additional counterterms have been worked out only for flat space. 
The necessity of additional counterterms ultimately derives from the fact that the Feynman 
propagator in the unitary gauge is {\it not} the Green's function,
but the two differ by a non-covariant local term. This difference between the two,
which does not usually appear in quantum field theories, follows from the rules of canonical 
quantization in the unitary gauge~\cite{Weinberg:1973ew,Greiner:1996zu}. 
If one insists on using just the covariant Green's function in the loop expansion as customary,
this is the reason behind the necessity of additional 
counterterms~\cite{Weinberg:1973ew,Weinberg:1973ua}.
When the distinction between the two is
maintained in the perturbative expansion, which can be 
inferred by using the Yang-Feldman 
equation~\cite{Yang:1950vi,Greiner:1996zu},
the additional counterterms should not be required. This is particularly important for the 
applications we have in mind, as it is not clear how to generalize the extra counterterms to
power-law inflation. That is why we pay great attention to the distinction between the
Feynman propagator and the Green's function in Sec.~\ref{sec: Vector field two-point functions},
and we carefully work out both, in addition to the Wightman functions.

Even though we believe that the unitary gauge propagators can be used in perturbative 
computations in power-law inflation, it would be 
beneficial to have the propagators
worked out in the~$R_\xi$ gauge as well. Those could then be applied to the studies
of massless vector fields as well, but even more importantly, we would have a free gauge-fixing
parameter which would allow to explicitly test the computed observables for gauge-dependence.
This is of a crucial importance for quantum 
field theory in cosmological spaces, where the~$S$-matrix either does not exist, 
or is of limited utility for the initial-value problems, and one is forced to construct observables
which are real-time expectation values,
for which the Schwinger-Keldysh 
formalism~\cite{Schwinger:1960qe,Keldysh:1964ud} 
is suitable. 
However, one big advantage of the~$R_\xi$ gauge is lost
when the scalar condensate acquires space-time dependence,~$\partial_\mu \overline{\phi}\!\neq\!0$. 
In that case the vector field no longer decouples
from the Goldstone boson~$\chi$~\cite{Mooij:2011fi,George:2012xs} at tree level, 
which complicates the analysis and requires the use of the mixed field (scalar-vector) propagators
in loop computations. Therefore, one would have to either come up with a modification of the
gauge-fixing functional~(\ref{gf functional}) which preserves the decoupling property, or
work out the mixed field propagators as well.

We should also mention that the unitary gauge is often referred to as the~$\alpha\!\to\!\infty$
limit of the~$R_\xi$ gauge~(\ref{gf functional}), in which the Goldstone boson~$\chi$
becomes infinitely massive and decouples. This indeed is true for the most part, except 
for a subtle, but important fact that taking this limit does not commute with loop
integrations in the perturbative expansion~\cite{Mooij:2011fi}. The origin of this issue can 
be traced to the difference between the Feynman propagator and the Green's function.
In fact, the limit~$\alpha\!\to\!\infty$ of the~$R_\xi$ gauge produces the Green's 
function of the unitary gauge, not the Feynman propagator.

\medskip

{\bf Related work.}
Not much is known about gauge field propagators in general cosmological spaces,
and in this regard the best understood is de Sitter space, 
which is the model space for cosmological inflation. 
There are several results in de Sitter space related to our work.
The massive vector field propagator in the unitary gauge
was constructed by Allen and Jacobson in Ref.~\cite{Allen:1985wd}.
Tsamis and Woodard~\cite{Tsamis:2006gj} constructed the massive 
vector propagator in de Sitter
in the Landau gauge limit of the~$R_\xi$ gauge, in which the propagator is 
exactly transverse. 
Fr{\"o}b and Higuchi~\cite{Frob:2013qsa} constructed the
massive vector propagator in de Sitter in the Stueckelberg model,
in the appropriate adaptation of the~$R_\xi$ gauge,
which was rederived  for a particular choice of the
gauge-fixing parameter in~\cite{Belokogne:2016dvd}.
The Stueckelberg model in de Sitter space can be seen to match the linearized Abelian Higgs 
model where the condensate takes a constant value, and thus reproduces the Allen-Jacobson
propagator in the unitary gauge limit.
The principal goal of this work is to generalize the above results and construct the massive vector
propagator for power-law inflation. The Tsamis-Woodard vector propagator~\cite{Tsamis:2006gj}
admits a regular massless limit, and is the only one that
has been successfully utilized in loop computations
for scalar electrodynamics~\cite{Prokopec:2006ue,Prokopec:2007ak,Gonzalez:2016jrn}, and for 
electromagnetism interacting with dynamical 
gravity~\cite{Glavan:2015ura,Glavan:2016bvp}.

\medskip

{\bf Outline.} The paper is organized as follows.
In sections~\ref{sec: FLRW and power-law inflation} 
and~\ref{sec: Abelian Higgs model in cosmological spaces}
we discuss the background fields in power-law inflation and introduce scalar electrodynamics,
the model for which the massive photon propagator is constructed in this work.  
 Since the cosmological literature is exiguous
regarding the delicate steps involved in the quantization of 
systems with constraints such as gauge theories, in 
Sec.~\ref{sec:  Abelian Higgs model in cosmological spaces} we present in some detail
the classical analysis of the constraints and construct the (extended) Hamiltonian,
the corresponding action, and calculate the Dirac brackets.
Next, in Sec.~\ref{sec: Condensate dynamics} we discuss the dynamics of the
scalar condensate, 
and in Sec.~\ref{sec: Scalar perturbations}
 the propagator for scalar perturbations is briefly discussed. 
In Sec.~\ref{sec: Dynamics of vector perturbations}
the Dirac-Bergmann 
quantization~\cite{Dirac:1950pj,Dirac:1951zz,Anderson:1951ta,Dirac:2011,Gitman:1990qh,Rothe:2010dzf} is applied to the model at hand, 
whereby a special attention is devoted to how to obtain 
the unitary gauge 
within that general procedure. 
The main result is obtained in Sec.~\ref{sec: Vector field two-point functions},
where we construct the {\it covariant vector propagator},
the Green function and the 
Wightman two-point function, 
which are the basic building blocks of perturbation theory in the 
Schwinger-Keldysh formalism~\cite{Schwinger:1960qe,Keldysh:1964ud}. 
We check our vector propagator by considering various limits, which
include Minkowski and de Sitter space, 
and compare with the available known results. 
The vector propagators are put to use in Sec.~\ref{sec: Field strength correlator}
where we compute the field strength correlators and discuss the behavior
of the equal-time electric and magnetic field correlators at super-Hubble separations.
Finally, in Sec.~\ref{sec: Discussion} we discuss our main results and outline future perspectives. 
Some important technical details are relegated to appendices.

\section{FLRW and power-law inflation}
\label{sec: FLRW and power-law inflation}

The spatially flat~\footnote{
The assumption of spatial flatness is
consistent with all current observations~\cite{Aghanim:2018eyx,Akrami:2018vks}. }
Friedmann-Lema{\^i}tre-Robertson-Walker (FLRW) space-time
is a spatially homogeneous and isotropic expanding space, given by
the invariant line element,
\begin{equation}
ds^2 = g_{\mu\nu}(x) dx^\mu dx^\nu
	= -dt^2 + a^2(t) d\vec{x}^{\,2} 
	= a^2(\eta) \bigl[ - dt^2 + d\vec{x}^{\,2} \bigr] \, ,
\end{equation}
where $t$ is the physical (comoving) time, $\eta$ is the conformal time, 
$\vec{x}\!=\!(x^1,x^2,\cdots,x^{D-1})$
denote the comoving spatial coordinates, and $a$ is the scale factor. 
In this work we prefer to use conformal time in which FLRW 
background is
given by a conformally flat metric,
\begin{equation}
g_{\mu\nu} = a^2(\eta) \, \eta_{\mu\nu} \, ,
\label{conformal gmn}
\end{equation}
where $\eta_{\mu\nu} \!=\! {\rm diag}(-1,1,\dots,1)$ is the $D$-dimensional
Minkowski metric. 
The rate of the expansion is encoded by the {\it conformal Hubble rate}, or by
the {\it physical Hubble rate}, respectively,
\begin{equation}
\mathcal{H} = \frac{\partial_0 a}{a} \, ,
\qquad \qquad H = \frac{\mathcal{H}}{a} \, ,
\end{equation}
where $\partial_0 \!=\! \partial/\partial x^0 \!=\! \partial/\partial\eta$ 
denotes a derivative with respect to conformal time.
We make frequent use of both Hubble rates. 
The acceleration rate is
conveniently parametrized by the {\it principal slow-roll parameter},
\begin{equation}
\epsilon = 1 - \frac{\partial_0 \mathcal{H}}{\mathcal{H}^2} \, ,
\label{epsilon: definition}
\end{equation}
which is related to the {\it deceleration parameter}~$q \!=\! \epsilon-1$.
When $\epsilon\!<\!1$ the universe is
accelerating, and when $\epsilon\!>\!1$ it is decelerating.

\begin{figure}[h!]
\centering
\begin{tikzpicture}

\draw[draw=white, fill=black!20] (0,-3) -- (3,0) -- (0,0) -- (0,-3);

\draw[draw=white, fill=black!20] (0,-3) -- (6,-9) -- (0,-9) -- (0,-3);

\draw[pattern=checkerboard, pattern color=black!10, draw=white]  
	(0,0) -- (8.8,0) -- (8.8,-8.8) -- cycle;

\draw[-triangle 45] (0,-9) -- (0,.8) ;
\node[align=left, above] at (0,0.8) {$\eta$} ;

\draw[-triangle 45] (0,-3) -- (9.4,-3) ;
\node[align=left, right] at (9.4,-3) {$\| \vec{x} \!-\! \vec{x}^{\,\prime} \|$} ;

\draw[very thick,dashed] (0,0) -- (8.8,0) ;
\node[align=left, left] at (0,0) {$\overline{\eta}$} ;

\node[align=left, left] at (0,-3) {$\eta'$} ;

\draw[very thick] (0,-3) -- (3,0) ;

\draw[very thick] (0,-3) -- (6,-9) ;


\draw[very thick, dotted] (0,0) -- (8.8,-8.8) ;

\end{tikzpicture}
\caption{
The conformal diagram of power-law inflation.
Thick solid lines represent the past and future null rays of 
some point~$(\eta',\vec{x}^{\,\prime})$, while the gray shaded regions
represent the past and future light-cones of that point. In the conformal coordinates
the past light-cone grows without limits, while the future light-cone 
is limited by~$\overline{\eta}$ from~(\ref{coordinates in FLRW spacetime}),
that corresponds to the asymptotic future.
Dotted line denotes the Hubble horizon, and the checkered region
the points that are causally disconnected from the
observer at~$\vec{x}^{\,\prime}$. 
}
\label{conformal diagram}
\end{figure}
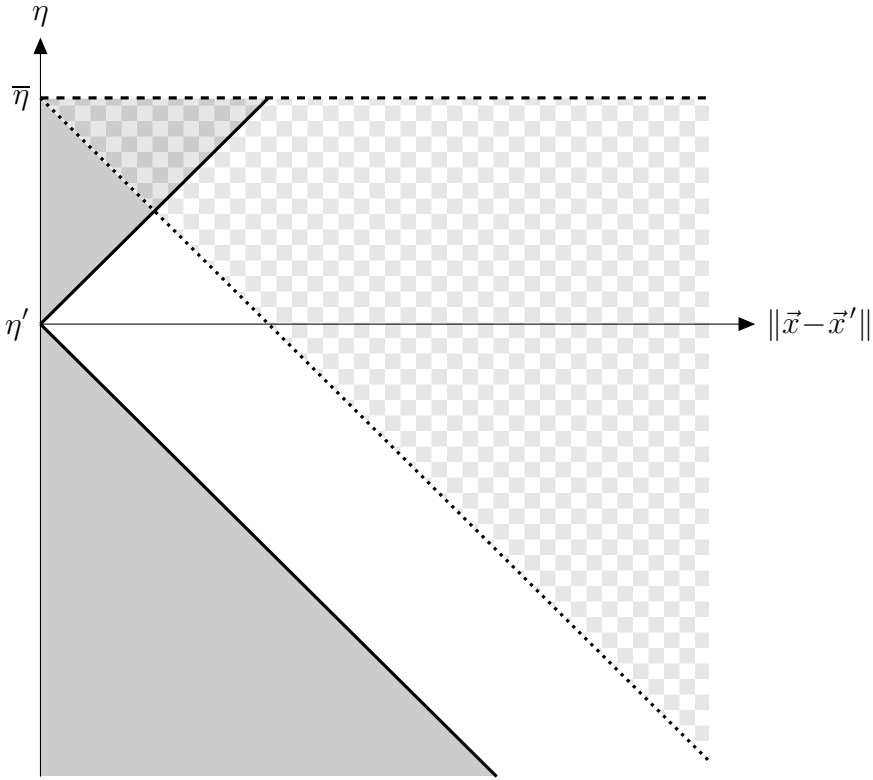

In this work we study the Abelian Higgs model in 
{\it power-law inflation}, which is characterized by
a constant $\epsilon$ parameter in the accelerating range,
\begin{equation}
\epsilon = {\tt const.}
\qquad \& \qquad 
0 \le \epsilon < 1
               \, \qquad\qquad (\text{\tt power-law\;\; inflation})
\, ,
\end{equation}
where~$\epsilon\!=\!0$ corresponds to de Sitter space-time
(more precisely to the Poincar\'e patch of de Sitter).
The scale factor and the conformal Hubble rate in power-law inflation are given by,
\begin{equation}
a = \bigl[ 1 -(1\!-\!\epsilon)H_0(\eta\!-\!\eta_0) \bigr]^{\frac{-1}{1-\epsilon}} \, ,
\qquad
\mathcal{H} = H_0 \bigl[ 1 -(1\!-\!\epsilon)H_0(\eta\!-\!\eta_0) \bigr]^{-1} \, ,
\qquad
\mathcal{H} a^{\epsilon-1} = H_0 \, ,
\label{scale factor and H in power-law expansion}
\end{equation}
where $\eta_0$ is some initial time, $H_0 \!=\! \mathcal{H}(\eta_0)$, 
and $a(\eta_0)\!=\!1$. The ranges of coordinates in power-law inflation are
\begin{equation}
-\infty < \eta < \eta_0 + \frac{1}{(1\!-\!\epsilon) H_0} \equiv \overline{\eta} \, ,
\qquad \qquad
-\infty < x^i < \infty \, ,
\label{coordinates in FLRW spacetime}
\end{equation}
and the conformal diagram of its causal structure is given in Fig.~\ref{conformal diagram}.
The curvature tensors~\footnote{
The conventions that we use 
are~${R^\alpha}_{\mu\beta\nu} \!=\! \partial_\beta \Gamma^\alpha_{\mu\nu}
	- \partial_\nu \Gamma^\alpha_{\mu \beta}
	+ \Gamma_{\mu\nu}^\rho \Gamma^\alpha_{\beta\rho}
	- \Gamma_{\mu\beta}^\rho \Gamma^{\alpha}_{\nu\rho} $
for the Riemann tensor, 
and~$\Gamma^\alpha_{\mu\nu} \!=\! \frac{1}{2} g^{\alpha\beta} ( \partial_\mu g_{\nu\beta}
	+ \partial_\nu g_{\mu\beta} - \partial_\beta g_{\mu\nu} )$ for the Christoffel symbol.} 
in power-law inflation are given by,
\begin{align}
R_{\mu\nu\rho\sigma} ={}&
	2 H^2 g_{\mu[\rho} g_{\sigma]\nu}
	+ 4 \epsilon H^2 \bigl( a^2 \delta^0_{[\mu} g_{\nu][\sigma} \delta^0_{\rho]} \bigr) \, ,
\\
R_{\mu\nu} ={}&
	(D\!-\!1\!-\!\epsilon) H^2 g_{\mu\nu}
	+ (D\!-\!2) \epsilon H^2 \bigl( a^2 \delta_\mu^0 \delta_\nu^0 \bigr) \, ,
\\
R ={}& 
	(D\!-\!1)(D\!-\!2\epsilon) H^2 \, ,
\end{align}

\medskip

We make a frequent use of the classical {\it distance function}
between two points $x$ and $x'$,
\begin{equation}
 y(x;x') = (1\!-\!\epsilon)^2 \mathcal{H} \mathcal{H}' \Delta x^2 \, ,
\label{y def}
\end{equation}
where $\Delta x^2 \!=\! \Delta x^2(x;x')$ 
is the geodesic distance in the conformal Minkowski space-time,
\begin{equation}
\Delta x^2 =\eta_{\mu\nu}(x\!-\!x')^\mu(x\!-\!x')^\nu
  =  \| \vec{x} \!-\! \vec{x}^{\,\prime} \|^2 - (\eta\!-\! \eta')^2
   \equiv  \| \Delta \vec{x} \|^2 - (\Delta\eta)^2
	  \, .
\end{equation}
In the de Sitter limit, where $\epsilon\!=\!0$, this function is related to the geodesic
distance $\ell(x;x')$ between two points $x$ and $x'$ 
as $y(x;x')|_{\epsilon=0}\!=\!4\sin^2\left[H_0\ell(x;x')/2\right]$. 
In power-law inflation this relation no longer holds,
but it is still very convenient to express our results in terms of $y(x;x')$.

In the quantum theory, the two-point functions will depend on the
distance function~(\ref{y def}), and their distributional character will be conveniently
encoded by suitable imaginary shifts of time in the distance function. 
All relevant two-point functions can be obtained from
the following $i\varepsilon$-prescriptions~\cite{Onemli:2002hr},
\begin{equation}
y_{\ss AB}(x;x') = (1\!-\!\epsilon)^2 \mathcal{H} \mathcal{H}'
	\Delta x^2_{\ss AB} \, ,
\qquad \qquad
A,B = \pm \, ,
\label{y AB}
\end{equation}
where the prescription lies entirely within the conformal Minkowski 
part,~\footnote{
Note there are different ways of defining~$i\varepsilon$-prescriptions
corresponding to given two-point functions, {\it e.g.}
\begin{equation*}
\Delta x^2_{\ss ++} = 
	\| \vec{x} \!-\! \vec{x}^{\,\prime} \|^2 - ( \eta\!-\! \eta' )^2 - i \varepsilon \, ,
\qquad \qquad
\Delta x^2_{\ss -+} = 
	\| \vec{x} \!-\! \vec{x}^{\,\prime} \|^2 - (\eta\!-\! \eta' )^2  - i \varepsilon \, {\rm sgn}(\eta\!-\!\eta') \, ,
\end{equation*}
as employed in {\it e.g.}~\cite{Tsamis:1992xa}. 
This reflects the fact that defining a distribution in terms of a limit of an
analytic function is not unique. One needs to be careful to use prescriptions consistently,
as some expressions involving derivatives may seem rather different in different prescriptions.
The prescriptions from~(\ref{X++}) and~(\ref{X-+}) we use here naturally arises when
the scalar propagator is constructed as a sum over modes~\cite{Janssen:2008px}. 
}
\begin{align}
&
\Delta x^2_{\ss ++} = 
	\| \vec{x} \!-\! \vec{x}^{\,\prime} \|^2 - \bigl( |\eta\!-\! \eta' | \!-\! i \varepsilon \bigr)^2 \, ,
&&
\Delta x^2_{\ss --} = \bigl(\Delta x^2_{\ss ++} \bigr)^* \, , \qquad
\label{X++}
\\
&
\Delta x^2_{\ss -+} = 
	\| \vec{x} \!-\! \vec{x}^{\,\prime} \|^2 - \bigl(\eta\!-\! \eta'  \!-\! i \varepsilon \bigr)^2 \, ,
&&
\Delta x^2_{\ss +-} = \bigl(\Delta x^2_{\ss -+} \bigr)^* \, , \qquad
\label{X-+}
\end{align}
The $(\pm)$ nomenclature is adapted to the Schwinger-Keldysh 
formalism~\cite{Schwinger:1960qe,Keldysh:1964ud,Berges:2004yj,NonEqLectures},
and refers to different types of two-point functions. Namely,~$(++)$ refers
to the Feynman, and $(--)$ to the Dyson propagator, while~$(-+)$ and~$(+-)$
refer to the positive- and negative-frequency Wightman functions, respectively.
Note that distance functions with different~$i\varepsilon$-prescriptions
are not all independent, in particular, 
\begin{equation}
y_{\ss ++} = \theta(\eta\!-\!\eta') \,y_{\ss -+} + \theta(\eta'\!-\!\eta) \, y_{\ss +-} \, .
\label{y decomposition}
\end{equation}
%

\section{Abelian Higgs model in cosmological spaces}
\label{sec: Abelian Higgs model in cosmological spaces}

In this section we consider non-minimally coupled
Abelian Higgs model on an expanding FLRW
background, and derive its canonical formulation. We discuss the 
gauge-fixing procedure according to the Dirac-Bergmann 
algorithm~\cite{Dirac:1950pj,Dirac:1951zz,Anderson:1951ta,Dirac:2011,Gitman:1990qh,Rothe:2010dzf},
and fix the {\it unitary gauge}. We then consider the perturbations in this
model around a background of expanding space-time with an evolving homogeneous and isotropic
scalar condensate.
The Hamiltonian operator for the linearized perturbations is identified,
as well as successively higher order interaction Hamiltonians.

\medskip

The action for the non-minimally coupled Abelian Higgs model we consider,
involving a vector field $A_\mu$ and a complex scalar field~$\Phi$, on a 
general curved background is given by
\begin{equation}
S[A_\mu, \Phi, \Phi^* ] = 
	\! \Lint d^{D\!}x \, \sqrt{-g} \,
	\biggl[
	-\frac{1}{4} g^{\mu\rho} g^{\nu\sigma} F_{\mu\nu} F_{\rho\sigma}
	\!- g^{\mu\nu} \bigl( D_\mu^* \Phi^* \bigr)
		\bigl( D_\nu \Phi \bigr)
	- \lambda \bigl( \Phi^* \Phi \bigr)^2
	\!- \xi R \Phi^* \Phi
	\biggr] 
 ,
\label{SQED action}
\end{equation}
where~$g_{\mu\nu}$ is the metric tensor of the background 
space-time, and~$g\!=\!{\rm det}(g_{\mu\nu})$ is the metric determinant.
Furthermore,~$F_{\mu\nu} \!=\! \partial_\mu A_\nu - \partial_\nu A_\mu$ is the vector 
field strength tensor, $D_\mu \!=\! \partial_\mu +iq A_\mu $ is the~$U(1)$
covariant derivative,~$q$ is the~$U(1)$ charge,~$\lambda$ is the 
scalar quartic self-coupling constant, and~$\xi$ is the dimensionless 
non-minimal coupling constant between the scalar and the Ricci scalar~$R$.
This action is invariant under the~$U(1)$ gauge transformations,
\begin{equation}
A_\mu \rightarrow A_\mu + \partial_\mu \Lambda \, ,
\qquad \qquad
\Phi \rightarrow e^{-iq\Lambda} \Phi \, ,
\qquad \qquad
\Phi^* \rightarrow e^{iq\Lambda} \Phi^* 
\, ,
\end{equation}
where $\Lambda$ is an arbitrary space-time dependent function.
The action~(\ref{SQED action}) specialized to the FLRW background reads,
\begin{align}
S[A_\mu, \Phi, \Phi^* ] ={}& \!
	\Lint d^{D\!}x \, a^{D-4} 
	\biggl[
	- \frac{1}{4} \eta^{\mu\rho} \eta^{\nu\sigma} F_{\mu\nu} F_{\rho\sigma}
	- a^2\eta^{\mu\nu} \bigl( D_\mu^* \Phi^* \bigr)
		\bigl( D_\nu \Phi \bigr)
\nonumber \\
&	\hspace{3cm}
	-  \lambda a^4   \bigl( \Phi^* \Phi \bigr)^2
	- \xi (D\!-\!1) (D\!-\!2\epsilon) a^2 \mathcal{H}^2  \Phi^* \Phi
	\biggr] \, .
\label{SQED in FLRW}
\end{align}
This model is expected to share some similarities with the Abelian Higgs model in 
flat space despite not containing the mass term, since in~$D\!=\!4$ 
the rescaled field~$\Phi/a$ is described by precisely the flat space action for
the Abelian Higgs model with a time-dependent 
mass~$\mathcal{M}^2 \!=\! (6 \xi\!-\!1) (2\!-\!\epsilon) \mathcal{H}^2$, which
encodes the cosmological expansion.

\subsection{Hamiltonian formulation and gauge-fixing}
\label{subsec: Hamiltonian formulation}

The Hamiltonian formulation of the Abelian Higgs model in~(\ref{SQED in FLRW}), 
and the subsequent imposition of the unitary gauge, is more transparent if we 
split the complex scalar into its real and imaginary 
components,~\footnote{ 
We employ the Cartesian coordinates for the complex scalar when discussing the unitary
gauge, rather than the more customary polar coordinates. This is to emphasize that the singularity at
the origin of the field space is not an artifact of the coordinate system used, but rather is a feature of the gauge condition itself.
}
\begin{equation}
\Phi = \frac{1}{\sqrt{2}} \bigl( \phi + i \chi \bigr) \, ,
\qquad \qquad
\Phi^* = \frac{1}{\sqrt{2}} \bigl( \phi - i \chi \bigr) \, ,
\label{complex field decomposition}
\end{equation}
in terms of which the FLRW action~(\ref{SQED in FLRW}) reads,
\begin{align}
S[A_\mu,\phi, \chi ] ={}& \!
	\Lint d^{D\!}x \, a^{D-4} 
	\biggl[
	- \frac{1}{4} \eta^{\mu\rho} \eta^{\nu\sigma} F_{\mu\nu} F_{\rho\sigma}
	- \frac{a^2}{2} \eta^{\mu\nu} \bigl( \partial_\mu \phi - q A_\mu \chi \bigr)
		\bigl( \partial_\nu \phi - q A_\nu \chi \bigr)
\nonumber \\
&	\hspace{1.cm}
	- \frac{a^2}{2} \eta^{\mu\nu} \bigl( \partial_\mu \chi + q A_\mu \phi \bigr)
		\bigl( \partial_\nu \chi + q A_\nu \phi \bigr)
	-   \frac{\lambda a^4 }{4} \bigl( \phi^2 \!+\! \chi^2 \bigr)^2
\nonumber \\
&
	- \frac{\xi}{2} (D\!-\!1) (D\!-\!2\epsilon) a^2 \mathcal{H}^2  \bigl( \phi^2 \!+\! \chi^2 \bigr)
	\biggr] \, .
\end{align}
When deriving the Hamiltonian formulation, it is also advantageous to decompose the
indices into temporal and spatial 
ones,~$A_\mu \!=\! (A_0, A_i)$ and~$\partial_\mu \!=\! (\partial_0, \partial_i)$,
\begin{align}
S[A_0,A_i, \phi, \chi ] ={}& \!
	\Lint d^{D\!}x \, a^{D-4} 
	\biggl[
	\frac{1}{2} F_{0i} F_{0i}
	- \frac{1}{4} F_{ij} F_{ij}
	+ \frac{a^2}{2} \bigl( \partial_0 \phi \!-\! q A_0 \chi \bigr)^2
	+ \frac{a^2}{2} \bigl( \partial_0 \chi \!+\! q A_0 \phi \bigr)^2
\nonumber \\
&	\hspace{-1cm}
	- \frac{a^2}{2} \bigl( \partial_i \phi \!-\! q A_i \chi \bigr)
		\bigl( \partial_i \phi \!-\! q A_i \chi \bigr)
	- \frac{a^2}{2} \bigl( \partial_i \chi \!+\! q A_i \phi \bigr)
		\bigl( \partial_i \chi \!+\! q A_i \phi \bigr)
\\
&	\hspace{-2cm}
	-   \frac{\lambda a^4 }{4} \bigl( \phi^2 \!+\! \chi^2 \bigr)^2
	- \frac{\xi}{2} (D\!-\!1) (D\!-\!2\epsilon) a^2 \mathcal{H}^2  \bigl( \phi^2 \!+\! \chi^2 \bigr)
	\biggr] \, ,
\nonumber
\end{align}
and since $\eta_{ij} \!=\! \delta_{ij}\!=\!\eta^{ij}$, 
there is no need to distinguish between 
raised and lowered indices. Henceforth we write all the quantities 
with lower indices, where the summation over repeated indices is implied.

Next we introduce the extended action~$\mathcal{S}$,
which is only linear in first time derivatives. This is accomplished by
promoting time derivatives to independent velocity 
fields,
\begin{equation}
\partial_0 A_0 \rightarrow V_0 \, ,
\qquad 
F_{0i} \rightarrow V_i \, ,
\qquad 
\bigl( \partial_0 \phi \!-\!  q A_0 \chi\bigr) \rightarrow v \, ,
\qquad 
\bigl( \partial_0 \chi \!+\!  q A_0 \phi \bigr) \rightarrow w \, , 
\end{equation}
and introducing the accompanying Lagrange 
multipliers,~$\Pi_0$, $\Pi_i$, $\pi$, and $\rho$, which ensure on-shell equivalence 
to the action~$S$ we started with,
\begin{align}
\MoveEqLeft[2]
\mathcal{S}\bigl[ A_0, V_0, \Pi_0, A_i, V_i, \Pi_i,
		\phi, v, \pi, \chi, w, \rho \bigr]
		=
			\Lint d^{D\!}x \, \Biggl\{ a^{D-4} 
	\biggl[
	\frac{1}{2} V_i V_i
	- \frac{1}{4} F_{ij} F_{ij}
\nonumber \\
&
	+ \frac{a^2}{2} \bigl( v^2 \!+\! w^2 \bigr)
	- \frac{a^2}{2} \bigl( \partial_i \phi \!-\! q A_i \chi \bigr)
		\bigl( \partial_i \phi \!-\! q A_i \chi \bigr)
	- \frac{a^2}{2} \bigl( \partial_i \chi \!+\! q A_i \phi \bigr)
		\bigl( \partial_i \chi \!+\! q A_i \phi \bigr)
\nonumber \\
&	
	-   \frac{\lambda a^4 }{4} \bigl( \phi^2 \!+\! \chi^2 \bigr)^2
	- \frac{\xi}{2} (D\!-\!1) (D\!-\!2\epsilon) a^2 \mathcal{H}^2  \bigl( \phi^2 \!+\! \chi^2 \bigr)
	\biggr] 
	+ \Pi_0 \bigl( \partial_0 A_0 \!-\! V_0 \bigr)
\nonumber \\
&	
	+ \Pi_i \bigl( F_{0i} \!-\! V_i \bigr)
	+ \pi \bigl( \partial_0 \phi \!-\! q A_0 \chi \!-\! v \bigr)
	+ \rho \bigl( \partial_0 \chi \!+\! q A_0 \phi \!-\! w \bigr)
	\Biggr\} \, .
\label{extended SQED action}
\end{align}
A variation of the extended action
with respect to the velocity fields
results in algebraic on-shell relations.~\footnote{
Note that we are employing Dirac's notation, which distinguishes 
{\it off-shell} (strong) equalities denoted by the ``='' symbol, and
{\it on-shell} (weak) equalities denoted by the ``$\approx$'' symbol.} We solve these for as many velocities 
as possible, which in this case means all except $V_0$,
\begin{align}
&
\frac{\delta \mathcal{S}}{\delta V_i} =
	a^{D-4} V_i - \Pi_i \approx 0
&\Rightarrow\qquad\qquad&
V_i \approx \overline{V}_i = a^{4-D} \Pi_i \, ,
\\
&
\frac{\delta \mathcal{S}}{\delta v} =
	a^{D-2} v - \pi \approx 0
&\Rightarrow\qquad\qquad&
v \approx \overline{v} = a^{2-D} \pi \, ,
\\
&
\frac{\delta \mathcal{S}}{\delta w} =
	a^{D-2} w - \rho \approx 0
&\Rightarrow\qquad\qquad&
w \approx \overline{w}= a^{2-D} \rho \, .
\end{align}
These on-shell relations are then plugged into the extended 
action~(\ref{extended SQED action}) as off-shell
equalities, which results in the canonical action~$\mathscr{S}$,
\begin{align}
\MoveEqLeft[4]
\mathscr{S} \bigl[ A_0, \Pi_0, A_i, \Pi_i,
		\phi, \pi, \chi, \rho, V_0 \bigr]
	\equiv \mathcal{S}\bigl[ A_0, V_0, \Pi_0, A_i, \overline{V}_i, \Pi_i,
		\phi, \overline{v}, \pi, \chi, \overline{w}, \rho \bigr]
\nonumber \\
&	=  \int\! d^{D\!}x \, \Bigl[ \Pi_0 \partial_0 A_0
	+ \Pi_i \partial_0 A_i + \pi \partial_0 \phi + \rho \partial_0 \chi
	- \mathscr{H} - V_0 \Psi_1 \Bigr] \, ,
\label{canonical action}
\end{align}
where the Hamiltonian density is
\begin{align}
\MoveEqLeft[2]
\mathscr{H} =
	\frac{a^{4-D}}{2} \Pi_i \Pi_i 
	+ \Pi_i \partial_i A_0 
	+ \frac{a^{D-4}}{4} F_{ij} F_{ij}
	+ \frac{a^{2-D}}{2} \bigl( \pi^2 \!+\! \rho^2 \bigr)
	+ q A_0 \bigl( \pi \chi - \rho \phi \bigr)
\nonumber \\
&	+ \frac{a^{D-2}}{2} \bigl( \partial_i \phi - q A_i \chi \bigr) 
		\bigl( \partial_i \phi - q A_i \chi \bigr)
	+ \frac{a^{D-2}}{2} \bigl( \partial_i \chi + q A_i \phi \bigr) 
		\bigl( \partial_i \chi + q A_i \phi \bigr)
\nonumber \\
&	+ \frac{\lambda a^D }{4} \bigl( \phi^2 \!+\! \chi^2 \bigr)^{\!2}
	+ \frac{\xi}{2} (D\!-\!1) (D\!-\!2\epsilon) a^{D-2} \mathcal{H}^2 
		\bigl( \phi^2 \!+\! \chi^2 \bigr) \, ,
\label{Hamiltonian density: 2}
\end{align}
and $\Psi_1 \!=\! \Pi_0$ is the {\it primary constraint}.
Note that in the canonical action~$V_0$ has an interpretation
of a Lagrange multiplier, as the variation
with respect to it generates the primary constraint.
The Poisson brackets follow from the symplectic part of the canonical 
action~(\ref{canonical action}),
the non-vanishing ones being,
\begin{align}
&
\bigl\{ A_0(\eta,\vec{x}) , \Pi_0(\eta,\vec{x}^{\,\prime}) \bigr\}
	= \delta^{D-1}(\vec{x} \!-\! \vec{x}^{\,\prime}) \, ,
&&
\bigl\{ A_i(\eta,\vec{x}) , \Pi_j(\eta,\vec{x}^{\,\prime}) \bigr\}
	= \delta_{ij} \delta^{D-1}(\vec{x} \!-\! \vec{x}^{\,\prime}) \, ,
\nonumber \\
&
\bigl\{ \phi(\eta,\vec{x}) , \pi(\eta,\vec{x}^{\,\prime}) \bigr\}
	= \delta^{D-1}(\vec{x} \!-\! \vec{x}^{\,\prime}) \, ,
&&
\bigl\{ \chi(\eta,\vec{x}) , \rho (\eta,\vec{x}^{\,\prime}) \bigr\}
	= \delta^{D-1}(\vec{x} \!-\! \vec{x}^{\,\prime}) \, . \quad
\label{Poisson}
\end{align}
Note that the Lagrange multiplier~$V_0$ is defined to have a vanishing Poisson bracket 
with all the canonical fields.
The dynamics is generated by the total Hamiltonian,
\begin{equation}
{\rm H}_{\rm tot} = \int\! d^{D-1}x \, \mathscr{H}_{\rm tot} \, ,
\qquad \qquad
\mathscr{H}_{\rm tot} = \mathscr{H} + V_0 \Psi_1 \, ,
\qquad
\end{equation}
producing the Hamilton's equations,
\begin{align}
&
\partial_0 A_0 \approx V_0 \, ,
\label{eom1}
\\
&
\partial_0 \Pi_0 \approx 
	\partial_i \Pi_i - q \bigl( \pi \chi - \rho \phi \bigr) \, ,
\label{eom2}
\\
&
\partial_0 A_i \approx 
	a^{4-D} \Pi_i + \partial_i A_0 \, ,
\label{eom3}
\\
&
\partial_0 \Pi_i \approx
	a^{D-4} \partial_j F_{ji}
	+ a^{D-2} q \bigl( \chi \partial_i \phi - \phi \partial_i \chi \bigr)
	- a^{D-2} q^2 \bigl( \phi^2 \!+\! \chi^2 \bigr) A_i \, ,
\label{eom4}
\\
&
\partial_0 \phi \approx
	a^{2-D} \pi + q A_0 \chi \, ,
\label{eom5}
\\
&
\partial_0 \pi \approx
	q A_0 \rho + a^{D-2} \bigl( \nabla^2 \phi 
		- 2 q A_i \partial_i \chi 
		- q \chi \partial_i A_i 
		- q^2 \phi A_i A_i \bigr)
\nonumber \\
&	\hspace{4.5cm}
	- \lambda a^D \bigl( \phi^2 \!+\! \chi^2 \bigr) \phi
	- \xi (D\!-\!1) (D\!-\!2\epsilon) a^{D-2} \mathcal{H}^2 \phi \, ,
\label{eom6}
\\
&
\partial_0 \chi \approx
	a^{2-D} \rho
	- q A_0 \phi \, ,
\label{eom7}
\\
&
\partial_0 \rho \approx
	- q A_0 \pi
	+ a^{D-2} \bigl( \nabla^2 \chi + 2 q A_i \partial_i \phi
		+ q \phi \partial_i A_i - q^2 \chi A_i A_i \bigr)
\nonumber \\
&	\hspace{4.5cm}
	- \lambda a^D \bigl( \phi^2 \!+\! \chi^2 \bigr) \chi
	- \xi (D\!-\!1)(D\!-\!2\epsilon) a^{D-2} \mathcal{H}^2 \chi \, ,
\label{eom8}
\end{align}
which are to be supplemented with the {\it primary constraint},
\begin{equation}
\Psi_1 = \Pi_0 \approx 0 \, .
\label{Psi1}
\end{equation}
Note that these are just the equations following from the canonical 
action~(\ref{canonical action}), and that the Poisson brackets are defined 
in a way that reproduces them.
The consistency of this primary constraint, {\it i.e.} its conservation
in time, generates a {\it secondary constraint},
\begin{equation}
\partial_0 \Psi_1 \approx
	\partial_i \Pi_i - q \bigl( \pi \chi - \rho \phi \bigr)
\qquad \quad
\Rightarrow 
\qquad \quad
\Psi_2 = \partial_i \Pi_i - q \bigl( \pi \chi - \rho \phi \bigr)   \approx 0 \, .
\label{Psi2}
\end{equation}
The conservation of the secondary constraint generates no further constraints,
rather it is conserved identically on-shell. All the Poisson brackets 
between the primary and the secondary constraints vanish,
\begin{equation}
\bigl\{ \Psi_I(\eta,\vec{x}) ,
	\Psi_J(\eta,\vec{x}^{\,\prime}) \bigr\} = 0 \, ,
\qquad \qquad I,J=1,2 \, ,
\end{equation}
and thus they form a complete set of first-class constraints.

In order to fully determine the dynamics of the system additional conditions have to 
be supplied by hand. Namely, we need two additional 
constraints,~$\Psi_3 \! \approx \! 0$ and~$\Psi_4 \! \approx \! 0$,
such that (i) they form
a complete set of second-class constraints with the two 
first-class constraints~(\ref{Psi1}) and~(\ref{Psi2}),
\begin{equation}
{\rm det} \bigl\{ \Psi_I(\eta,\vec{x}) , \Psi_J(\eta,\vec{x}^{\,\prime}) \bigr\}
	 \not\approx 0 \, ,\qquad \qquad I,J = 1,\dots,4 \, ,
\end{equation}
and that (ii) they are consistent with the equations of motion~(\ref{eom1})--(\ref{eom8}). 
Satisfying the first
condition is straightforward, but the second condition is more subtle, and 
in general does not allow us to chose the two gauge-fixing conditions
independently.
In this work we chose the {\it unitary gauge}, defined by
requiring the scalar to be real,
\begin{align}
\Psi_3 =\frac{1}{\sqrt{2}} \, {\rm Im}(\Phi) =  \chi \approx 0 \, .
\label{Psi3}
\end{align}
It is sufficient to specify just this condition in order to fully fix the gauge.
This is true because the second gauge condition is supplied by the conservation of the first one,
required for consistency with the equations
of motion,
\begin{equation}
\partial_0 \Psi_3 \approx 
	a^{2-D} \rho
	- q A_0 \phi
\qquad \quad
\Rightarrow
\qquad \quad
\Psi_4 =
	a^{2-D} \rho
	- q A_0 \phi 
	\approx 0 \, .
\label{Psi4}
\end{equation}
The conservation of $\Psi_4$ generates no further constraints.
The Poisson brackets between all four constraints ($I,J \!=\! 1,2,3,4$)
form a matrix,
\begin{equation}
\bigl\{ \Psi_I(\eta,\vec{x}) , \Psi_{\! J}(\eta,\vec{x}^{\,\prime}) \bigr\}
	\approx \!
\begin{pmatrix}
0	&	0	&	0	&	q \phi
\\
0	&	0	&	- q \phi	&	-  a^{2-D} q \pi
\\
0	&	q \phi	&	0	&	a^{2-D}
\\
- q \phi	&	\, a^{2-D} q \pi	&	- a^{2-D}	&	0
\end{pmatrix} 
\delta^{D-1}(\vec{x} \!-\! \vec{x}^{\,\prime}) \, ,
\label{matrix of constraints}
\end{equation}
with a non-vanishing determinant,
\begin{equation}
{\rm det} \,\bigl\{ \Psi_I(\eta,\vec{x}) , \Psi_J(\eta,\vec{x}^{\,\prime}) \bigr\}
	\approx ( q \phi )^{4} \times 
	\delta^{D-1}(\vec{x}\!-\! \vec{x}^{\,\prime}) 
\, .
\end{equation}
This matrix is therefore invertible,
\begin{equation}
\bigl\{ \Psi_I(\eta,\vec{x}) , \Psi_{\! J}(\eta,\vec{x}^{\,\prime}) \bigr\}^{\! -1} \!
	\approx  \!
\begin{pmatrix}
0	&	a^{2-D}	&	a^{2-D} q\pi	&	- q \phi
\\
- a^{2-D}	&	0	&	q \phi	&	0
\\
- a^{2-D} q\pi	&	- q \phi&	0	&	0
\\
q \phi	&	0	&	0	&	0
\end{pmatrix} \!
\frac{ 1 }{ ( q \phi)^2} \, \delta^{D-1}(\vec{x} \!-\! \vec{x}^{\,\prime}) \, ,
\end{equation}
and thus the four constraints -- two first-class ones and two gauge conditions --
form a complete set of second-class constraints.
It is important to note that the unitary gauge is singular for $\phi\!=\!0$.
In a quantized theory this is known to lead to some
non-perturbative illnesses of the gauge~\cite{Anderberg:2008gs}. 
However, for situations where $\phi$ develops a non-vanishing
classical condensate, as is the case considered in this work, the unitary
gauge suffices.

Having specified the gauge conditions, we may proceed, according 
to~\cite{Dirac:2011,Gitman:1990qh},
to determine the Dirac 
brackets,~\footnote{
We denote the Dirac
brackets by $\bigl( \hspace{-0.7mm}\bigr| \cdot \, , \, \cdot  \bigl| \hspace{-0.7mm}\bigr)$
instead of the common symbol $\bigl\{ \cdot \, , \, \cdot \bigr\}_{\rm  D}$, 
which we find less distinguished and aesthetically less pleasing.}
which are defined for two functions of canonical variables as,
\begin{align}
\MoveEqLeft[1]
\dl {\rm A}(\eta, \vec{x}) , {\rm B}(\eta, \vec{x}^{\,\prime}) \dr =
	\bigl\{ {\rm A}(\eta, \vec{x}) , {\rm B}(\eta, \vec{x}^{\,\prime}) \bigr\}
\\
&	- \int\! d^{D-1} z \, d^{D-1}z' \,
	\bigl\{ {\rm A}(\eta, \vec{x}) , \Psi_I(\eta,\vec{z}) \bigr\}
	\bigl\{ \Psi_I(\eta,\vec{z}) , \Psi_J(\eta,\vec{z}^{\,\prime}) \bigr\}^{-1}
	\bigl\{ \Psi_J(\eta,\vec{z}^{\,\prime} ) , {\rm B}(\eta,\vec{x}^{\,\prime}) \bigr\} \, ,
\nonumber 
\end{align}
where the summation over the repeated constraint indices $I,J$ is implied.
The Dirac brackets between the canonical variables for
 the case at hand  are given in Table~\ref{Dirac brackets}.
\begin{table}[h!]
\renewcommand{\arraystretch}{1.7}
\begin{tabular}{| >{\centering}p{0.9cm} | >{\centering}p{1.3cm} >{\centering}p{1.3cm} >{\centering}p{1.3cm} >{\centering}p{1.3cm} >{\centering}p{1.3cm} >{\centering}p{1.3cm} >{\centering}p{1.3cm} >{\centering\arraybackslash}p{1.3cm} | }
\hline
	&	$A_0(x')$
	&	$\Pi_0(x')$
	&	$A_j(x')$
	&	$\Pi_j(x')$
	&	$\phi(x')$
	&	$\pi(x')$
	&	$\chi(x')$
	&	$\rho(x')$
\\
\hline
$A_0(x)$
	&	$0$
	&	$0$
	&	$\frac{a^{2-D}}{(q\phi)^2} \partial_j$
	&	$0$
	&	$0$
	&	$\frac{2a^{2-D} \partial_i \Pi_i}{q^2 \phi^3}$
	&	$0$
	&	$0$
\\
$\Pi_0(x)$
	&	$0$
	&	$0$
	&	$0$
	&	$0$
	&	$0$
	&	$0$
	&	$0$
	&	$0$
\\
$A_i(x)$
	&	$\partial_i \frac{a^{2-D}}{(q\phi)^2}$
	&	$0$
	&	$0$
	&	$\delta_{ij}$
	&	$0$
	&	$0$
	&	$0$
	&	$\partial_i \frac{1}{q\phi}$
\\
$\Pi_i(x)$
	&	$0$
	&	$0$
	&	$- \delta_{ij}$
	&	$0$
	&	$0$
	&	$0$
	&	$0$
	&	$0$
\\
$\phi(x)$ 
	&	$0$
	&	$0$
	&	$0$
	&	$0$
	&	$0$
	&	$1$
	&	$0$
	&	$0$
\\
$\pi(x)$
	&	$\ - \frac{2a^{2-D} \partial_i \Pi_i}{q^2 \phi^3}$
	&	$0$
	&	$0$
	&	$0$
	&	$-1$
	&	$0$
	&	$0$
	&	$- \frac{\partial_i \Pi_i}{q \phi^2}$
\\
$\chi(x)$
	&	$0$
	&	$0$
	&	$0$
	&	$0$
	&	$0$
	&	$0$
	&	$0$
	&	$0$
\\
$\rho(x)$
	&	$0$
	&	$0$
	&	$\frac{1}{q\phi} \partial_i $
	&	$0$
	&	$0$
	&	$\frac{\partial_i \Pi_i}{q \phi^2}$
	&	$0$
	&	$0$
\\
\hline
\end{tabular}
\caption{
Coefficient operators~${\rm C}_{m,n}$ from Dirac brackets between canonical fields,
$\dl {\rm A}_m(\eta, \vec{x}) , {\rm B}_n(\eta, \vec{x}^{\,\prime}) \dr = 
	{\rm C}_{m,n} \!\times i \delta^{D-1}(\vec{x} \!-\! \vec{x}^{\,\prime}) $,
where~${\rm A}_m$ is the entry from the leftmost column, ${\rm B}_n$
the entry from the topmost row,
and~${\rm C}_{m,n}$ are the entries from the table. Unprimed derivatives are with respect to
unprimed coordinates, and act on everything to the right of them.}
\label{Dirac brackets}
\end{table}
They serve the same role as Poisson brackets,
with an important difference that the constraints may be imposed as off-shell 
equalities before computing any of the brackets, which corresponds to solving for the
constraints. We use the four 
conditions~(\ref{Psi1}),~(\ref{Psi2}),~(\ref{Psi3}), and~(\ref{Psi4}) to solve
for four canonical fields,
\begin{equation}
\Pi_0 = 0 \, ,
\qquad \quad
\rho = - \frac{(\partial_i \Pi_i)}{q \phi } \, ,
\qquad \quad
\chi = 0 \, ,
\qquad \quad
A_0 = - \frac{a^{2-D} (\partial_i \Pi_i)}{(q \phi)^2 } \, ,
\label{elimination}
\end{equation}
so that we can define the {\it reduced phase space} Hamiltonian,
\begin{align}
\MoveEqLeft[2]
{\rm H}_{\rm red} = \! \Lint d^{D-1}x \, \biggl[
	\frac{a^{4-D}}{2} \Pi_i \Pi_i 
	+ \frac{a^{2-D}}{2} \frac{(\partial_i \Pi_i) (\partial_j \Pi_j) }{(q\phi)^2}
	+ \frac{a^{D-4}}{4} F_{ij} F_{ij}
	+\frac{a^{D-2}}{2} (q\phi)^2 A_i A_i
\nonumber \\
&	\hspace{0.cm}
	+ \frac{a^{2-D}}{2} \pi^2 
	+ \frac{a^{D-2}}{2} (\partial_i \phi) (\partial_i \phi)
	+ \frac{\lambda a^D }{4} \phi^4
	+ \frac{\xi}{2} (D\!-\!1) (D\!-\!2\epsilon) a^{D-2} \mathcal{H}^2 \phi^2 \biggr] 
\, ,
\label{H red}
\end{align}
When relations~(\ref{elimination}) are inserted into the Dirac brackets
from Table~\ref{Dirac brackets}, 
one obtains the reduced phase space Dirac brackets which, together with the Hamiltonian
${\rm H}_{\rm red}$ in~(\ref{H red}), generate the reduced phase space dynamics.

\subsection{Perturbations}
\label{subsec: Perturbations}

We are interested in studying the (quantum) fluctuations in the
Abelian Higgs model in a situation where the scalar field
develops
a non-vanishing time-dependent condensate $\overline{\phi}(\eta)$.
In the canonical formulation, the scalar conjugate momentum then also has a 
non-vanishing condensate~$\overline{\pi}(\eta)$, since they both have to
satisfy Hamilton's equations generated by the reduced phase space Hamiltonian~(\ref{H red}),
\begin{equation}
\partial_0 \overline{\phi} = a^{2-D} \overline{\pi} \, ,
\qquad \qquad
\partial_0 \overline{\pi} 
= - a^D \lambda \,\overline{\phi}^3
	- \xi (D\!-\!1) (D\!-\!2\epsilon) a^{D-2} \mathcal{H}^2 \overline{\phi} \, .
\label{background equations}
\end{equation}
No such condensate is assumed for the vector field, even though those
have been considered in literature, most notably in the context of vector inflationary 
models~\cite{Golovnev:2008cf,Golovnev:2008hv,Dimopoulos:2006ms,Dimopoulos:2008yv}.
The Hamiltonian encoding the dynamics of the fluctuations around such a
classical configuration is obtained by shifting the variables of the scalar field
and its conjugate momentum by their condensates,
\begin{equation}
\phi(\eta,\vec{x}) = \overline{\phi}(\eta) + \varphi(\eta,\vec{x}) \, ,
\qquad \qquad
\pi(\eta,\vec{x}) = \overline{\pi}(\eta) + p(\eta,\vec{x}) \, ,
\label{defining condensate fields}
\end{equation}
which then naturally organizes the reduced Hamiltonian according to the powers of 
the fluctuation fields,
\begin{align}
{\rm H}_{\rm red} = {\rm H}^{\ss (0)}  +
	{\rm H}^{\ss (2)} + {\rm H}^{\ss (3)} + {\rm H}^{\ss (4)} + \dots \, ,
\label{H perts}
\end{align}
where ${\rm H}^{\ss (0)}$ denotes 
the Hamiltonian for the condensate.
At the linear level (quadratic Hamiltonian) the scalar and the vector field perturbations
decouple,
\begin{equation}
{\rm H}^{\ss (2)} =  {\rm H}^{\ss (2)}_\varphi + {\rm H}^{\ss (2)}_A \, ,
\end{equation}
where the scalar part is
\begin{equation}
{\rm H}^{\ss (2)}_\varphi = \Lint d^{D-1}x \, \biggl[
	\frac{a^{2-D}}{2} p^2
	+ \frac{a^{D-2}}{2} (\partial_i \varphi) (\partial_j \varphi)
	+ \frac{a^D}{2} \Bigl( 3\lambda \overline{\phi}^{\,2}
		+ \xi (D\!-\!1) (D\!-\!2\epsilon) H^2 \Bigr) \varphi^2
	\biggr] \, ,
\label{H phi}
\end{equation}
and the vector part is
\begin{equation}
{\rm H}^{\ss (2)}_A = \Lint d^{D-1}x \, \biggl[
	\frac{a^{4-D}}{2} \Pi_i \Pi_i
	+ \frac{a^{2-D}}{2 (q\overline{\phi})^2 } (\partial_i \Pi_i) (\partial_j \Pi_j)
	+ \frac{a^{D-4}}{4} F_{ij} F_{ij}
	+ \frac{a^{D-2}}{2}  (q\overline{\phi})^2 A_i A_i
	\biggr] \, .
\label{H A}
\end{equation}
At higher orders the perturbations couple.
For the cubic Hamiltonian we have,
\begin{equation}
{\rm H}^{\ss(3)} = \Lint d^{D-1}x \, \biggl[
	- \frac{ a^{2-D} }{(q\overline{\phi})^2} 
		\Bigl( \frac{\varphi}{\overline{\phi}} \Bigr) (\partial_i \Pi_i ) (\partial_j \Pi_j)
	+ a^{D-2} (q\overline{\phi})^2 \Bigl( \frac{\varphi}{\overline{\phi} } \Bigr) A_i A_i
	+ a^D \lambda \overline{\phi} \, \varphi^3
	\biggr] \, ,
\label{cubic interactions in unitary gauge}
\end{equation}
for the quartic one,
\begin{equation}
{\rm H}^{\ss(4)} = \Lint d^{D-1}x \, \biggl[
        \frac{3}{2} \frac{a^{2-D}}{(q\overline{\phi})^2} 
        	\Bigl( \frac{\varphi}{\overline{\phi} } \Bigr)^{\!2} 
        		(\partial_i \Pi_i ) (\partial_j \Pi_j)
	+ \frac{a^{D-2}}{2} q^2 \varphi^2 A_i A_i
	+ \frac{a^D}{4} \lambda \varphi^4
	\biggr] \, ,
\label{quartic interactions in unitary gauge}
\end{equation}
and for the higher order ones,
\begin{equation}
{\rm H}^{\ss(n)} = \Lint d^{D-1}x \, \biggl[
	\frac{(-1)^n}{2}(n\!-\!1)  \frac{a^{2-D}}{(q\overline{\phi})^{2}} 
	\Bigl( \frac{\varphi}{\overline{\phi} } \Bigr)^{\!n} 
	             (\partial_i \Pi_i ) (\partial_j \Pi_j)
	\biggr] \, ,
\qquad \quad (n\geq 5)
\, .
\label{higher order interactions in unitary gauge}
\end{equation}
It is worth noting that all orders of the reduced phase space Hamiltonians above
are reproduced by the following action.
\begin{align}
S\bigl[ A_\mu, \phi \bigr] ={}&
	 \Lint d^{D\!}x \, \sqrt{-g} \, \biggl[
	- \frac{1}{4} g^{\mu\rho} g^{\nu\sigma} F_{\mu\nu} F_{\rho\sigma}
	- \frac{1}{2} (q\phi)^2 g^{\mu\nu} A_\mu A_\nu
\nonumber \\
&	\hspace{4cm}
	- \frac{1}{2} g^{\mu\nu} (\partial_\mu \phi) (\partial_\nu \phi)
	- \frac{\lambda}{4} \phi^4
	- \frac{\xi}{2} R \phi^2
	\biggr] \, ,
\label{dS unitary gauge}
\end{align}
after expanding it in the perturbations,~$\phi\!=\!\overline{\phi} \!+\! \varphi$.
The infinite tower of interactions~(\ref{higher order interactions in unitary gauge}) 
between the physical degrees of freedom in the canonical formulation
is thus translated into a finite number of interactions in the gauge-fixed covariant
formulation, at the expense of re-introducing what is now effectively an auxiliary
field,~$A_0\!=\!-a^{2-D}(\partial_i \Pi_i)/(q\phi)^2$. This auxiliary field is encoded in the
Proca-like action above as a second-class constraint.


\subsection{Quantization}
\label{subsec: Quantization}

The quantization of the dynamics is performed in the reduced phase 
space formulation in the unitary gauge in three steps:
\begin{itemize}
\item[(i)]
All the real field perturbations are promoted to Hermitian field operators,
\begin{align}
&
\varphi(x) \rightarrow \hat{\varphi}(x) \, ,
&&
\hat{\varphi}^{\,\dag}(x) = \hat{\varphi}(x) \, ,
\label{promotion 1}
\\
&
p(x) \rightarrow \hat{p}(x) \, ,
&&
\hat{p}^{\,\dag}(x) = \hat{p}(x) \, ,
\\
&
A_i(x) \rightarrow \hat{A}_i(x) \, ,
&&
\hat{A}_i^{\,\dag}(x) = \hat{A}_i(x) \, ,
\\
&
\Pi_i(x) \rightarrow \hat{\Pi}_i(x) \, ,
&&
\hat{\Pi}_i^{\,\dag}(x) = \hat{\Pi}_i(x) \, .
\qquad
\label{promotion 4}
\end{align}

\item[(ii)]
The Dirac brackets are promoted to commutators,
$\bigl( \hspace{-0.7mm}\bigr| \cdot \, , \, \cdot  \bigl| \hspace{-0.7mm}\bigr) \rightarrow (- i) \, \bigl[ \, \cdot \, , \, \cdot \, \bigr]$, and from the Table~\ref{Dirac brackets}
we see that the non-vanishing ones are,
\begin{equation}
\bigl[ \hat{\phi}(\eta,\vec{x}) ,  \hat{p}(\eta,\vec{x}^{\,\prime})\bigr]
	= i \delta^{D-1}(\vec{x} \!-\! \vec{x}^{\,\prime}) \, ,
\qquad
\bigl[ \hat{A}_i(\eta,\vec{x}) ,  \hat{\Pi}_j(\eta,\vec{x}^{\,\prime})\bigr]
	= \delta_{ij} \, i \delta^{D-1}(\vec{x} \!-\! \vec{x}^{\,\prime}) \, .
\label{fundamental commutators}
\end{equation}

\item[(iii)]
The dynamics is given by the quantum Hamiltonian operator, which is obtained
from the classical Hamiltonian~(\ref{H perts}) by substituting 
classical fields by field operators as 
in~(\ref{promotion 1})--(\ref{promotion 4}), and 
symmetrizing the products of non-commuting field operators 
if necessary.
Here we contend ourselves with studying just the linear fluctuations
whose dynamics is given by Hamiltonians~(\ref{H phi}) and~(\ref{H A}), 
whose quantized versions are,
\begin{equation}
\hat{\rm H}^{\ss (2)}_{\varphi} = \! \Lint d^{D-1}x \, \biggl[
	\frac{a^{2-D}}{2} \hat{p}^{\,2} 
	+ \frac{a^{D-2}}{2} ( \partial_i \hat{\varphi} \, )
		( \partial_i \hat{\varphi} \, )
	+ \frac{a^D}{2} \Bigl( 3\lambda \overline{\phi}^{\,2}
		+ \xi (D\!-\!1) (D\!-\!2\epsilon) H^2 \Bigr)  \hat{\varphi}^{\,2}
	\biggr] \, ,
\label{H2 s}
\end{equation}
for the scalar sector, and
\begin{equation}
\hat{\rm H}^{\ss (2)}_A = \! \Lint d^{D-1}x \, \biggl[
	\frac{a^{4-D}}{2} \hat{\Pi}_i \hat{\Pi}_i 
	+ \frac{a^{2-D}}{2 (q\overline{\phi})^2 }  (\partial_i \hat{\Pi}_i) (\partial_j \hat{\Pi}_j) 
	+ \frac{a^{D-4}}{4} \hat{F}_{ij} \hat{F}_{ij}
	+ \frac{a^{D-2}}{2}  (q\overline{\phi})^2
		\hat{A}_i \hat{A}_i	
	\biggr] \, ,
\label{H2A}
\end{equation}
for the vector sector. Interactions between the fields are then governed by the quantized version
of the interaction 
Hamiltonians~(\ref{cubic interactions in unitary gauge})--(\ref{higher order interactions in unitary gauge}).

\end{itemize}
%


\section{Condensate dynamics }
\label{sec: Condensate dynamics}

In this work we consider the condensate~$\overline{\phi}$ of the complex scalar
to be a spectator, in the sense that it does not influence the space-time dynamics.
The power-law inflation space-time is assumed to be driven by some other 
field.~\footnote{
For example, an inflaton field~$\phi$ with an exponential
potential~$U(\phi)\!=\! U_0 \exp\bigl( - \beta \, \phi/M_{\ss \rm P} \bigr)$
yields,
in its attractor regime, a power-law inflation 
with~$\epsilon\!=\!\beta^2/2$, and~$\phi \!\propto\! \ln(a)$;
see {\it e.g.}~\cite{La:1989za}.
}
The dynamics of the homogeneous and isotropic spectator condensate~$\overline{\phi}$ of
our Abelian Higgs model satisfies first order dynamical 
equations~(\ref{background equations}) in FLRW space-times.
In power-law inflation, where $\epsilon \!=\! \text{const.}$, the two equations
combine into a single second order one,
\begin{equation}
\partial_0^2 \, \overline{\phi} 
	+ (D\!-\!2) \mathcal{H} \partial_0 \overline{\phi}
	+ a^2 \Bigl[ \lambda \overline{\phi}^{\,2 }
		+ \xi (D\!-\!1)(D\!-\!2\epsilon) H^2 \Bigr] \overline{\phi}  = 0 \, .
\end{equation}
We are interested in a particular attractor behaviour, where the
condensate scales as the physical Hubble rate,
\begin{equation}
\overline{\phi} 
	= \biggl( \frac{\overline{\phi}_0}{H_0} \biggr) H
	=\overline{\phi}_0 \, a^{-\epsilon} \, ,
\label{background scaling}
\end{equation}
where the amplitude~$\overline{\phi}_0$ is a constant.
Such attractor solutions in power-law inflation have been studied 
already in the context of a self-interacting scalar~\cite{Janssen:2009pb},
and the Yukawa model~\cite{Koksma:2009tc}.
 The attractor behaviour of 
this solution is manifested in its independence of initial conditions.
The assumed scaling~(\ref{background scaling}) 
yields a cubic equation for the condensate amplitude,
\begin{equation}
\biggl[ 
	- \epsilon (D \!-\!1\!-\!2\epsilon) 
	+ \xi (D\!-\!1)(D\!-\!2\epsilon)
	+ \lambda \biggl( \frac{\overline{\phi}_0}{H_0} \biggr)^{\!\!2} \,
	\biggr] \overline{\phi}_0 = 0 \, ,
\end{equation}
which can be seen as a dynamical generalization of the static situation in 
flat space, where one assumes the condensate is in a global minimum of the 
potential. It has a~{\it symmetric} solution~$\overline{\phi}_0 \!=\! 0$, and
two~{\it symmetry-breaking attractor} solutions,
\begin{equation}
\frac{\overline{\phi}_0}{H_0} = \pm
	\sqrt{ \frac{1}{\lambda} \Bigl[ \epsilon (D \!-\!1\!-\!2\epsilon)
	- \xi (D\!-\!1)(D\!-\!2\epsilon)  \Bigr] } \, .
\label{condensate}
\end{equation}
These symmetry-breaking solutions exist only if the non-minimal 
coupling from the action~(\ref{SQED action}) satisfies,
\begin{equation}
\xi < \frac{\epsilon(D\!-\!1\!-\!2\epsilon)}{(D\!-\!1)(D\!-\!2\epsilon)} \, ,
\label{attractor existence}
\end{equation}
and otherwise only the symmetric solution exists. In the symmetric 
solution the scalar condensate vanishes, which
is a singular point of the unitary gauge, and in the rest of the paper we
assume that the condition~(\ref{attractor existence}) is fulfilled.

\medskip

The remainder of the paper is devoted to computing the dynamics and
two-point functions of scalar and vector perturbations around the 
symmetry-breaking attractor solution~(\ref{condensate}) 
for the scalar condensate in power-law inflation.

\section{Scalar perturbations}
\label{sec: Scalar perturbations}

The dynamics of the linear scalar perturbations is encoded in the Hamiltonian~(\ref{H phi}),
and is fully captured by the scalar two-point function,
\begin{equation}
i \bigl[ \tensor*[^{\ss \! - \!}]{\Delta}{^{\ss \! + \!}} \bigr] (x;x')
	= \bigl\langle \Omega \bigr| \hat{\varphi}(x) \, \hat{\varphi}(x') \bigl| \Omega \bigr\rangle \, .
\end{equation}
Solving for the dynamics of the scalar field operators and computing the
expectation value above would be straightforward. However, it 
becomes even simpler upon realizing
that the Hamiltonian~(\ref{H phi}) arises
from considering a massless non-minimally coupled scalar,~\footnote{
This correspondence no longer holds if the 
 scalar perturbation couples to the inflaton condensate, 
rather than the spectator condensate.
Typically the inflaton scales as~$\overline{\phi}\!\propto\! \ln(a)$,
and not as~$\overline{\phi}\!\propto\! H$ as the spectator does. For an example how such 
cases can be treated semi-analytically 
see {\it e.g.} Ref.~\cite{Prokopec:2003iu}.
}
\begin{equation}
S_\varphi^{\ss (2)}[\varphi] = \Lint d^{D\!}x \, \sqrt{-g} \,
	\biggl[
	- \frac{1}{2} g^{\mu\nu} (\partial_\mu \varphi) (\partial_\nu \varphi)
	- \frac{1}{2} \xi_{\ss \rm S} R \, \varphi^2 
	\biggr] \, ,
\label{equivalent scalar action}
\end{equation}
in power-law inflationary space-time, with the effective non-minimal coupling given by,
\begin{equation}
\xi_{\ss \rm S} = 
	\xi 
	+ \frac{3\lambda }{(D\!-\!1)(D\!-\!2\epsilon) } \biggl( \frac{\overline{\phi}_0}{H_0} \biggr)^{\!\!2}
	= - 2 \xi + \frac{3\epsilon(D\!-\!1\!-\!2\epsilon)}{(D\!-\!1)(D\!-\!2\epsilon)} \, .
\label{non-minimal coupling}
\end{equation}
Therefore, the dynamics of linear scalar perturbations in our Abelian Higgs model
in power-law inflation is 
equivalent to the massless non-minimally coupled scalar with the non-minimal 
coupling~$\xi_{\rm \ss S}$ in~(\ref{non-minimal coupling}).
The two-point functions for such a scalar with respect to the
Chernikov-Tagirov-Bunch-Davies (CTBD) state~\cite{Chernikov:1968zm,Bunch:1978yq}, 
{\it i.e.} the state that minimizes energy
mode-per-mode in the asymptotic past, has been worked out 
in~\cite{Janssen:2008px,Janssen:2008dw}, and we simply quote it here,
\begin{equation}
i \bigl[ \tensor*[^{\ss \! A \!}]{\Delta}{^{\ss \! B \!}} \bigr] (x;x')
	= (aa')^{-\frac{(D-2)\epsilon}{2}} \times \mathcal{F}_{\nu_{\ss \rm S}} (y_{\ss AB}) \, ,
	\qquad\qquad
	 (A,B= +,-) \, ,
\label{scalar 2-pt}
\end{equation}
where we define the {\it rescaled propagator function},
\begin{equation}
\mathcal{F}_\lambda(y) =
	\frac{[(1\!-\!\epsilon)H_0]^{D-2}}{(4\pi)^{D/2}}
	\frac{\Gamma\bigl( \tfrac{D-1}{2} \!+\! \lambda \bigr) \, 
			\Gamma\bigl( \tfrac{D-1}{2} \!-\! \lambda \bigr)}
		{\Gamma\bigl( \tfrac{D}{2} \bigr)} \times
	{}_2F_1 \Bigl( \bigl\{\tfrac{D-1}{2} \!+\! \lambda , \tfrac{D-1}{2} \!-\! \lambda \bigr\} , 
		\bigl\{ \tfrac{D}{2} \bigr\} , 1 \!-\! \tfrac{y}{4} \Bigr) \, ,
\label{F def}
\end{equation}
whose properties are given in 
Appendix~\ref{app: Rescaled propagator function F_nu}.
The argument~$y$ of the rescaled propagator function in~(\ref{scalar 2-pt}) 
is the distance function defined in~(\ref{y AB}), and its index is
\begin{equation}
\nu_{\ss \rm S} = \sqrt{ \biggl( \frac{D\!-\!1\!-\!\epsilon}{2(1\!-\!\epsilon)} \biggr)^{\! 2}
	- \frac{(D\!-\!1)(D\!-\!2\epsilon) \xi_{\ss \rm S} }{(1\!-\!\epsilon)^2} } \, .
\end{equation}
The CTBD state in power-law inflation is known to be IR divergent for certain ranges 
of non-minimal couplings~\cite{Ford:1977in,Janssen:2009nz}, in which case the 
IR sector of the state needs to be modified and the two-point 
function contains additional terms. However, the IR divergence is absent from the CTBD
state if the following condition on the non-minimal coupling is 
satisfied~\cite{Ford:1977in,Janssen:2009nz},
\begin{equation}
\xi_{\ss \rm S} > \frac{\epsilon (D\!-\!2) 
	\bigl[ 2(D\!-\!1) \!-\! D\epsilon \bigr]}{4 (D\!-\!1)(D\!-\!2\epsilon)} \, ,
\end{equation}
which here translates into a condition on the non-minimal coupling of the Abelian Higgs model,
\begin{equation}
\xi < \frac{\epsilon (D\!-\!1\!-\!2\epsilon)}{(D\!-\!1) (D\!-\!2\epsilon)}
	- (D\!-\!4)  \times \frac{\epsilon \bigl[ 2(D\!-\!1) - (D\!+\!2) \epsilon \bigr]}{8(D\!-\!1)(D\!-\!2\epsilon)} \, .
\end{equation}
In~$D\!=\!4$ this condition matches precisely the independent  
condition we found for the existence of the attractor 
solution for the condensate~(\ref{attractor existence}). 
Therefore, we conclude that the attractor condensate
in power-law inflation {\it always} induces an effective non-minimal coupling~$\xi_{\ss\rm S}$
for the scalar perturbation which regulates its IR behaviour.

\section{Dynamics of vector perturbations}
\label{sec: Dynamics of vector perturbations}

The dynamics of linearized vector perturbations in the Abelian Higgs model in FLRW
space-time is given by the Hamiltonian~(\ref{H A}). When the background is specialized 
to power-law inflation, and the scalar condensate is in the attractor 
regime~(\ref{background scaling}),
this system is equivalent at the linear level to the non-minimally coupled vector field
given by the action,
\begin{equation}
S^{\ss (2)}_{\ss A}[A_\mu,\phi] = 
	\Lint d^{D\!}x \, \sqrt{-g} \, \biggl[
	- \frac{1}{4} g^{\mu \rho} g^{\nu \sigma} F_{\mu\nu} F_{\rho\sigma}
	- \frac{1}{2} \xi_{\ss \rm V} R \,g^{\mu\nu} \! A_\mu A_\nu
	\biggr] \, ,
\label{non-minimal vector action}
\end{equation}
where the effective non-minimal coupling of the vector is given by,
\begin{equation}
\xi_{\ss \rm V} = \frac{(q \overline{\phi}_0)^2}
                                 {(D\!-\!1)(D\!-\!2\epsilon)H_0^2}
= \frac{(q \overline{\phi}\,)^2}
                                 {(D\!-\!1)(D\!-\!2\epsilon)H^2}
\, .
\label{xi V}
\end{equation}
The equivalence of action~(\ref{non-minimal vector action}) with the Hamiltonian 
in~(\ref{H A}) can be demonstrated by deriving the canonical formulation, 
following the procedure of
Sec.~\ref{subsec: Hamiltonian formulation}. The difference is that here one finds two 
second-class constraints, which essentially amount to the first and the last equality 
in~(\ref{elimination}). In other words, the first and last equality in~(\ref{elimination})
that eliminates~$A_0$ and~$\Pi_0$ from the phase space in the unitary gauge can be used 
to reintroduce the eliminated fields, allowing us to
write the covariant action~(\ref{non-minimal vector action}). 
This equivalence is not
crucial, and we make no further use of it, but it does provide a more intuitive
picture and establishes a connection with the way the unitary gauge is usually treated.

\medskip

In this section we solve for the dynamics of the vector field 
operators~$\hat{A}_i$ and~$\hat{\Pi}_i$ of the reduced phase space.
Their dynamics is generated by the Hamiltonian operator~(\ref{H2A}),
specialized to power-law inflation and 
the attractor solution for scalar condensate,
\begin{equation}
\hat{\rm H}^{\ss (2)}_A = \! \Lint d^{D-1}x  \Biggl[
	\frac{a^{4-D}}{2} \hat{\Pi}_i \hat{\Pi}_i 
	+ \frac{a^{2-D}}{2 (q\overline\phi)^2}  (\partial_i \hat{\Pi}_i) (\partial_j \hat{\Pi}_j) 
	+ \frac{a^{D-4}}{4} \hat{F}_{ij} \hat{F}_{ij}
	+ \frac{a^{D-2}}{2} (q\overline\phi)^2
		\hat{A}_i \hat{A}_i	
	\Biggr] \, ,
\label{H2A spec}
\end{equation}
which generates the equations of motion,
\begin{eqnarray}
\partial_0 \hat{A}_i &=&
	a^{4-D} \hat{\Pi}_i 
	- \frac{a^{2-D}}{(q\overline\phi)^2} 
	\, \partial_i (\partial_j \hat{\Pi}_j) 
\, ,
\label{vector eom 1}
\\
\partial_0 \hat{\Pi}_i &=&
	a^{D-4} \partial_j \hat{F}_{ji}
	 - a^{D-2} (q\overline\phi)^2 \hat{A}_i \, .
\label{vector eom 2}
\end{eqnarray}
It is convenient to decompose the 
fields into transverse and longitudinal sectors, 
and to consider the equations of motion in
a spatial Fourier space. In the following sections we first perform these transformations
and then proceed to solve the equations for each sector.

\subsection{Scalar-vector decomposition}
\label{subsec: Scalar-vector decomposition}

The field operators~$\hat{A}_i$ and~$\hat{\Pi}_i$ are decomposed into 
transverse and longitudinal parts,
\begin{equation}
\hat{A}_i = \hat{A}_i^{\ss \rm \, T} + \hat{A}_i^{\ss \rm \, L} \, ,
\qquad \qquad
\hat{\Pi}_i = \hat{\Pi}_i^{\ss \rm \,T} + \hat{\Pi}_i^{\ss \rm \, L} \, ,
\label{T L decomposition}
\end{equation}
where the longitudinal parts,
\begin{equation}
\hat{A}_i^{\ss \rm \, L} (\eta,\vec{x} \hspace{0.3mm} ) =
	\mathbb{P}_{ij}^{\ss \rm \, L}(\vec{x} \hspace{0.3mm} ) \, 
	\hat{A}_j (\eta,\vec{x} \hspace{0.3mm} ) \, ,
\qquad \qquad
\hat{\Pi}_i^{\ss \rm \, L} (\eta,\vec{x} \hspace{0.3mm} ) =
	\mathbb{P}_{ij}^{\ss \rm \, L}(\vec{x} \hspace{0.3mm} ) \, 
	\hat{\Pi}_j (\eta,\vec{x} \hspace{0.3mm} ) \, ,
\label{T L decomposition: 2}
\end{equation}
and the transverse parts,
\begin{equation}
\hat{A}_i^{\ss \rm \, T} (\eta,\vec{x} \hspace{0.3mm} ) =
	\mathbb{P}_{ij}^{\ss \rm \, T}(\vec{x} \hspace{0.3mm} ) \, 
	\hat{A}_j (\eta,\vec{x} \hspace{0.3mm} ) \, ,
\qquad \qquad
\hat{\Pi}_i^{\ss \rm \, T} (\eta,\vec{x} \hspace{0.3mm} ) =
	\mathbb{P}_{ij}^{\ss \rm \, T}(\vec{x} \hspace{0.3mm} ) \, 
	\hat{\Pi}_j (\eta,\vec{x} \hspace{0.3mm} )
	 \, ,
\end{equation}
are written with the help of the longitudinal and transverse projectors, respectively,
\begin{equation}
\mathbb{P}_{ij}^{\ss \rm \, L}(\vec{x} \hspace{0.3mm} ) = \frac{\partial_i \partial_j}{\nabla^2} \, ,
\qquad\qquad
\mathbb{P}_{ij}^{\ss \rm \, T}(\vec{x} \hspace{0.3mm} ) = \delta_{ij} 
	- \frac{\partial_i \partial_j}{\nabla^2} \, ,
\label{spatial  longitudinal and transverse projectors}
\end{equation}
where $\nabla^2\!=\!\delta_{ij}\partial_i \partial_j$ 
is the Laplace operator.
These projection operators are idempotent,
$\mathbb{P}_{ij}^{\ss \rm \, L}(\vec{x} \hspace{0.3mm} ) 
	\, \mathbb{P}_{jk}^{\ss \rm \, L}(\vec{x} \hspace{0.3mm} )
 	\!=\! \mathbb{P}_{ik}^{\ss \rm \, L}(\vec{x} \hspace{0.3mm} )$,
$\mathbb{P}_{ij}^{\ss \rm \, T}(\vec{x} \hspace{0.3mm} ) 
	\,\mathbb{P}_{jk}^{\ss \rm \, T}(\vec{x} \hspace{0.3mm} )
	\! =\! \mathbb{P}_{ik}^{\ss \rm \, T}(\vec{x} \hspace{0.3mm} )$, and 
orthogonal,~$\mathbb{P}_{ij}^{\ss \rm \, T}(\vec{x} \hspace{0.3mm} ) 
	\, \mathbb{P}_{jk}^{\ss \rm \, L}(\vec{x} \hspace{0.3mm} )
	\!=\! \mathbb{P}_{ij}^{\ss \rm \, L}(\vec{x} \hspace{0.3mm} ) 
	\, \mathbb{P}_{jk}^{\ss \rm \, T}(\vec{x} \hspace{0.3mm} )
	\! =\! 0$.
Both the transverse and the longitudinal fields are Hermitian, and the 
non-vanishing commutators of the decomposed fields are,
\begin{align}
&
\bigl[ \hat{A}_i^{\ss \rm \, T}(\eta,\vec{x} \hspace{0.3mm} ) ,
	\hat{\Pi}_j^{\ss \rm \, T}(\eta,\vec{x}^{\,\prime}) \bigr] =
	\mathbb{P}_{ij}^{\ss \rm \, T}(\vec{x} \hspace{0.3mm} ) \,
	i \delta^{D-1}(\vec{x} \!-\! \vec{x}^{\,\prime}) \, ,
\\
&
\bigl[ \hat{A}_i^{\ss \rm \, L}(\eta,\vec{x} \hspace{0.3mm} ) ,
	\hat{\Pi}_j^{\ss \rm \, L}(\eta,\vec{x}^{\,\prime}) \bigr] =
	\mathbb{P}_{ij}^{\ss \rm \, L}(\vec{x} \hspace{0.3mm} ) \,
	i \delta^{D-1}(\vec{x} \!-\! \vec{x}^{\,\prime}) \, .
\label{vector field commutator for the reduced space:L+T}
\end{align}
The dynamics of the transverse and longitudinal sectors decouple at the
linear level. This is due to the splitting of the Hamiltonian~(\ref{H2A spec})
into two disjoint parts,
\begin{equation}
\hat{\rm H}^{\ss (2)}_A = \hat{\rm H}^{\ss \rm \, T}
	+ \hat{\rm H}^{\ss \rm \, L} \, ,
\end{equation}
where the transverse part,
\begin{equation}
\hat{\rm H}^{\ss \rm \, T} = \Lint d^{D-1}x \, \biggl[
	\frac{a^{4-D}}{2} \hat{\Pi}_i^{\ss \rm \, T} \hat{\Pi}_i^{\ss \rm \, T}
	+ \frac{a^{D-4}}{2} \bigl( \partial_i \hat{A}_j^{\ss \rm \, T} \bigr)
		\bigl( \partial_i \hat{A}_j^{\ss \rm \, T} \bigr)
	+ \frac{a^{D-2}}{2} (q\overline\phi\,)^2 
		\hat{A}_i^{\ss \rm \, T} \hat{A}_i^{\ss \rm \, T}
	\biggr] \, ,
\label{transverse Hamiltonian: position space}
\end{equation}
generates the equations of motion of the transverse sector,
\begin{align}
\partial_0 \hat{A}_i^{\ss \rm \, T} ={}&
	a^{4-D} \hat{\Pi}_i^{\ss \rm \, T} \, ,
\label{T eom 1}
\\
\partial_0 \hat{\Pi}_i^{\ss \rm \,T} ={}&
	a^{D-4} \nabla^2\hat{A}_i^{\ss \rm \, T}
	- a^{D-2} (q\overline\phi\,)^2 \hat{A}_i^{\ss \rm \, T} \, ,
\label{T eom 2}
\end{align}
and the longitudinal part,
\begin{equation}
\hat{\rm H}^{\ss \rm \, L} = \Lint d^{D-1}x \, \biggl[
	\frac{ a^{4-D}}{2} \hat{\Pi}_i^{\ss \rm \, L} \hat{\Pi}_i^{\ss \rm \, L}
	+ \frac{a^{2-D}}{2 (q\overline\phi\,)^2 }
		\bigl( \partial_i \hat{\Pi}_i^{\ss \rm \, L} \bigr)
		\bigl( \partial_j \hat{\Pi}_j^{\ss \rm \, L} \bigr)
	+ \frac{a^{D-2}}{2} (q\overline\phi\,)^2 
		\hat{A}_i^{\ss \rm \, L} \hat{A}_i^{\ss \rm \, L}
	\biggr] \, ,
\label{longitudinal Hamiltonian: position space}
\end{equation}
generates the equations of motion of the longitudinal sector,
\begin{align}
\partial_0 \hat{A}_i^{\ss \rm \, L} ={}&
	a^{4-D} \hat{\Pi}_i^{\ss \rm \, L}
	- \frac{a^{2-D}}{(q\overline\phi\,)^2 } 
		\nabla^2 \hat{\Pi}_i^{\ss \rm \, L} \, ,
\label{L eom 1}
\\
\partial_0 \hat{\Pi}_i^{\ss \rm \, L} ={}&
	- a^{D-2} (q\overline\phi\,)^2 
		\hat{A}_i^{\ss \rm \, L} \, ,
\label{L eom 2}
\end{align}
where we made use of
$\partial_i \partial_j \hat{\Pi}_j^{\ss \rm \, L} \!=\! \nabla^2 \hat{\Pi}_i^{\ss \rm \, L}$.

\subsection{Fourier decomposition}
\label{subsec: Fourier decomposition}

Homogeneity and isotropy of the FLRW space-time and the scalar condensate
suggest that it is convenient to introduce the spatial Fourier decomposition
of the field operators,
\begin{align}
\hat{A}_i^{\ss \rm \, T}(\eta, \vec{x} \hspace{0.3mm} ) ={}&
	a^{- \frac{D-4}{2}} \! \Lint \frac{d^{D-1}k}{(2\pi)^{\frac{D-1}{2}}} \,
	e^{i \vec{k} \cdot \vec{x}} 
	\sum_{\sigma=1}^{D-2} \varepsilon_i(\sigma, \vec{k} \hspace{0.3mm} ) \,
	\hat{\mathcal{A}}_{{\ss \rm T}, \sigma}(\eta,\vec{k} \hspace{0.3mm} ) \, ,
\label{FT1}
\\
\hat{\Pi}_i^{\ss \rm \, T}(\eta, \vec{x} \hspace{0.3mm} ) ={}&
	a^{ \frac{D-4}{2}} \! \Lint \frac{d^{D-1}k}{(2\pi)^{\frac{D-1}{2}}} \,
	e^{i \vec{k} \cdot \vec{x}} 
	\sum_{\sigma=1}^{D-2} \varepsilon_i(\sigma, \vec{k} \hspace{0.3mm} ) \,
	\hat{\pi}_{{\ss \rm T}, \sigma}(\eta,\vec{k} \hspace{0.3mm} ) \, ,
\label{FT2}
\\
\hat{A}_i^{\ss \rm \, L}(\eta, \vec{x} \hspace{0.3mm} ) ={}&
	a^{- \frac{D-4}{2}} \Bigl( \frac{\mathcal{H}}{H_0} \Bigr)^{\! -1} \!
	\Lint \frac{d^{D-1}k}{(2\pi)^{\frac{D-1}{2}}} \,
	e^{i \vec{k} \cdot \vec{x}} \,
	\Bigl( - \frac{i k_i}{k} \Bigr) \,
	\hat{\mathcal{A}}_{\ss \rm L} (\eta,\vec{k} \hspace{0.3mm} ) \, ,
\label{FT3}
\\
\hat{\Pi}_i^{\ss \rm \, L}(\eta, \vec{x} \hspace{0.3mm} ) ={}&
	a^{\frac{D-4}{2}} \Bigl( \frac{\mathcal{H}}{H_0} \Bigr)
	\! \Lint \frac{d^{D-1}k}{(2\pi)^{\frac{D-1}{2}}} \,
	e^{i \vec{k} \cdot \vec{x}} \,
	\Bigl( - \frac{i k_i}{k} \Bigr) \,
	\hat{\pi}_{\ss \rm L}(\eta,\vec{k} \hspace{0.3mm} ) 
\, ,
\label{FT4}
\end{align}
where~$k\!=\!\| \vec{k} \|$, and
where the  $(D\!-\!2)$  transverse polarization 
vectors~$\varepsilon_i(\sigma, \vec{k} \hspace{0.3mm} )$ satisfy, 
\begin{align}
&
k_i \,\varepsilon_i(\sigma, \vec{k} \hspace{0.3mm} ) = 0 \, ,
&&
\varepsilon_i^*(\sigma, \vec{k} \hspace{0.3mm}) = \varepsilon_i(\sigma, -\vec{k} \hspace{0.3mm} ) \, ,
\nonumber \\
& 
\varepsilon_i^*(\sigma, \vec{k} \hspace{0.3mm}) 
	\, \varepsilon_i(\sigma^\prime, \vec{k} \hspace{0.3mm} )
	= \delta_{\sigma\sigma^\prime} \, ,
&&
\sum_{\sigma=1}^{D-2}
	\varepsilon_i^*(\sigma, \vec{k} \hspace{0.3mm} ) \, 
		\varepsilon_j(\sigma, \vec{k} \hspace{0.3mm}) 
 	= \delta_{ij}-\frac{k_ik_j}{k^2} \, ,
\label{vector polarization identities}
\end{align}
with the summation over the repeated spatial indices implied.
Note the different powers of the scale factor and the conformal Hubble rate 
taken out of the integrals in~(\ref{FT1})--(\ref{FT4}), which is done to simplify the
momentum space equations of motion. 
The hermiticity of the field operators~(\ref{promotion 1})--(\ref{promotion 4})
in momentum space takes the form,
\begin{align}
&
\hat{\mathcal{A}}_{{\ss \rm T}, \sigma}^{\, \dag}(\eta,\vec{k} \hspace{0.3mm} )
	= \hat{\mathcal{A}}_{{\ss \rm T}, \sigma}(\eta,-\vec{k} \hspace{0.3mm} ) \, ,
&&
\hat{\pi}_{{\ss \rm T}, \sigma}^{\, \dag}(\eta,\vec{k} \hspace{0.3mm} )
	= \hat{\pi}_{{\ss \rm T}, \sigma}(\eta,-\vec{k} \hspace{0.3mm} ) \, ,
\\
&
\hat{\mathcal{A}}_{\ss \rm L}^{\, \dag}(\eta,\vec{k} \hspace{0.3mm} )
	= \hat{\mathcal{A}}_{\ss \rm L}(\eta,-\vec{k} \hspace{0.3mm} ) \, ,
&&
\hat{\pi}_{\ss \rm L}^{\, \dag}(\eta,\vec{k} \hspace{0.3mm} )
	= \hat{\pi}_{\ss \rm L}(\eta,-\vec{k} \hspace{0.3mm} ) \, ,
\label{dependencies of momentum space operators}
\end{align}
and the non-vanishing commutators~(\ref{vector field commutator for the reduced space:L+T})
 of these momentum space field operators are,
\begin{align}
&
\bigl[ \, \hat{\mathcal{A}}_{{\ss \rm T},\sigma}(\eta,\vec{k} \hspace{0.3mm} ) ,
	 \hat{\pi}_{{\ss \rm T},\sigma'}(\eta,\vec{k}^{\,\prime}) \bigr]
	 = \delta_{\sigma \sigma'} \, i \delta^{D-1}(\vec{k} \!+\! \vec{k}^{\,\prime})
 \, ,
 \label{transverse momentum commutator}
\\
&
\bigl[ \, \hat{\mathcal{A}}_{\ss \rm L}(\eta,\vec{k} \hspace{0.3mm} ) ,
	 \hat{\pi}_{\ss \rm L}(\eta,\vec{k}^{\,\prime}) \bigr]
	 = i \delta^{D-1}(\vec{k} \!+\! \vec{k}^{\,\prime}) \, .
\label{longitudinal momentum commutator}
\end{align}
The transverse sector Hamiltonian~(\ref{transverse Hamiltonian: position space})
 expressed in momentum space reads,
\begin{equation}
\hat{\rm H}^{\ss \rm \, T} \!= \!\! \Lint d^{D-1}k \, \sum_{\sigma=1}^{D-2}
	\biggl[ \frac{1}{2} \hat{\pi}_{{\ss \rm T} , \sigma}^{\, \dag}
		\hat{\pi}_{{\ss \rm T} , \sigma}
	\!+\! \frac{1}{4} (D\!-\!4) \mathcal{H} 
		\bigl\{ \hat{\pi}_{{\ss \rm T}, \sigma}^{\, \dag} ,
		 \hat{\mathcal{A}}_{{\ss \rm T}, \sigma} \bigr\}
	\!+\! \frac{1}{2} \Bigl( k^2 \!+\! a^2(q\overline\phi\,)^2  \Bigr)
		\hat{\mathcal{A}}_{{\ss \rm T},\sigma}^{\, \dag }
			\hat{\mathcal{A}}_{{\ss \rm T}, \sigma}
	\biggr] \, ,
\label{transverse Hamiltonian: momentum space}
\end{equation}
where~$\{ \hat{\rm A}, \hat{\rm B} \} \!=\! \hat{\rm A} \hat{\rm B} + \hat{\rm B} \hat{\rm A}$ 
denotes the anti-commutator,
and it generates the momentum space equations of motion of the
transverse sector [{\it cf.} Eqs.~(\ref{T eom 1})--(\ref{T eom 2})],
\begin{align}
\partial_0 \hat{\mathcal{A}}_{{\ss \rm T}, \sigma }={}&
	\hat{\pi}_{{\ss \rm T}, \sigma }
	+ \frac{1}{2}(D\!-\!4) \mathcal{H} 
		\hat{\mathcal{A}}_{{\ss \rm T}, \sigma } \, ,
\label{T momentum eom 1}
\\
\partial_0 \hat{\pi}_{{\ss \rm T}, \sigma }={}&
	- \frac{1}{2}(D\!-\!4) \mathcal{H} \hat{\pi}_{{\ss \rm T}, \sigma }
	- \Bigl( k^2 \!+\! a^2(q\overline\phi\,)^2 \Bigr)
		\hat{\mathcal{A}}_{{\ss \rm T}, \sigma }
\, .
\label{T momentum eom 2}
\end{align}
The longitudinal sector Hamiltonian (\ref{longitudinal Hamiltonian: position space})
 in momentum space reads,
\begin{equation}
\hat{\rm H}^{\ss \rm \, L} \!= \! \! \Lint d^{D-1}k  \biggl[
	\frac{1}{2 (q\overline\phi_0)^2 } 
	\Bigl( k^2 \!+\!  a^2(q\overline\phi\,)^2  \Bigr)
		\hat{\pi}_{\ss \rm L}^{\, \dag} \hat{\pi}_{\ss \rm L}
	+ \frac{1}{4} (D\!-\!2\!-\!2\epsilon) \mathcal{H}
		\bigl\{ \hat{\pi}_{\ss \rm L}^{\, \dag} ,
			\hat{\mathcal{A}}_{\ss \rm L} \bigr\}
	+  \frac{1}{2} (q\overline\phi_0)^2  \hat{\mathcal{A}}_{\ss \rm L}^{\, \dag}
		\hat{\mathcal{A}}_{\ss \rm L}
	\biggr]  ,
\label{longitudinal Hamiltonian: momentum space}
\end{equation}
and generates the momentum space equations of motions of the longitudinal sector
[{\it cf.} Eqs.~(\ref{L eom 1})--(\ref{L eom 2})],
\begin{align}
\partial_0 \hat{\mathcal{A}}_{\ss \rm L}={}&
\frac{1}{(q\overline\phi_0)^2 } 
    \Bigl( k^2 \!+\!  a^2(q\overline\phi\,)^2 \Bigr)
          \hat{\pi}_{{\ss \rm L}}
	+ \frac{1}{2}(D\!-\!2\!-\!2\epsilon) \mathcal{H} 
		\hat{\mathcal{A}}_{{\ss \rm L}} \, ,
\label{L momentum eom 1}
\\
\partial_0 \hat{\pi}_{\ss \rm L }={}&
	- \frac{1}{2}(D\!-\!2\!-\!2\epsilon) \mathcal{H} \hat{\pi}_{{\ss \rm L}}
	- (q\overline\phi_0)^2  \hat{\mathcal{A}}_{{\ss \rm L}} 
\,.
\label{L momentum eom 2}
\end{align}
Note that the momentum space Hamiltonians~(\ref{transverse Hamiltonian: momentum space})
and~(\ref{longitudinal Hamiltonian: momentum space}) are not merely 
field transformations of the position space Hamiltonians~(\ref{transverse Hamiltonian: position space})
and~(\ref{longitudinal Hamiltonian: position space}), but contain additional
parts originating from the explicit time dependence of the Fourier 
transforms~(\ref{FT1})--(\ref{FT4}).


\subsection{Dynamics of the transverse sector}
\label{subsec: Dynamics of the transverse sector}

The two first-order equations~(\ref{T momentum eom 1})
and~(\ref{T momentum eom 2}) for the transverse fields can be combined into
a second order equation for the transverse vector potential, whose solution then determines
the transverse canonical momentum,
\begin{align}
\Bigl[ \partial_0^2 + k^2 
	- \bigl( \nu^2 \!-\! \tfrac{1}{4} \bigr) (1\!-\!\epsilon)^2 \mathcal{H}^2
	\Bigr] \hat{\mathcal{A}}_{{\ss \rm T}, \sigma} ={}& 0 \, ,
\label{2nd order eom A_T}
\\
\hat{\pi}_{{\ss \rm T}, \sigma} ={}& 
	\Bigl[ \partial_0 - \tfrac{1}{2}(D\!-\!4) \mathcal{H} \Bigr] 
	\hat{\mathcal{A}}_{{\ss \rm T},\sigma} \, ,
\end{align}
where we have introduced the index for the vector field mode functions,
\begin{equation}
\nu = \sqrt{\biggl( \frac{D\!-\!3\!-\!\epsilon}{2(1\!-\!\epsilon)} \biggr)^{\!\!2} 
	- \frac{(q\overline\phi_0)^2}{(1\!-\!\epsilon)^2H_0^2} } \, ,
\label{index nu}
\end{equation}
which is either purely real and positive-semidefinite, or purely imaginary,
depending on the value of the condensate
parameter~$(q\overline{\phi}_0)^2/H_0^2$.
Equation~(\ref{2nd order eom A_T}) is of the form typical for 
scalar mode functions in power-law inflation. Its general solutions 
and their properties are given
in Appendix~\ref{app: Mode functions}. Here we give the solutions as,
\begin{align}
\hat{\mathcal{A}}_{{\ss \rm T}, \sigma}(\eta,\vec{k} \hspace{0.3mm} ) ={}&
	U_{\nu}(\eta,k) \, \hat{b}_{{\ss \rm T},\sigma}(\vec{k} \hspace{0.3mm} )
	+ U^*_{\nu}(\eta,k) \, \hat{b}_{{\ss \rm T},\sigma}^{\,\dag} (-\vec{k} \hspace{0.3mm} ) \, ,
\label{A_T solution}
\\
\hat{\pi}_{{\ss \rm T}, \sigma}(\eta,\vec{k} \hspace{0.3mm}) ={}&
	\Bigl[ \partial_0 - \tfrac{1}{2}(D\!-\!4)\mathcal{H} \Bigr]
	  U_{\nu}(\eta,k)\, \hat{b}_{{\ss \rm T},\sigma} (\vec{k} \hspace{0.3mm} )
	+ \Bigl[ \partial_0 - \tfrac{1}{2}(D\!-\!4)\mathcal{H} \Bigr]
		U^*_{\nu}(\eta,k) \,\hat{b}_{{\ss \rm T},\sigma}^{\, \dag} (-\vec{k} \hspace{0.3mm} )
\, ,
\label{Pi_T solution}
\end{align}
where the mode function~$U_\nu$ given in~(\ref{U def}) is,
\begin{equation}
U_\nu(\eta,k) = e^{\frac{i\pi}{4}(2\nu+1)}
	e^{\frac{-ik}{(1-\epsilon)\mathcal{H}_0}}
	\sqrt{\frac{\pi}{4(1\!-\! \epsilon)\mathcal{H}}} \
	H^{\ss (1)}_{\nu} \biggl( \frac{k}{(1\!-\!\epsilon)\mathcal{H}} \biggr) \, ,
\end{equation}
and where~$H_\nu^{\ss(1)}$ is the Hankel function of the first kind.
The time-independent
operators~$\hat{b}_{{\ss \rm T},\sigma}^{\, \dag}(\vec{k})$
and~$ \hat{b}_{{\ss \rm T},\sigma} (\vec{k})$ are introduced as 
initial conditions for field operators, and are chosen such that they 
satisfy the  commutation relations for creation and annihilation operators,
\begin{equation}
\bigl[ \, \hat{b}_{{\ss \rm T},\sigma} (\vec{k} \hspace{0.3mm} ) , 
	\hat{b}_{{\ss \rm T},\sigma' }^{\, \dag} (\vec{k}^{\,\prime} ) \bigr]
	= \delta_{\sigma\sigma'} \, \delta^{D-1}(\vec{k} \!-\! \vec{k}^{\,\prime}) \, ,
\label{canonical commutation relations for the creation and annihilation operators: T}
\end{equation}
owing to the fact that the mode functions are normalized to the Wronskian
given in~(\ref{Wronskian}),
\begin{equation}
U_\nu(\eta,k) \, \partial_0 U_\nu^*(\eta,k)
	- U_\nu^*(\eta,k) \, \partial_0 U_\nu(\eta,k) = i \, ,
\end{equation}
and that the transverse field operators commute as in~(\ref{transverse momentum commutator}).
A more detailed discussion of the meaning of these operators is postponed until 
Sec.~\ref{subsec: Choice of the state}.
When  expressed in terms of the creation and annihilation operators,  
the transverse 
sector Hamiltonian~(\ref{transverse Hamiltonian: momentum space}) reads,
\begin{align}
&
\hspace{-2mm}
\hat{\rm H}_{\ss \rm \, T} \!= \!\!
	\Lint d^{D-1}k \sum_{\sigma=1}^{D-2} 
	\frac{1}{2} \Biggl\{ \biggl[ \bigl| \partial_0 U_\nu \bigr|^2
		\!+ \Bigl( k^2 \!+\! a^2(q\overline\phi\,)^2\!-\! 
		\tfrac{1}{4}(D\!-\!4) \mathcal{H}^2 \Bigr)
			\bigl| U_\nu\bigr|^2 \biggr]
			\bigl\{ \hat{b}_{{\ss \rm T},\sigma} (\vec{k} \hspace{0.3mm} ) ,
                                     \hat{b}_{{\ss \rm T},\sigma} (\vec{k} \hspace{0.3mm} ) \bigr\}
\nonumber 
\\
&	\hspace{2.cm}
+ \biggl[ \bigl( \partial_0 U_\nu \bigr)^2
		+ \Bigl( k^2 \!+\! a^2(q\overline\phi\,)^2\!-\! 
		\tfrac{1}{4}(D\!-\!4) \mathcal{H}^2 \Bigr)
			\bigl( U_\nu\bigr)^2 \biggr] \,
				\hat{b}_{{\ss \rm T},\sigma} (\vec{k} \hspace{0.3mm}) \,
                                     \hat{b}_{{\ss \rm T},\sigma} (-\vec{k} \hspace{0.3mm})
\nonumber \\
&	\hspace{1.cm}
+ \biggl[ \bigl( \partial_0 U_\nu^* \bigr)^2
		+ \Bigl( k^2 \!+\! a^2(q\overline\phi\,)^2 \!-\! 
		\tfrac{1}{4}(D\!-\!4) \mathcal{H}^2 \Bigr)
			\bigl( U_\nu^* \bigr)^2 \biggr] \,
				\hat{b}_{{\ss \rm T},\sigma}^{\, \dag} (\vec{k} \hspace{0.3mm} ) \,
                                     \hat{b}_{{\ss \rm T},\sigma}^{\, \dag} (-\vec{k} \hspace{0.3mm} )
	\Biggr\} \, .
\label{transverse Hamiltonian: momentum space: 2}
\end{align}
The second and third line contain operators that induce a squeezing of the 
state~\cite{Albrecht:1992kf}, which are absent in flat space.

\subsection{Dynamics of the longitudinal sector}
\label{subsec: Dynamics of the longitudinal sector}

The two first order equations~(\ref{L momentum eom 1})
and~(\ref{L momentum eom 2}) for the longitudinal fields are combined into 
a single second order equation for the longitudinal canonical momentum
with the longitudinal vector potential being determined by its solution,
\begin{align}
\Bigl[ \partial_0^2 + k^2 
	- \bigl( \nu^2 \!-\! \tfrac{1}{4} \bigr) (1\!-\!\epsilon)^2\mathcal{H}^2 \Bigr]
	\hat{\pi}_{\ss \rm L} ={}& 0 \, ,
\label{pi_L eq}
\\
\hat{\mathcal{A}}_{\ss \rm L} ={}& 
	- \frac{1}{(q\overline\phi_0)^2} 
	\Bigl[ \partial_0 + \tfrac{1}{2}(D\!-\!2\!-\!2\epsilon) \mathcal{H} \Bigr]
	\hat{\pi}_{\ss \rm L} \, ,
\label{AL in terms of PiL}
\end{align}
where the index $\nu$ was already defined in~(\ref{index nu}). 
Equation~(\ref{pi_L eq}) is the same as equation~(\ref{2nd order eom A_T})
we had solved in the preceding section, so we have for the solutions,
\begin{align}
\hat{\pi}_{\ss \rm L}(\eta,\vec{k} \hspace{0.3mm} ) ={}&
	(q\overline\phi_0) \, U_{\nu}(\eta,k) \, \hat{b}_{\ss \rm L}( \vec{k} \hspace{0.3mm} )
	+
	(q\overline\phi_0)
 \, U^*_{\nu}(\eta,k) \, \hat{b}_{\ss \rm L}^{\, \dag} ( -\vec{k} \hspace{0.3mm} ) \, ,
\label{longitudinal model function: solution}
\\
\hat{\mathcal{A}}_{\ss \rm L}(\eta,\vec{k} \hspace{0.3mm} ) ={}& \!
	- \frac{1}{(q\overline\phi_0)}
		\Bigl[ \partial_0 + \tfrac{1}{2}(D\!-\!2\!-\!2\epsilon) \mathcal{H} \Bigr] 
			\Bigl[U_{\nu}(\eta,k)\, \hat{b}_{\ss \rm L}(\vec{k} \hspace{0.3mm} )
			\!+\!U^*_{\nu}(\eta,k) \, \hat{b}_{\ss \rm L}^{\, \dag} ( -\vec{k} \hspace{0.3mm} ) 
			\Bigr] \, .
\label{longitudinal model function: solution 2}
\end{align}
The normalization in the solutions above was chosen such that
the time-independent 
operators~$\hat{b}_{\ss \rm L}^{\, \dag}( \vec{k})$
and~$\hat{b}_{\ss \rm L}( \vec{k})$ 
satisfy the commutation relations for creation/annihilation operators,
\begin{equation}
\bigl[ \, \hat{b}_{\ss \rm L}(\vec{k} \hspace{0.3mm} ) , 
	\hat{b}_{\ss \rm L}^{\, \dag} (\vec{k}^{\,\prime}) \bigr]
	= \delta^{D-1} (\vec{k} \!-\! \vec{k}^{\,\prime}) 
\, .
\label{canonical commutation relations for the creation and annihilation operators: L}
\end{equation}
The Hamiltonian operator of the longitudinal 
sector~(\ref{longitudinal Hamiltonian: momentum space}) expressed in terms
of the creation and annihilation operators is,
\begin{align}
&
\hspace{-2mm}
\hat{\rm H}_{\ss \rm \, L} = \!
	\Lint d^{D-1}k \, \, \frac{1}{2} \Biggl\{
	\biggl[ \bigl| \partial_0 U_\nu\bigr|^2
		+ \Bigl( k^2 \!+\! a^2(q\overline\phi\,)^2 
			\!-\! \tfrac{1}{4} (D\!-\!2\!-\!2\epsilon)^2 \mathcal{H}^2 \Bigr)
				\bigl| U_\nu \bigr|^2 \biggr]
					\bigl\{ \hat{b}_{{\ss \rm L}}^{\, \dag} (\vec{k} \hspace{0.3mm} ) ,
						\hat{b}_{{\ss \rm L}} (\vec{k} \hspace{0.3mm} ) \bigr\}
\nonumber \\
&	\hspace{2cm}
+	\biggl[ \bigl( \partial_0 U_\nu \bigr)^2
		+ \Bigl( k^2 \!+\! a^2(q\overline\phi\,)^2 
			\!-\! \tfrac{1}{4} (D\!-\!2\!-\!2\epsilon)^2 \mathcal{H}^2 \Bigr)
				\bigl( U_\nu \bigr)^2 \biggr]
				\hat{b}_{{\ss \rm L}} (\vec{k} \hspace{0.3mm} ) \,
                                     \hat{b}_{{\ss \rm L}} (-\vec{k} \hspace{0.3mm} )
\nonumber \\
&	\hspace{0.8cm}
+	\biggl[ \bigl( \partial_0 U_\nu^* \bigr)^2
		+ \Bigl( k^2 \!+\! a^2(q\overline\phi\,)^2  
			\!-\! \tfrac{1}{4} (D\!-\!2\!-\!2\epsilon)^2 \mathcal{H}^2 \Bigr)
				\bigl( U_\nu^* \bigr)^2 \biggr]
				\hat{b}_{{\ss \rm L}}^{\, \dag} (\vec{k} \hspace{0.3mm} ) \,
                                     \hat{b}_{{\ss \rm L}}^{\, \dag} (-\vec{k} \hspace{0.3mm} )
	\Biggr\} \, .
\label{long mom H}
\end{align}

\bigskip

\section{Vector field two-point functions}
\label{sec: Vector field two-point functions}

The linearized vector perturbations are assumed to be in some Gaussian 
state~$\bigl|\Omega\bigr\rangle$, in which case their dynamics is fully captured 
by the {\it Wightman two-point function},
\begin{equation}
	\bigl\langle \Omega \bigr| \hat{A}_\mu(x) \, \hat{A}_\nu(x') \bigl| \Omega \bigr\rangle \, ,
\label{2pt def}
\end{equation}
In this section we first properly define various two-point functions, and derive some
of their properties independent of the choice of the state. We then proceed to construct the
quantum state of interest, and to compute the associated two-point functions. The section closes
by discussing various limits of the computed two-point functions.

\subsection{Equations of motion for two-point functions}
\label{subsec: Equations of motion for two-point functions}

The meaning of the definition of the covariant Wightman two-point function
given in~(\ref{2pt def}) at the beginning of this section is not immediately clear,
as it contains field operators corresponding to the temporal component of the vector 
potential, which we have explicitly eliminated in the gauge-fixing procedure 
in Sec.~\ref{subsec: Hamiltonian formulation}. 
Even though this field component is redundant in the unitary gauge, there is
a lot to be gained by adopting a covariant representation. To this end the 
temporal component~$\hat{A}_0$ of the vector potential can be restored, not as an
independent field, but rather through the very expression~(\ref{elimination}) 
used to eliminate it in the first place,
\begin{equation}
\hat{A}_0 = - \frac{a^{2-D}}{(q\overline{\phi})^2 } 
	\bigl( \partial_i \hat{\Pi}_i^{\ss \rm \, L} \bigr) \, .
\label{A0 operator}
\end{equation}
It follows from this definition and the equation of motion~(\ref{L eom 2})
that the covariant vector potential operator is transverse in the following sense,
\begin{equation}
g^{\mu\nu} \nabla_{\!\mu} \Bigl[ (q\overline{\phi})^2 \hat{A}_\nu \Bigr] = 0 \, .
\label{operator transversality}
\end{equation}
Restoring the temporal component allows us to write the operator equations of 
motion~(\ref{vector eom 1})--(\ref{vector eom 2}) 
in a more familiar form,
as covariant second  order equations,
\begin{equation}
{\mathcal{D}_{\mu}}^\nu \hat{A}_\nu(x) = 0 \, ,
\qquad \qquad
\mathcal{D}_{\mu\nu} = 
	g_{\mu\nu} \square
	- \nabla_{\!\nu} \nabla_{\!\mu} 
	- g_{\mu\nu} (q \overline{\phi})^2 
	\, .
\label{operator equation of motion}
\end{equation}
This equation is written in a covariant form, but strictly speaking
we have derived it only for FLRW backgrounds with homogeneous and isotropic condensates.
However, it is reasonable to posit that it holds in general space-times and 
for general condensates. In the remainder of this section, whenever a covariant expression appears, 
strictly speaking it should be considered as 
evaluated in a FLRW space-time.
Even though we have restored covariance to the equations of motion,
the commutators between the covariant vector field components 
do not take on a covariant appearance,
\begin{align}
&
\bigl[ \hat{A}_0(\eta,\vec{x} \hspace{0.3mm} ) , \hat{A}_i(\eta,\vec{x}^{\,\prime}) \bigr]
	= \frac{a^{2-D}}{(q\overline{\phi})^2} \,
		\partial_i \, i\delta^{D-1}(\vec{x} \!-\! \vec{x}^{\,\prime}) \, ,
\label{com1}
\\
&
\bigl[ \hat{A}_0(\eta,\vec{x} \hspace{0.3mm} ) , \partial_0 \hat{A}_0(\eta, \vec{x}^{\,\prime}) \bigr]
	= - \frac{a^{2-D}}{(q\overline{\phi})^2} \nabla^2 i \delta^{D-1}(\vec{x}\!-\!\vec{x}^{\,\prime}) \, ,
\label{com2}
\\
&
\bigl[ \hat{A}_i(\eta,\vec{x} \hspace{0.3mm} ) , \partial_0 \hat{A}_0(\eta,\vec{x}^{\,\prime}) \bigr]
	= \partial_0 \biggl( \frac{a^{2-D}}{(q\overline{\phi})^2 } \biggr)
		\partial_i \, i \delta^{D-1}(\vec{x} \!-\! \vec{x}^{\,\prime}) \, ,
\label{com3}
\\
&
\bigl[ \hat{A}_i(\eta,\vec{x} \hspace{0.3mm} ) , \partial_0 \hat{A}_j(\eta,\vec{x}^{\,\prime}) \bigr]
	= a^{2-D} \biggl( g_{ij} 
		- \frac{\partial_i \partial_j}{(q\overline{\phi})^2} \biggr)
		i \delta^{D-1}(\vec{x} \!-\! \vec{x}^{\,\prime}) \, .
\label{com4}
\end{align}
These were computed using the fundamental commutators~(\ref{fundamental commutators}) 
of the reduced phase space, the operator  equations of 
motion~(\ref{vector eom 1})--(\ref{vector eom 2}), and the definition~(\ref{A0 operator})
for the temporal component of the vector potential.

\medskip

In what follows we define different two-point functions that are needed 
for the time dependent perturbation theory in the Schwinger-Keldysh
({\it in-in}, {closed-time-path}) formalism.
The {\it positive-} and the {\it negative-frequency Wightman functions} are respectively defined 
by~\footnote{
The $\pm$ labeling conventions for two-point functions are
adopted from the Schwinger-Keldysh formalism, 
see {\it e.g.}~\cite{Berges:2004yj,NonEqLectures}. },
\begin{align}
&
i\bigl[ \tensor*[^{\ss - \!}_\mu]{\Delta}{^{\ss \! +}_\nu} \bigr](x;x')
	= \bigl\langle \Omega \bigr| \hat{A}_\mu(x) \, \hat{A}_\nu(x') \bigl| \Omega \bigr\rangle \, ,
\label{positive frequency Wightman function}
\\
&
i\bigl[ \tensor*[^{\ss + \!}_\mu]{\Delta}{^{\ss \! -}_\nu} \bigr](x;x')
	= \bigl\langle \Omega \bigr| \hat{A}_\nu(x') \, \hat{A}_\mu(x) \bigl| \Omega \bigr\rangle
	= \Bigl\{ i\bigl[ \tensor*[^{\ss - \!}_\mu]{\Delta}{^{\ss \! +}_\nu} \bigr](x;x') \Bigr\}^{\!*} \, ,
\label{negative frequency Wightman function}
\end{align}
and the~{\it Feynman (time-ordered)} and the {\it Dyson (anti-time-ordered)} propagators are,
\begin{align}
i\bigl[ \tensor*[^{\ss + \!}_\mu]{\Delta}{^{\ss \! +}_\nu} \bigr](x;x')
	={}& \bigl\langle \Omega \bigr| 
		\mathcal{T} \hat{A}_\mu(x) \, \hat{A}_\nu(x') \bigl| \Omega \bigr\rangle
\nonumber \\
&
	= \theta(\eta\!-\!\eta') \, i\bigl[ \tensor*[^{\ss - \!}_\mu]{\Delta}{^{\ss \! +}_\nu} \bigr](x;x')
		+  \theta(\eta'\!-\!\eta) \, i\bigl[ \tensor*[^{\ss + \!}_\mu]{\Delta}{^{\ss \! -}_\nu} \bigr](x;x')
	  \, ,
\label{Feynman propagator}
\\
i\bigl[ \tensor*[^{\ss - \!}_\mu]{\Delta}{^{\ss \! -}_\nu} \bigr](x;x')
	={}& \bigl\langle \Omega \bigr| 
		\widetilde{\mathcal{T}} \hat{A}_\nu(x') \, \hat{A}_\mu(x) \bigl| \Omega \bigr\rangle
	= \Bigl\{ i\bigl[ \tensor*[^{\ss + \!}_\mu]{\Delta}{^{\ss \! +}_\nu} \bigr](x;x') \Bigr\}^{\!*} \, ,
\label{Dyson propagator}
\end{align}
The operators~$\mathcal{T}$ and~$\widetilde{\mathcal{T}}$ denote time ordering 
and anti-time-ordering, respectively, of the field operators in the product.

The transversality~(\ref{operator transversality}) of the vector potential field operator
implies some properties of the two-point functions which are independent of the state. Namely,
the Wightman function is transverse on both legs,
\begin{equation}
\nabla^{\mu} \Bigl\{
	(q\overline{\phi})^2 \,
	i \bigl[ \tensor*[_\mu^{\ss - \!}]{\Delta}{_\nu^{\ss +}} \bigr](x;x') 
	\Bigr\}= 0 \, ,
\qquad \qquad
\nabla'^{\sigma} \Bigl\{
	(q\overline{\phi}^{\,\prime} )^2 \,
	i \bigl[ \tensor*[_\mu^{\ss - \!}]{\Delta}{_\nu^{\ss +}} \bigr](x;x') 
	\Bigr\} = 0 \, ,
\label{Wightman transversality}
\end{equation}
but the Feynman propagator is not,
\begin{align}
&
\nabla^{\mu} \Bigl\{
	(q\overline{\phi})^2 \,
	 i \bigl[ \tensor*[_\mu^{\ss + \!}]{\Delta}{_\nu^{\ss \! +}} \bigr](x;x') 
	 \Bigr\}
=
	\delta_\nu^i \partial_i' \, \frac{ i\delta^{D}(x \!-\! x') }{\sqrt{-g} } \, ,
\label{non-transverse 1}
\\
&
\nabla'^{\nu} \Bigl\{
	(q\overline{\phi}^{\,\prime} )^2 \,
	 i \bigl[ \tensor*[_\mu^{\ss + \!}]{\Delta}{_\nu^{\ss \! +}} \bigr](x;x') 
	 \Bigr\}
=
		\delta_\mu^i \partial_i  \, \frac{ i\delta^{D}(x \!-\! x') }{\sqrt{-g} } \, .
\label{non-transverse 2}
\end{align}
The non-transversality of the latter one follows from its definition~(\ref{Feynman propagator}) 
that includes a time-ordered product, and from the non-vanishing commutator~(\ref{com1}).
The non-covariance of these relations might seem disconcerting, even though they 
are dictated by the rules of canonical quantization. However, there seems to be 
no {\it a priori} reason to expect that a time-ordered product, which singles out the time coordinate,
must respect covariance.

The equations of motion for the two-point functions follow from the 
equation of motion for the vector potential field operator~(\ref{operator equation of motion}).
The Wightman function satisfies homogeneous equations on both legs,
\begin{equation}
{\mathcal{D}_\mu}^\rho \,
	i \bigl[ \tensor*[^{\ss - \!}_\rho]{\Delta}{^{\ss \! +}_\nu} \bigr] (x;x') = 0 \, ,
 \qquad\qquad
{\mathcal{D}'_\nu}^\sigma \,
	i \bigl[ \tensor*[^{\ss - \!}_\mu]{\Delta}{^{\ss \! +}_\sigma} \bigr] (x;x') = 0 
\, ,
\label{EOM: Wightman functions}
\end{equation}
while the Feynman propagator satisfies inhomogeneous ones,
\begin{align}
{\mathcal{D}_\mu}^\rho \,
	i \bigl[ \tensor*[^{\ss + \!}_\rho]{\Delta}{^{\ss \! +}_\nu} \bigr] (x;x') 
	={}&
	\Bigl( g_{\mu\nu} + a^2 \delta_\mu^0 \delta_\nu^0 \Bigr) \frac{i \delta^D(x\!-\!x')}{\sqrt{-g}}
\label{EOM: Feynman propagator 1} \\
&	
	- \biggl( \delta_\mu^0 \delta_\nu^0 \, \nabla^2
		+ \delta_\mu^i \delta_\nu^0  \bigl[ \partial_0 \!+\! (D\!-\!2) \mathcal{H} \bigr] \partial_i \biggr)
			\biggl[ \frac{ 1}{ (q\overline{\phi})^2 }\frac{i \delta^D(x\!-\!x')}{\sqrt{-g}} \biggr] \, ,
\nonumber
\\
{\mathcal{D}'_\nu}^\sigma \,
	i \bigl[ \tensor*[^{\ss + \!}_\mu]{\Delta}{^{\ss \! +}_\sigma} \bigr] (x;x') 
	={}&
	\Bigl( g_{\mu\nu} + a^2 \delta_\mu^0 \delta_\nu^0 \Bigr) \frac{i \delta^D(x\!-\!x')}{\sqrt{-g}}
\label{EOM: Feynman propagator 2} \\
&	
	- \biggl( \delta_\mu^0 \delta_\nu^0 \, \nabla'^2
		+ \delta_\mu^i \delta_\nu^0  \bigl[ \partial_0' \!+\! (D\!-\!2) \mathcal{H}' \bigr] \partial_i' \biggr)
			\biggl[ \frac{ 1}{ (q\overline{\phi})^2 }\frac{i \delta^D(x\!-\!x')}{\sqrt{-g}} \biggr] \, ,
\nonumber
\end{align}
where the sources originate from the time-ordering in the definition, and 
from the non-vanishing commutators~(\ref{com1})--(\ref{com4}). The non-covariant sources
for the Feynman propagator might seem even more disconcerting than the 
non-transversality relations~(\ref{non-transverse 1})--(\ref{non-transverse 2}),
but we stress again that this is dictated by the rules of canonical quantization,
and is in fact well known in the Proca model in flat space~\cite{Greiner:1996zu}.
There is an important observation to be made here, that the Feynman propagator 
is {\it not} the Green's function of the theory.

\medskip

A Green's function ought to satisfy equations of motion with simpler covariant sources.
Such a function can be constructed {\it from} the Feynman propagator
by adding to it an additional local piece,~\footnote{
Since this Green's function is constructed from the Feynman propagator we 
label it with the same~$(++)$ Schwinger-Keldysh polarities.
}
\begin{equation}
i \bigl[ \tensor*[_{\mu}^{\ss + }]{G}{^{\ss +}_\nu} \bigr] (x;x')
	\equiv 
	i \bigl[ \tensor*[_{\mu}^{\ss + \! }]{\Delta}{^{\ss \! +}_\nu} \bigr] (x;x')
	+ \frac{a^2\delta_\mu^0 \delta_\nu^0}{(q \overline{\phi})^2} \,
		\frac{i \delta^D(x\!-\!x')}{\sqrt{-g} } \, .
\label{Green's function}
\end{equation}
We emphasize that this is not a modification of the definition of the time ordering operation.
There is nothing wrong with the Feynman propagator defined as an expectation
value of the time-ordered product. It just happens that the Feynman propagator
is {\it not} a Green's function for the differential operator~${\mathcal{D}_\mu}^\nu$.
The Green's function in~(\ref{Green's function}) is not transverse, but in a covariant manner,
\begin{align}
&
\nabla^{\mu} \Bigl\{
	(q\overline{\phi})^2 \,
	 i \bigl[ \tensor*[_\mu^{\ss + }]{G}{_\nu^{\ss +}} \bigr](x;x') 
	 \Bigr\}
=
	\partial_\nu' \frac{ i\delta^{D}(x \!-\! x') }{\sqrt{-g} } \, ,
\label{Green transversality}
\\
&\hspace{4cm}
\nabla'^{\nu} \Bigl\{
	(q\overline{\phi}^{\,\prime} )^2 \,
	 i \bigl[ \tensor*[_\mu^{\ss + }]{G}{_\nu^{\ss +}} \bigr](x;x') 
	 \Bigr\}
=
		\partial_\mu  \frac{ i\delta^{D}(x \!-\! x') }{\sqrt{-g} } \, .
\nonumber
\end{align}
and the equations of motion it satisfies are covariant as well,
\begin{equation}
{\mathcal{D}_\mu}^\rho \,
	i \bigl[ \tensor*[_{\rho}^{\ss + }]{G}{^{\ss +}_\nu} \bigr] (x;x')
	= g_{\mu\nu} \frac{i \delta^D(x\!-\!x')}{\sqrt{-g}} \, ,
\qquad \quad
{\mathcal{D}'_\nu}^\sigma \,
	i \bigl[ \tensor*[_{\mu}^{\ss + }]{G}{^{\ss +}_\sigma} \bigr] (x;x')
	= g_{\mu\nu} \frac{i \delta^D(x\!-\!x')}{\sqrt{-g}} \, .
\label{Green EOM}
\end{equation}
It is worth pointing out that the retarded Green's function is constructed 
from the Wightman function~(\ref{negative frequency Wightman function}), 
and the~$(++)$ Green's function in~(\ref{Green's function}),
\begin{equation}
i \bigl[ \tensor*[_{\mu}]{G}{_\nu^{\ss \rm R}} \bigr](x;x')
	= i \bigl[ \tensor*[_{\mu}^{\ss +}]{G}{_\nu^{\ss +}} \bigr](x;x')
		- i \bigl[ \tensor*[_{\mu}^{\ss + \!}]{\Delta}{_\nu^{\ss \! -}} \bigr](x;x') \, .
\end{equation}
and not from the Feynman propagator~(\ref{Feynman propagator}).
It is this retarded Green's function that 
describes causal propagation of (quantum) field perturbations.
This distinction between the Feynman propagator and the Green's function
in the unitary gauge 
of scalar electrodynamics (and in Proca theory), 
that results from canonical quantization,
is well established in flat space~\cite{Weinberg:1973ew,Greiner:1996zu}, and has also
been noted in de Sitter space~\cite{Cotaescu:2009yc}.

\medskip

The distinction between the Feynman propagator and the~$(++)$ Green's function
bares implications for the Feynman rules of perturbation theory, since in the diagrammatic expansion there is a difference
between the vector line descending from an actual expectation value of a product of
operators, and a line descending from inverting the kinetic operator in the Yang-Feldman equation.
If one insists on using only the Green's function or only the Feynman propagator
in perturbation theory in the unitary gauge, then the local terms by which they differ
must be taken into account. This is the reason for the additional counterterms
that have to be added to the effective action in the unitary gauge to restore 
renormalizability~\cite{Weinberg:1973ew,Weinberg:1973ua,Appelquist:1973ms}.
If the distinction between
the Feynman propagator and the Green's function is maintained,
the additional counterterms should not be necessary. 
One should use the 
Yang-Feldman equation~\cite{Yang:1950vi,Greiner:1996zu}
to uniquely determine which
two-point function appears
in the perturbative expansion.

\subsection{Choice of the state}
\label{subsec: Choice of the state}

The two-point functions depend on the physical state of the system~$\bigl|\Omega\bigr\rangle$,
which needs to be specified before computing them.
The evolution of vector perturbations at leading order is governed by linear equations,
and therefore it is natural to assume their quantum state to be Gaussian, since the evolution
does not change this property. A pure Gaussian state~$\bigl|\Omega\bigr\rangle$ 
with a vanishing condensate is in general defined as being annihilated by 
linear combinations of creation and annihilation operators of the transverse and the 
longitudinal sectors,
for each mode~$\vec{k}$ and helicity~$\sigma$,
\begin{equation}
\Bigl[ \alpha_{ {\ss\rm T},\sigma} (k ) \,
	\hat{b}_{ {\ss\rm T},\sigma}(\vec{k}\hskip 0.3mm)
	+ \beta_{ {\ss\rm T},\sigma} (k  ) \,
	\hat{b}_{ {\ss\rm T}, \sigma}^{\, \dag}(\vec{k} \hspace{0.3mm} ) \Bigr] 
	\bigl| \Omega \bigr\rangle = 0 \, ,
\qquad \Bigl[ \alpha_{\ss \rm L} (k) \,
	\hat{b}_{ \ss\rm L}(\vec{k} \hspace{0.3mm} )
	+ \beta_{\ss \rm L} (k) \,
	\hat{b}_{\ss\rm L}^{\,\dag}(\vec{k} \hspace{0.3mm} ) \Bigr] 
	\bigl| \Omega \bigr\rangle = 0 \, ,
\label{Bogolyubov transformations}
\end{equation}
where we have assumed the operators from different sectors and of different helicities
do not mix, which is preserved by the 
dynamics on cosmological backgrounds.
The coefficients in the two conditions above can be seen to be
the Bogolyubov coefficients, which have to satisfy,
\begin{equation}
\bigl| \alpha_{ {\ss\rm T},\sigma}(k) \bigr|^2
	- \bigl| \beta_{ {\ss\rm T},\sigma}(k) \bigr|^2 = 1 \, ,
\qquad \quad
\bigl| \alpha_{ \ss\rm L}(k) \bigr|^2
	- \bigl| \beta_{ \ss\rm L} (k) \bigr|^2 = 1 \, , 
\end{equation}
on the account of commutators of creation/annihilation operators.
In general they depend on the momentum vector, but in the spirit of
homogeneity and isotropy of the cosmological setting we assume they
depend just on the modulus of the momentum,~$k\!=\!\| \vec{k}\|$.
Specifying Bogolyubov coefficients completely determines the Gaussian 
state.~\footnote{
Here we only work with pure Gaussian states. The most general Gaussian states are
mixed states; see {\it e.g.} Ref.~\cite{NonEqLectures} for 
how to define general mixed Gaussian states.
}

In accelerating cosmological space-times, 
such as power-law inflation,
the physically distinct state is the one which minimizes the energy mode-per-mode
in the asymptotic past, which is the analogue of the Chernikov-Tagirov-Bunch-Davies (CTBD) 
state in de Sitter space~\cite{Chernikov:1968zm,Bunch:1978yq}. 
Determining the Bogolyubov coefficients corresponding to this condition
is accomplished by taking the expectation value of the transverse and longitudinal
sector Hamiltonians~(\ref{transverse Hamiltonian: momentum space: 2}) 
and~(\ref{long mom H}), and minimizing their values 
mode-by-mode
(as described in detail in Ch.~6 of Ref.~\cite{Mukhanov:2007zz}). The result of this
tedious exercise is simply~$\alpha_{{\ss \rm T},\sigma} \!=\! \alpha_{\ss \rm L}\!=\!1$
and~$\beta_{{\ss \rm T},\sigma} \!=\! \beta_{\ss \rm L}\!=\!0$, so that the 
CTBD state we consider here is defined by,
\begin{equation}
\hat{b}_{ {\ss\rm T},\sigma}(\vec{k} \hspace{0.3mm} )
	\bigl| \Omega \bigr\rangle = 0 \, ,
\qquad \quad
\hat{b}_{ \ss\rm L}(\vec{k} \hspace{0.3mm} )
	\bigl| \Omega \bigr\rangle = 0 \, ,
\qquad
\qquad
 \forall \vec k,\sigma \, .
 \label{CTBD definition}
\end{equation}
These conditions uniquely define the state, 
and the solutions for the vector potential field 
operators~(\ref{A_T solution})--(\ref{Pi_T solution}) 
and~(\ref{longitudinal model function: solution})--(\ref{longitudinal model function: solution 2})
now completely determine 
the Wightman function~(\ref{2pt def}) given at the
beginning of the section, and other related two-point functions.

\subsection{Two-point functions as mode sums}
\label{subsec: Two-point function as mode sums}

In this section we make use of explicit solutions for the field operators
found in Sec.~\ref{sec: Dynamics of vector perturbations} to compute 
the covariant Wightman two-point function,
introduced in~(\ref{positive frequency Wightman function}), for the
state defined in Sec.~\ref{subsec: Choice of the state}.
The Feynman propagator then follows from its
definition~(\ref{Feynman propagator}) in terms of the Wightman functions,
and the Green's function from~(\ref{Green's function}).
As the temporal and longitudinal components of the vector potential are
expressed in terms of the longitudinal momentum
according to~(\ref{L eom 2}) and~(\ref{A0 operator}),
\begin{equation}
\hat{A}_0 = 
- \frac{a^{2-D}}
    {(q\overline\phi\,)^2} 
               \bigl( \partial_i \Pi_i^{\ss \rm L} \bigr) \, ,
\qquad \qquad
\hat{A}_i^{\ss \rm L} = 
- \frac{a^{2-D}}
    {(q\overline\phi\,)^2} 
     \bigl( \partial_0 \Pi_i^{\ss \rm L} \bigr) \, ,
\label{operator re-intro}
\end{equation}
for the purpose of computing the Wightman function it is enough to compute the
two-point functions of the transverse vector potential, and the longitudinal canonical
momentum. Using the Fourier decompositions~(\ref{FT1}) and~(\ref{FT4})  of the said 
field operators, and their solutions~(\ref{A_T solution})
and~(\ref{longitudinal model function: solution}), the required two-point functions
are expressed as,
\begin{align}
&
\bigl\langle \Omega \bigr|  \hat{A}_i^{\ss \rm \, T} (\eta,\vec{x} \hspace{0.3mm} )
	\hat{A}_j^{\ss \rm \, T} (\eta',\vec{x}^{\,\prime}) \bigl| \Omega \bigr\rangle
	= (aa') \, \mathbb{P}_{ij}^{\ss \rm \, T} (\vec{x} \hspace{0.3mm} ) \,
		i \bigl[ \tensor[^{ \ss \! - \!}]{\Delta}{^{\ss \! + \!} } \bigr]_\nu (x;x') \, ,
\label{A_T A_T}
\\
&
\bigl\langle \Omega \bigr|  \hat{\Pi}_i^{\ss \rm \, L} (\eta,\vec{x} \hspace{0.3mm} ) \,
	\hat{\Pi}_j^{\ss \rm \, L} (\eta',\vec{x}^{\,\prime}) 
		\bigl| \Omega \bigr\rangle
	=
 (aa')^{D-2} (q\overline\phi)\,(q\overline\phi\,') 
 \,
		\mathbb{P}_{ij}^{\ss \rm \, L} (\vec{x} \hspace{0.3mm} ) \,
		 i \bigl[ \tensor[^{ \ss \! - \!}]{\Delta}{^{\ss \! + \!} } \bigr]_\nu (x;x') 
\, ,
\label{Pi_L Pi_L}
\end{align}
where only a single integral over modes appears,
\begin{equation}
i \bigl[ \tensor[^{ \ss \! - \!}]{\Delta}{^{\ss \! + \!}} \bigr]_\nu (x;x')
	= (aa')^{-\frac{D-2}{2}} \! \! \Lint \frac{d^{D-1}k}{(2\pi)^{\frac{D-1}{2}}} \,
	e^{i \vec{k} \cdot\vec{x}} \,
	U_\nu(\eta,k) \, U_\nu^*(\eta', k) 
\, ,
\label{I intro}
\end{equation}
and where the parameter~$\nu$ is defined in~(\ref{index nu}).
As long as~$\nu^2\!<\! (D\!-\!1)^2/4$, this integral is IR finite and can be recognized as the Wightman 
function for a scalar field in spaces of 
constant~$\epsilon$~\cite{Janssen:2008dw},
with an effective non-minimal coupling~$\xi_{\rm eff}\!=\! \frac{(D-1-\epsilon)^2 - 4 (1-\epsilon)^2 \nu^2}{4 (D-1)(D-2\epsilon)}$.
The inequality condition in our case reads,
\begin{equation}
\frac{(q\overline\phi_0)^2}{H_0^2} 
> - (1\!-\!\epsilon)^2 \Bigl( 1 \!-\! \tfrac{(D-4)\epsilon}{2(1-\epsilon)} \Bigr)
	\Bigl( 2 \!+\! \tfrac{(D-4)(2-\epsilon)}{2(1-\epsilon)} \Bigr) \, .
\end{equation}
and is always satisfied in~$D\!=\!4$,
since the scalar condensate parameter~$(q \overline{\phi}_0)^2/H_0^2$
is always positive, and the right hand side always negative.
The scalar condensate thus regulates 
the infrared of the vector perturbations as well as the scalar ones.
We have already encountered the integral over 
modes in~(\ref{I intro}) in Sec.~\ref{sec: Scalar perturbations}, 
and its solution is
\begin{equation}
i \bigl[ \tensor[^{ \ss  \! - \!}]{\Delta}{^{\ss \! + \! }} \bigr]_\nu (x;x')
	= (aa')^{- \frac{(D-2)\epsilon}{2}} \times \mathcal{F}_\nu (y_{\ss -+}) 
\, ,
\label{scal prop}
\end{equation}
where the rescaled propagator function was defined in~(\ref{F def}),
and $y_{\ss -+}$ was introduced in~(\ref{y AB}).
The way to evaluate the integral in~(\ref{I intro}) and to obtain the
above solution is described in considerable detail
in Ref.~\cite{Janssen:2008px},
and therefore we do not repeat it here.
The two-point functions from~(\ref{A_T A_T}) and~(\ref{Pi_L Pi_L}) are
now expressed as,
\begin{align}
&
\bigl\langle \Omega \bigr|  \hat{A}_i^{\ss \rm \, T} (\eta,\vec{x} \hspace{0.3mm} ) \,
	\hat{A}_j^{\ss \rm \, T} (\eta',\vec{x}^{\,\prime}) \bigl| \Omega \bigr\rangle
	= (aa')^{1 - \frac{(D-2)\epsilon}{2}} \, 
		\mathbb{P}_{ij}^{\ss \rm T} (\vec{x} \hspace{0.3mm} ) \,
		\mathcal{F}_\nu (y_{\ss -+})  \, ,
\label{A_T A_T:2}
\\
&
\bigl\langle \Omega \bigr|  \hat{\Pi}_i^{\ss \rm \, L} (\eta,\vec{x} \hspace{0.3mm} ) \,
	\hat{\Pi}_i^{\ss \rm \, L} (\eta',\vec{x}^{\,\prime}) 
		\bigl| \Omega \bigr\rangle
	= (aa')^{D-2-\frac{D}{2}\epsilon} 
	   (q \overline{\phi}_0)^2 \,
		\mathbb{P}_{ij}^{\ss \rm \, L} (\vec{x} \hspace{0.3mm}) \,
		 \mathcal{F}_\nu (y_{\ss -+}) 
\, .
\label{Pi_L Pi_L:2}
\end{align}
Computing the covariant Wightman 
function of the vector potential field is now
a straightforward matter of writing out its components according to~(\ref{operator re-intro}) 
and plugging in the two-point functions~(\ref{A_T A_T:2})--(\ref{Pi_L Pi_L:2}),
\begin{align}
i \bigl[ \tensor*[_{0}^{\ss - \!}]{\Delta}{_{0}^{\ss \! +} } \bigr] (x;x') ={}&
	\bigl\langle \Omega \bigr| \hat{A}_0(\eta,\vec{x} \hspace{0.3mm} )
		\hat{A}_0(\eta', \vec{x}^{\,\prime}) \bigl| \Omega \bigr\rangle
\! = \!
\frac{(aa')^{2-D}}{(q \overline{\phi}\,)^2
(q \overline{\phi} \,' )^2}  
\,
		\partial_i \partial_j'
	\bigl\langle \Omega \bigr|  \hat{\Pi}_i^{\ss \rm \, L} (\eta,\vec{x} \hspace{0.3mm} )
		\hat{\Pi}_j^{\ss \rm \, L} (\eta',\vec{x}^{\,\prime}) \bigl| \Omega \bigr\rangle
\nonumber \\
={}&	- \frac{(aa')^{-\frac{(D-4)\epsilon}{2}} }
      { (q \overline{\phi}_0)^2 } \nabla^2 
	\mathcal{F}_\nu(y_{\ss \rm -+}) \, ,
\label{00 solution}
\\
i \bigl[ \tensor*[_{0}^{\ss - \!}]{\Delta}{_{i}^{\ss \! +} } \bigr] (x;x') ={}&
	\bigl\langle \Omega \bigr| \hat{A}_0(\eta,\vec{x} \hspace{0.3mm} )
		\hat{A}_i(\eta', \vec{x}^{\,\prime}) \bigl| \Omega \bigr\rangle
\!=\! 
\frac{(aa')^{2-D} }{ (q \overline{\phi}\,)^2(q \overline{\phi}\,')^2} \,
	\partial_j	 \partial_0'
	\bigl\langle \Omega \bigr|  \hat{\Pi}_j^{\ss \rm \, L} (\eta,\vec{x} \hspace{0.3mm})
		\hat{\Pi}_i^{\ss \rm \, L} (\eta',\vec{x}^{\,\prime})
		\bigl| \Omega \bigr\rangle
\nonumber \\
={}&	- \frac{  (aa')^{2-D+2\epsilon} }{ (q \overline{\phi}_0)^2} \,
	\partial_i' \partial_0' \Bigl[ (aa')^{D-2 - \frac{D\epsilon}{2}} 
		\mathcal{F}_\nu(y_{\ss \rm -+}) \Bigr] \, ,
\label{0i solution}
\\
i \bigl[ \tensor*[_{i\,}^{\ss - \!}]{\Delta}{_{j}^{\ss \! +} } \bigr] (x;x') ={}&
	\bigl\langle \Omega \bigr| \hat{A}_i(\eta,\vec{x} \hspace{0.3mm})
		\hat{A}_j(\eta', \vec{x}^{\,\prime}) \bigl| \Omega \bigr\rangle
\nonumber \\
& \hspace{-2.cm} =
 \bigl\langle \Omega \bigr| \hat{A}_i^{\ss \rm \, T}(\eta,\vec{x} \hspace{0.3mm} )
		\hat{A}_j^{\ss\rm \, T}(\eta', \vec{x}^{\,\prime}) 
			\bigl| \Omega \bigr\rangle
	+ \frac{(aa')^{2-D} }{(q \overline{\phi}\,)^2(q \overline{\phi}\,')^2}   \,
		\partial_0 \partial_0'
	\bigl\langle \Omega \bigr|  \hat{\Pi}_i^{\ss \rm \, L} (\eta,\vec{x} \hspace{0.3mm} )
	\hat{\Pi}_j^{\ss \rm \, L} (\eta',\vec{x}^{\,\prime})
		\bigl| \Omega \bigr\rangle
\nonumber \\
& \hspace{-2.cm} 
	=	(aa')^{1 - \frac{(D-2)\epsilon}{2} } \delta_{ij} \mathcal{F}_\nu(y_{\ss -+})
\nonumber \\
&
	+ \frac{(aa')^{ - \frac{(D-4)\epsilon}{2} } }
	      { (q \overline{\phi}_0)^2} \,
	\frac{\partial_i \partial_j'}{\nabla^2} \biggl\{
	\Bigl[ \tfrac{1}{4}(D\!-\!3\!-\!\epsilon)^2 \!-\! (1\!-\!\epsilon)^2 \nu^2 \Bigr] 
		\mathcal{H} \mathcal{H}' \mathcal{F}_\nu(y_{\ss -+})
\nonumber \\
&	\hspace{2cm}
	- (aa')^{2-D+\frac{D\epsilon}{2} } \partial_0 \partial_0'
		\Bigl[ (aa')^{D-2 - \frac{D\epsilon}{2}} \mathcal{F}_\nu(y_{\ss -+}) \Bigr]
	\biggr\} \, .
\label{ij solution}
\end{align}
The component $i \bigl[ \tensor*[_{i \, }^{\ss - \!}]{\Delta}{_{0}^{\ss \! +} } \bigr] (x;x')$
is obtained from~(\ref{0i solution}) by complex conjugation. 
Furthermore, using the identity~(\ref{important identity}) 
from Appendix~\ref{app: Useful identity for F_nu}
for the special case~$\lambda \!=\! \nu$ 
and~$\rho \!=\! D\!-\!2 \!-\! (D\epsilon)/2$,
\begin{align}
\MoveEqLeft[14]
\partial_0 \partial_0' \Bigl[ (aa')^{D-2-\frac{D\epsilon}{2}} \mathcal{F}_\nu (y)\Bigr]
	=
	(aa')^{D-2-\frac{D\epsilon}{2}} \biggl\{
	\Bigl[ \tfrac{1}{4}(D\!-\!3\!-\!\epsilon)^2 \!-\! (1\!-\!\epsilon)^2 \nu^2 \Bigr] 
		\mathcal{H} \mathcal{H}'
\mathcal{F}_\nu
\nonumber \\
&	
	- \frac{1}{2} \nabla^2 I 
		\Bigl[ (2\!-\!y) \frac{\partial \mathcal{F}_\nu}{\partial y}
		\!-\! \tfrac{1}{2}\bigl( D\!+\!\tfrac{D-4}{1-\epsilon} \bigr) 
			\mathcal{F}_\nu \Bigr]
	\biggr\} \, ,
\label{used identity}
\end{align}
where~$I[f(y)] \!=\! \int^y \!dy'\, f(y')$ denotes the primitive function,
allows us to eliminate the inverse Laplacian in the $(ij)$ component~(\ref{ij solution}),
which now reads,
\begin{align}
\MoveEqLeft[4]
i \bigl[ \tensor*[_{i\,}^{\ss - \!}]{\Delta}{_{j}^{\ss \! +} } \bigr] (x;x') =
	(aa')^{1 - \frac{(D-2)\epsilon}{2} } \delta_{ij} \mathcal{F}_\nu(y_{\ss -+})
\label{ij solution 2}\\
&
	+\frac{(aa')^{ - \frac{(D-4)\epsilon}{2} \!\!} }{ 2 (q\overline{\phi}_0)^2 } \,
	\partial_i \partial_j' 
		I \Bigl[ \bigl( 2\!-\! y_{\ss -+} \bigr) 
		\frac{\partial \mathcal{F}_\nu(y_{\ss -+})}{\partial y }
		\!-\! \Bigl(2 + \tfrac{(D-4)(2-\epsilon)}{2(1-\epsilon)} \Bigr) 
			\mathcal{F}_\nu(y_{\ss -+}) \Bigr] 
\, .
\nonumber 
\end{align}
%

\subsection{Covariantizing two-point functions}
\label{subsec: Covariantizing two-point functions}

Expressions~(\ref{00 solution}),~(\ref{0i solution}),
and~(\ref{ij solution 2}) represent the 
solutions for the covariant 
positive-frequency Wightman 
function of vector perturbations in the unitary gauge. 
For practical computations, however, it is advantageous to have it written in 
a manifestly covariant form, which we derive in this section.
Subsequently, we derive the covariant form for the Feynman propagator,
and for the~($++$) Green's function.
The covariant form is not expressible solely in terms of
the distance function $y$ from~(\ref{y def}) (as it would be in the de Sitter limit),
but rather we need to introduce two other bi-local 
quantities,~\footnote{
These quantities can be used to express the cosmological plasma 4-vectors,
\begin{equation*}
u_\mu(\eta) =\frac{\partial_\mu u(x;x') }{(1\!-\!\epsilon)H(\eta)}
       = \frac{\partial_\mu v(x; x')}{(1\!-\!\epsilon)H(\eta)}
\,,\qquad  
u_\nu(\eta') = \frac{\partial'_\nu u(x;x') }{(1\!-\!\epsilon)H(\eta')}
          = -\frac{\partial'_\nu v(x; x')}{(1\!-\!\epsilon)H(\eta')}
\,,\quad  
\end{equation*}
which are canonically normalized,~$g^{\mu\nu} u_\mu u_\nu\!=\!-1$,
and the energy-momentum tensor of the classical cosmological fluid can 
be expressed in terms of them,~$T_{\mu\nu}\!=\! (p\!+\!\rho) u_\mu u_\nu \!+\! p g_{\mu\nu}$.
}
\begin{equation}
u(x;x') = (1\!-\!\epsilon) \ln(aa') \, ,
\qquad \qquad
v(x; x') = (1\!-\!\epsilon) \ln(a/a') \, .
\label{u and v def}
\end{equation}
The appropriate tensor basis in which to express the Wightman function is,
\begin{align}
\MoveEqLeft[3]
i \bigl[ \tensor*[^{\ss - \!}_{\mu}]{\Delta}{^{\ss \! +}_\nu} \bigr] (x;x') =
	\bigl( \partial_\mu \partial_\nu' y_{\ss -+} \bigr) \times \mathcal{C}_1\bigl( y_{\ss -+},u \bigr)
	+ \bigl( \partial_\mu y_{\ss -+} \bigr) \bigl( \partial_\nu' y_{\ss -+} \bigr) 
		\times \mathcal{C}_2 \bigl(  y_{\ss -+}, u \bigr)
\label{covariant form}
\\
&	+ \Bigl[ \bigl( \partial_\mu y_{\ss -+} \bigr) \bigl( \partial_\nu' u \bigr) 
		+ \bigl( \partial_\mu u \bigr) \bigl( \partial_\nu' y_{\ss -+} \bigr) \Bigr]
		\times \mathcal{C}_3 \bigl(  y_{\ss -+}, u \bigr)
	+ \bigl( \partial_\mu u \bigr) \bigl( \partial_\nu' u \bigr) \times 
		\mathcal{C}_4 \bigl( y_{\ss -+}, u \bigr) \, , \qquad
\nonumber 
\end{align}
where~$\mathcal{C}_i$'s are the scalar structure functions.
Determining the four scalar structure functions is a matter of comparing 
the form above with expressions~(\ref{00 solution}),~(\ref{0i solution}), and~(\ref{ij solution 2})
for the components of the Wightman function.
In order to facilitate the comparison we first rewrite the components of the
covariant expression~(\ref{covariant form}) as,
\begin{align}
i \bigl[ \tensor*[^{\ss - \!}_{i\,}]{\Delta}{^{\ss \! +}_{j}} \bigr] (x;x') ={}&
	\! - 2 \delta_{ij} (1\!-\!\epsilon)^2 H_0^2 \, e^u \,
	\Bigl\{ \mathcal{C}_1 \bigl( y_{\ss -+},u \bigr) \!-\! I[\mathcal{C}_2]\bigl( y_{\ss -+},u \bigr) \Bigr\}
	+ \partial_i \partial_j' I^2[\mathcal{C}_2] \bigl( y_{\ss -+},u \bigr)
 \, ,
\label{reform 1}
\\
i \bigl[ \tensor*[^{\ss - \!}_{0}]{\Delta}{^{\ss \! +}_{i}} \bigr] (x;x') ={}&
	(1\!-\!\epsilon) H_0 \, e^{\frac{u+v}{2} } \bigl( \partial_i' \, y_{\ss -+} \bigr) 
\nonumber
\\
&	\hspace{1cm}
	\times \biggl\{
	\mathcal{C}_1\bigl( y_{\ss -+},u \bigr) 
		+  \Bigl[ 2 e^{-v} - (2\!-\!y) \Bigr] \mathcal{C}_2 \bigl( y_{\ss -+},u \bigr)
		+ \mathcal{C}_3\bigl( y_{\ss -+},u \bigr)  \biggr\} \, ,
\label{reform 2}
\\
i \bigl[ \tensor*[^{\ss - \!}_{0}]{\Delta}{^{\ss \! +}_{0}} \bigr] (x;x') ={}&
	(1\!-\!\epsilon)^2 H_0^2 \, e^u \biggl\{
	\bigl[ 4 \ch(v) \!-\! \bigl( 2\!-\!y_{\ss -+} \bigr) \bigr] \mathcal{C}_1 \bigl( y_{\ss -+},u \bigr)
\nonumber \\
&	\hspace{3.cm}
	- \Bigl[ \bigl( 4y_{\ss -+} \!-\! y_{\ss -+}^2 \bigr) 
		+ 4 \ch(v) \, \bigl( 2\!-\!y_{\ss -+} \bigr) - 8 \, \Bigr] \mathcal{C}_2 \bigl( y_{\ss -+},u \bigr)
\nonumber \\
&	\hspace{3.cm}
	+ 2 \Bigl[ 2 \ch(v) - \bigl( 2\!-\! y_{\ss -+} \bigr) \Bigr] \mathcal{C}_3 \bigl( y_{\ss -+},u \bigr)
		+ \mathcal{C}_4 \bigl( y_{\ss -+},u \bigr) \biggr\} \, .
\label{reform 3}
\end{align}
This form is obtained by using the chain rule, and the partial 
derivatives~(\ref{der1}--\ref{der7}) from 
Appendix~\ref{app: Useful identity for F_nu}.
We also rewrite~(\ref{00 solution}),~(\ref{0i solution}), 
and~(\ref{ij solution 2})
in the analogous form in terms of $u$ and $v$ variables,
\begin{align}
i \bigl[ \tensor*[^{\ss - \!}_{i\,}]{\Delta}{^{\ss \! +}_{j}} \bigr] (x;x') ={}&
	- 2 \delta_{ij} (1\!-\!\epsilon)^2 H_0^2 \, e^u \times
	\frac{e^{-\frac{(D-4)\epsilon}{2(1-\epsilon)} u}}
	 { 2 (q\overline{\phi}_0)^2 } 
	 \biggl\{
\Bigl[ \nu^2 - \Bigl( \tfrac{D-3-\epsilon}{2(1-\epsilon)} \Bigr)^{\!\!2} \, \Bigr]
	\, \mathcal{F}_\nu(y_{\ss -+})
	\biggr\}
\nonumber 
\\
&	\hspace{-1.5cm}
	+ \frac{e^{-\frac{(D-4)\epsilon}{2(1-\epsilon)} u} }
	           {2 (q\overline{\phi}_0)^2 } 
	\partial_i \partial_j' \,
		I \Bigl[ \bigl( 2\!-\!y_{\ss -+} \bigr) \tfrac{\partial \mathcal{F}_\nu(y_{\ss -+})}{\partial y_{\ss -+}}
		- \Bigl(\tfrac{2(D-2)-D\epsilon}{2(1-\epsilon)} \Bigr) 
			\mathcal{F}_\nu (y_{\ss -+}) \Bigr] \, ,
\\
i \bigl[ \tensor*[^{\ss - \!}_{0}]{\Delta}{^{\ss \! +}_{i}} \bigr] (x;x') ={}&
	(1\!-\!\epsilon)H_0 \, e^{\frac{u+v}{2}} \bigl( \partial_i' \, y_{\ss -+} \bigr) \,
\\
&	\hspace{-1.5cm}
	\times \frac{e^{-\frac{(D-4)\epsilon}{2(1-\epsilon)} u} }
	       { (q\overline{\phi}_0)^2 }
	\biggl\{
	\Bigl[ \bigl( 2\!-\! y_{\ss -+} \bigr) e^{-v} \!-\! 2 \Bigr] \frac{\partial^2}{\partial y_{\ss -+}^2} 
		- \Bigl( \tfrac{2(D-1) - (D+2)\epsilon }{2 (1-\epsilon)} \Bigr) e^{-v}
			\frac{\partial}{\partial y_{\ss -+}} 
	\biggr\} \mathcal{F}_\nu (y_{\ss -+}) \, ,
\nonumber 
\\
i \bigl[ \tensor*[^{\ss - \!}_{0}]{\Delta}{^{\ss \! +}_{0}} \bigr] (x;x') ={}&
	(1 \!-\! \epsilon)^2 H_0^2 \, e^u \times
\nonumber 
\\
&	\hspace{-1.5cm}
	\times
	\frac{e^{- \frac{(D-4)\epsilon}{2(1-\epsilon)} u}}
	     { (q\overline{\phi}_0)^2 }
	\biggl\{ 4 \Bigl[ \bigl( 2\!-\! y_{\ss -+} \bigr) - 2 \ch(v) \Bigr] \frac{\partial^2}{\partial y_{\ss -+}^2}
	- 2 (D\!-\!1) \frac{\partial}{\partial y_{\ss -+}} \biggr\}
	 \mathcal{F}_\nu(y_{\ss -+}) \, .
\end{align}
Now the first two structure functions~$\mathcal{C}_1$ 
and~$\mathcal{C}_2$ follow from comparing the 
expressions for the $(ij)$ component,~$\mathcal{C}_3$ 
follows from comparing the~$(0i)$ component, and ~$\mathcal{C}_4$
from comparing the $(00)$ component,
\begin{align}
\mathcal{C}_1(y,u) ={}&
	\frac{e^{- \frac{(D-4)\epsilon}{2(1-\epsilon)} u}}
	 {2( q\overline\phi_0 )^2 } \times
	\biggl[ - \bigl( 4y\!-\!y^2 \bigr) \frac{\partial^2}{\partial y^2}
		- (D\!-\!1) (2\!-\!y) \frac{\partial}{\partial y}
\nonumber \\
&	\hspace{5.cm}
		- \tfrac{(D-4) \epsilon }{2(1-\epsilon)} 
			\Bigl( 2 + \tfrac{(D-4) (2-\epsilon) }{2(1-\epsilon)} \Bigr)
		\biggr] \mathcal{F}_\nu(y) \, ,
\label{C1 solution}
\\
\mathcal{C}_2(y,u)  ={}&
	\frac{e^{- \frac{(D-4)\epsilon}{2(1-\epsilon)} u}}
	{2( q\overline\phi_0 )^2} \!\times\!
		\biggl[ (2\!-\!y) \frac{\partial^2}{\partial y^2}
		- \Bigl( 3 + \tfrac{(D-4) (2-\epsilon)}{2(1-\epsilon)} \Bigr)
			\frac{\partial}{\partial y} \biggr]
		\mathcal{F}_\nu(y) \, ,
\label{C2 solution}
\\
\mathcal{C}_3(y,u)  ={}&
	\frac{e^{- \frac{(D-4)\epsilon}{2(1-\epsilon)} u}}
	{2( q\overline\phi_0 )^2} \times
	\tfrac{(D-4) \epsilon }{2(1-\epsilon)}
	\biggl[
	- (2\!-\!y) \frac{\partial}{\partial y}
		+ \Bigl( 2 + \tfrac{(D-4)(2-\epsilon)}{2(1-\epsilon)}  \Bigr)
	\biggr]  \mathcal{F}_\nu(y) \, ,
\label{C3 solution}
\\
\mathcal{C}_4(y,u) ={}&
	\frac{e^{- \frac{(D-4)\epsilon}{2(1-\epsilon)} u}}
	{2( q\overline\phi_0 )^2} \times
	\tfrac{(D-4) \epsilon }{2(1-\epsilon)}
	\biggl[
	\bigl( 4 y\!-\!y^2 \bigr) \frac{\partial}{\partial y}
		+ \Bigl( 2 + \tfrac{(D-4) (2-\epsilon)}{2(1-\epsilon)} \Bigr) (2\!-\!y)
	\biggr]  \mathcal{F}_\nu(y) \, ,
\label{C4 solution}
\end{align}
where we have used Eq.~(\ref{F equation}) in order to 
put $\mathcal{C}_1$ in the form~(\ref{C1 solution}).
The above solutions for the four structure functions now specify the desired
covariant positive-frequency Wightman function in~(\ref{covariant form})
that we were after.
Note that we could have assumed that the scalar structure functions in the covariant
{\it Ansatz}~(\ref{covariant form}) depends on variable~$v$ as well, but we would 
have obtained the same solution above.
The negative-frequency Wightman function is now obtained from the positive-frequency
one by complex conjugation, which simply amounts to switching the~$(-+)$ prescription 
in the distance function~$y$ into the~$(+-)$ one.

Any Wightman function has to satisfy the transversality relations~(\ref{Wightman transversality}),
and the homogeneous equations of motion~(\ref{EOM: Wightman functions}). Demonstrating
that our solution for the Wightman function satisfies these conditions is a straightforward matter
of applying identities~(\ref{derivative reduction 1})--(\ref{derivative reduction 2})
to reduce the derivative order and the contraction identities from Table~\ref{tensor contractions}
in Appendix~\ref{app: Various derivative identities}.
To demonstrate the equation of motion, in addition the differential 
equation~(\ref{F equation}) satisfied by the rescaled propagator function must be used.

\bigskip

The most straightforward way to obtain the Feynman propagator is to use
its definition in terms of the positive- and negative-frequency Wightman functions,
\begin{equation}
i \bigl[ \tensor*[_{\mu}^{\ss + \!}]{\Delta}{_\nu^{\ss \! +}} \bigr] (x;x')
	= \theta(\eta\!-\!\eta') \, i \bigl[ \tensor*[_{\mu}^{\ss - \!}]{\Delta}{_\nu^{\ss \! +}} \bigr] (x;x')
		+ \theta(\eta'\!-\!\eta) \, i \bigl[ \tensor*[_{\mu}^{\ss + \!}]{\Delta}{_\nu^{\ss \! -}} \bigr] (x;x') \, .
\label{step Feynman}
\end{equation}
We wish to write the Feynman propagator in the covariant form of~(\ref{covariant form}) 
with the~$i\varepsilon$-prescription changed to the~$(++)$ one from~(\ref{X++}).
In order to arrive at that form, we need to commute the step functions
in~(\ref{step Feynman}) through the derivatives in the tensor structures 
in~(\ref{covariant form}), so that we can change the~$i\varepsilon$-prescription in the
basis tensors using,
\begin{equation}
y_{\ss ++} = \theta(\eta\!-\!\eta') \,y_{\ss -+} + \theta(\eta'\!-\!\eta) \, y_{\ss +-} \, .
\end{equation}
For the last three terms from~(\ref{covariant form})  this exercise 
simply amounts to substituting all the~$(-+)$ prescriptions with the~$(++)$
ones. For the first term from~(\ref{covariant form}) this is more subtle, 
as the result of the commuting gives,
\begin{align}
\MoveEqLeft[2]
\theta(\eta\!-\!\eta') \, \bigl( \partial_\mu \partial'_\nu y_{\ss -+} \bigr) 
	\mathcal{C}_1\bigl( y_{\ss -+} , u \bigr)
+ \theta(\eta'\!-\!\eta) \, \bigl( \partial_\mu \partial'_\nu y_{\ss +-} \bigr) 
	\mathcal{C}_1\bigl( y_{\ss +-} , u \bigr)
\label{modified tensor}
\\
&
	=
	\Bigl[ \bigl( \partial_\mu \partial'_\nu y_{\ss ++} \bigr)
		+ \delta_\mu^0 \delta_\nu^0 (1\!-\!\epsilon)^2 \mathcal{H}^2 
			\! \times \! i \varepsilon \delta(\eta\!-\!\eta') \Bigr]
	\mathcal{C}_1\bigl( y_{\ss ++} , u \bigr)
	\equiv
	\overline{\bigl( \partial_\mu \partial'_\nu y_{\ss -+} \bigr) }
	\mathcal{C}_1\bigl( y_{\ss ++} , u \bigr) \, .
\nonumber 
\end{align}
The extra term that comes together with the tensor structure above seems irrelevant at 
a first glance, on the account of containing the infinitesimal~$i\varepsilon$. However,
given that the most singular term in the first scalar structure function~(\ref{C1 solution})
is~$\propto y_{\ss ++}^{D/2}$ we absolutely~{\it must} keep the extra term, as it harbors
a delta-function according to the identity,
\begin{equation}
i \varepsilon \delta(\eta\!-\!\eta') \times \Bigl( \frac{y_{\ss ++}}{4} \Bigr)^{\!\! - \frac{D}{2}}
	= \bigl[ (1\!-\!\epsilon) H \bigr]^{-D} \, \frac{(4\pi)^{D/2}}{\Gamma\bigl( \frac{D}{2} \bigr)}
		\times \frac{i \delta^D(x\!-\!x')}{\sqrt{-g}} \, .
\label{fundamental delta text}
\end{equation}
Therefore, the Feynman propagator is obtained from the Wightman 
function~(\ref{covariant form}) by switching all the Schwinger-Keldysh polarity
labels from~$(-+)$ to~$(++)$, {\it and} by using the modified first tensor structure 
from~(\ref{modified tensor}),
\begin{align}
\MoveEqLeft[3]
i \bigl[ \tensor*[^{\ss + \!}_{\mu}]{\Delta}{^{\ss \! +}_\nu} \bigr] (x;x') =
	\overline{\bigl( \partial_\mu \partial'_\nu y_{\ss ++} \bigr)} 
	\times \mathcal{C}_1\bigl( y_{\ss ++},u \bigr)
	+ \bigl( \partial_\mu y_{\ss ++} \bigr) \bigl( \partial_\nu' y_{\ss ++} \bigr) 
		\times \mathcal{C}_2 \bigl(  y_{\ss ++}, u \bigr)
\label{Feynman propagator solution}
 \\
&	+ \Bigl[ \bigl( \partial_\mu y_{\ss ++} \bigr) \bigl( \partial_\nu' u \bigr) 
		+ \bigl( \partial_\mu u \bigr) \bigl( \partial_\nu' y_{\ss ++} \bigr) \Bigr]
		\times \mathcal{C}_3 \bigl(  y_{\ss ++}, u \bigr)
	+ \bigl( \partial_\mu u \bigr) \bigl( \partial_\nu' u \bigr) \times 
		\mathcal{C}_4 \bigl( y_{\ss ++}, u \bigr) \, . \qquad
\nonumber
\end{align}
The ($++$) Green's function is now constructed from the above Feynman one,
and with the use of the identity~(\ref{fundamental delta text}) for the delta-function,
\begin{align}
\MoveEqLeft[3]
i \bigl[ \tensor*[^{\ss + }_{\mu}]{G}{^{\ss  +}_\nu} \bigr] (x;x') =
	\bigl( \partial_\mu \partial_\nu' y_{\ss ++} \bigr) \times \mathcal{C}_1\bigl( y_{\ss ++},u \bigr)
	+ \bigl( \partial_\mu y_{\ss ++} \bigr) \bigl( \partial_\nu' y_{\ss ++} \bigr) 
		\times \mathcal{C}_2 \bigl(  y_{\ss ++}, u \bigr)
\label{Green function}
\\
&	+ \Bigl[ \bigl( \partial_\mu y_{\ss ++} \bigr) \bigl( \partial_\nu' u \bigr) 
		+ \bigl( \partial_\mu u \bigr) \bigl( \partial_\nu' y_{\ss ++} \bigr) \Bigr]
		\times \mathcal{C}_3 \bigl(  y_{\ss ++}, u \bigr)
	+ \bigl( \partial_\mu u \bigr) \bigl( \partial_\nu' u \bigr) \times 
		\mathcal{C}_4 \bigl( y_{\ss ++}, u \bigr)  \, . \qquad
\nonumber 
\end{align}
Note that it is the ($++$) Green's function that we obtain from the Wightman function by the
na\^ive switch of the polarity labels, and not the Feynman propagator.

Checking that the Feynman propagator satisfies the correct transversality 
relations~(\ref{non-transverse 1})--(\ref{non-transverse 2}), and the correct equations
of motion~(\ref{EOM: Feynman propagator 1})--(\ref{EOM: Feynman propagator 2})
is a tedious task of pedantically applying identities provided in 
Appendix~\ref{app: Various derivative identities}.
To confirm the transversality relations we first use 
identities~(\ref{++ derivative 1})--(\ref{++ derivative 5}) to isolate the 
delta functions in time descending from time derivatives acting on the~($++$)
$i\varepsilon$-prescription, and to reduce the derivative order in the tensor structures.
Then we use the identities from Table~\ref{tensor contractions} to perform the 
necessary tensor contractions. Lastly, we use the solutions for the scalar structure 
functions, together with the identities~(\ref{covariant delta 1})--(\ref{covariant delta 6})
to recognize the space-time delta functions. To confirm the equations of motion satisfied
by the Feynman propagator, we follow the same sequence of identities, with the
additions of using the differential equation for the rescaled propagator function~(\ref{F equation})
in the end. Had we not been careful about the definition of the tensor
structures and the~$i\varepsilon$-prescriptions utilized in them, we would not have reproduced
the correct transversality relations and equations of motion dictated by the canonical quantization.

\subsection{Various limits}
\label{subsec: Various limits}

Anticipating future computations, this section provides various limits 
of the Wightman function~(\ref{covariant form}) and
the Feynman propagator~(\ref{Feynman propagator solution}),
with scalar structure functions given in 
(\ref{C1 solution})--(\ref{C4 solution}), namely the
de Sitter limit, the flat space limit, and the coincidence limit.
 Computing the latter two involves examining the two-point function
around $y\!\sim\!0$, hence it is very convenient to have a power series
representation of the rescaled propagator function~$\mathcal{F}_\nu(y)$ 
that is derived in Appendix~\ref{app: Rescaled propagator function F_nu},
\begin{align}
\MoveEqLeft[0.2]
\mathcal{F}_\nu(y) = \frac{[ (1\!-\!\epsilon) H_0]^{D-2}}{(4\pi)^{D/2}} \,
	\Gamma\bigl( \tfrac{D-2}{2} \bigr) \times \Biggl\{
	\Bigl( \frac{y}{4} \Bigr)^{\! - \frac{D-2}{2}}
	+ \frac{\Gamma\bigl( \frac{4-D}{2} \bigr)}
		{\Gamma\bigl( \frac{1}{2} \!+\! \nu \bigr) \,
			\Gamma\bigl( \frac{1}{2} \!-\! \nu \bigr)} \times
\label{Fnu: Taylor series}
\\
&	\times \!\sum_{n=0}^{\infty} \biggl[
	\frac{\Gamma\bigl( \frac{3}{2}  \!+\! \nu \!+\! n \bigr) \,
			\Gamma\bigl( \frac{3}{2}  \!-\! \nu \!+\! n \bigr) }
		{\Gamma\bigl( \frac{6-D}{2} \!+\! n \bigr) \, (n\!+\!1)!}
		\Bigl( \frac{y}{4} \Bigr)^{\! n - \frac{D-4}{2}}
	- \frac{\Gamma\bigl( \frac{D-1}{2} \!+\! \nu \!+\! n \bigr) \,
			\Gamma\bigl( \frac{D-1}{2} \!-\! \nu \!+\! n \bigr) }
		{\Gamma\bigl( \frac{D}{2} \!+\! n \bigr) \, n!}
		\Bigl( \frac{y}{4} \Bigr)^{\! n}
	\biggr]
	\Biggr\} \, .
\nonumber 
\end{align}
The limit~$D\!\rightarrow\!4$ of the two-point functions~(\ref{covariant form})
and~(\ref{Feynman propagator solution})
is trivial, and only the first two structure functions contribute to it.
The limit of vanishing condensate (massless limit) is not accessible due to the unitary 
gauge being singular in 
this limit.

\subsubsection{De Sitter limit}
\label{subsubsec: De Sitter limit}

The de Sitter limit of the two-point 
functions~(\ref{covariant form}),~(\ref{Feynman propagator solution}),
and~(\ref{Green function})
simply amounts to setting~$\epsilon\!=\!0$ 
everywhere, including the distance function~(\ref{y AB}). 
The last two scalar structure functions  vanish in this limit, and
the two-point functions take a manifestly de Sitter-invariant form,
\begin{align}
i \bigl[ \tensor*[^{\ss -\! }_\mu]{\Delta}{^{\ss \! +}_\nu} \bigr] (x;x') 
	\xrightarrow{\epsilon\rightarrow0}
	\bigl( \partial_\mu \partial_\nu' y_{\ss -+} \bigr) \times \mathcal{C}^{\ss\rm dS}_1(y_{\ss -+})
+ \bigl( \partial_\mu y_{\ss -+} \bigr) \bigl( \partial_\nu' y_{\ss -+} \bigr) 
		\times \mathcal{C}^{\ss\rm dS}_2(y_{ \ss -+}) 
	\, ,
\label{Wightman function: de Sitter}
\\
i \bigl[ \tensor*[^{\ss + \! }_\mu]{\Delta}{^{\ss \! +}_\nu} \bigr] (x;x') 
	\xrightarrow{\epsilon\rightarrow0}
	\overline{\bigl( \partial_\mu \partial_\nu' y_{\ss ++} \bigr)} 
		\times \mathcal{C}^{\ss\rm dS}_1(y_{\ss ++})
+ \bigl( \partial_\mu y_{\ss ++} \bigr) \bigl( \partial_\nu' y_{\ss ++} \bigr) 
		\times \mathcal{C}^{\ss\rm dS}_2(y_{ \ss ++}) 
	\, ,
\label{Feynman function: de Sitter}
\\
i \bigl[ \tensor*[^{\ss + }_\mu]{G}{^{\ss +}_\nu} \bigr] (x;x') 
	\xrightarrow{\epsilon\rightarrow0}
	\bigl( \partial_\mu \partial_\nu' y_{\ss ++} \bigr) \times \mathcal{C}^{\ss\rm dS}_1(y_{\ss ++})
+ \bigl( \partial_\mu y_{\ss ++} \bigr) \bigl( \partial_\nu' y_{\ss ++} \bigr) 
		\times \mathcal{C}^{\ss\rm dS}_2(y_{ \ss ++}) 
	\, ,
\label{Green function: de Sitter}
\end{align}
where the two structure functions are,
\begin{align}
\mathcal{C}_1^{\ss \rm dS}(y) ={}&
	\lim_{\epsilon\to0} \mathcal{C}_1(y,u)
	=
\frac{1}{2 (q\overline{\phi}_0)^2 } \biggl[
	- (4y \!-\! y^2) \frac{\partial^2}{\partial y^2}
	- (D\!-\!1) (2\!-\!y) \frac{\partial}{\partial y}
	\biggr] \mathcal{F}^{\ss\rm dS}_{\nu_0}(y) \, ,
\label{de Sitter limit: unitary gauge 1} \\
\mathcal{C}_2^{\ss \rm dS}(y)  ={}&
	\lim_{\epsilon\to0} \mathcal{C}_2(y,u)
	=
	 \frac{1}{2 (q\overline{\phi}_0)^2 } 
	 \biggl[
	(2 \!-\! y) \frac{\partial^2}{\partial y^2}
	- (D\!-\!1) \frac{\partial}{\partial y}
	\biggr] \mathcal{F}^{\ss\rm dS}_{\nu_0}(y) \, ,
\label{de Sitter limit: unitary gauge 2}
\end{align}
where $\mathcal{F}^{\ss\rm dS}_{\nu_0}(y) $ is
obtained from $\mathcal{F}_{\lambda}(y) $
defined in Eq.~(\ref{F def}) 
by setting $\epsilon=0$ and $\lambda\rightarrow \nu_0$,
and where the index~(\ref{index nu})
 simplifies to,
\begin{equation}
\nu_0 = \sqrt{\biggl( \frac{D\!-\!3}{2} \biggr)^{\!\!2} 
	- \biggl( \frac{q\overline{\phi}_0}{H_0} \biggr)^{\!\!2} } 
\, ,
\qquad \qquad
\biggl( \frac{q \overline{\phi}_0}{H_0} \biggr)^{\!\!2}
	= - D(D\!-\!1) \frac{q^2 \xi }{\lambda} \, .
\label{index nu: dS}
\end{equation}
This implies that in de Sitter we need $\xi<0$,
such that the scalar potential has a Mexican hat form,
which harbors a symmetry-breaking minimum.
The transversality relations (\ref{Wightman transversality})--(\ref{non-transverse 2}) 
and~(\ref{Green transversality}),
and the equations of motion (\ref{EOM: Wightman functions})--(\ref{EOM: Feynman propagator 2}) 
and (\ref{Green EOM}) have a straightforward 
de Sitter limit  as well.

\subsubsection{Flat space limit}
\label{subsubsec: Flat space limit}

The na\^ive flat space limit corresponds to taking~$H_0 \!\to \! 0$. However,
that would correspond to the singular point of the unitary gauge in flat space,
as in that limit the scalar condensate~(\ref{condensate}) also vanishes.
In order to obtain a meaningful expression we need to consider 
the formal flat space limit where the non-minimal 
coupling reduces to the tachyonic mass 
term,~$\xi R \!\xrightarrow{H_0\to0} \! \mathcal{M}^2 < 0$,
which thus produces the usual symmetry-breaking potential.
Therefore, the flat space limit for the photon two-point function needs to be taken as,
\begin{equation}
( q\overline{\phi}_0 )^{2}
	\xrightarrow{H_0\rightarrow0} m^2 > 0.
\end{equation}
In the flat space limit we have for the variables~(\ref{y def}) and~(\ref{u and v def}),
\begin{align}
&
y(x;x') \;\stackrel{H_0\rightarrow0}{\sim} \;
	(1\!-\!\epsilon)^2 H_0^2 (\Delta x^2) \, ,
\\
&
u(x;x')  \;\stackrel{H_0\rightarrow0}{\sim} \;
	(1\!-\!\epsilon) H_0 (\eta\!+\!\eta'\!-\!2\eta_0) \, ,
\\
& 
v(x;x')  \;\stackrel{H_0\rightarrow0}{\sim} \;
	(1\!-\!\epsilon) H_0 (\eta\!-\!\eta') \, ,
\end{align}
for the tensor basis,
\begin{align}
\overline{ \bigl( \partial_\mu \partial_\nu' y_{\ss ++} \bigr) } ,
\bigl( \partial_\mu \partial_\nu' y_{\ss -+} \bigr) 
	& \;\stackrel{H_0\rightarrow0}{\sim} \;
	- 2 \bigl[ (1\!-\!\epsilon) H_0 \bigr]^2 \eta_{\mu\nu} \, ,
\label{flat tensor basis 1} 
 \\
\bigl( \partial_\mu y_{\ss AB} \bigr) \bigl( \partial_\nu' y_{\ss AB} \bigr) 
	& \;\stackrel{H_0\rightarrow0}{\sim} \;
	\bigl[ (1\!-\!\epsilon) H_0 \bigr]^4
		\bigl( \partial_\mu \Delta x_{\ss AB}^2 \bigr)
			\bigl( \partial'_\nu\Delta x_{\ss AB}^\sigma \bigr) \, ,
\label{flat tensor basis 2}
  \\
\Bigl[ \bigl( \partial_\mu y_{\ss AB} \bigr) \bigl( \partial_\nu' u \bigr) 
		\!+\! \bigl( \partial_\mu u \bigr) \bigl( \partial_\nu' y_{\ss AB} \bigr) \Bigr]
	&  \;\stackrel{H_0\rightarrow0}{\sim} \;
		\bigl[ (1\!-\!\epsilon) H_0 \bigr]^3
	       \Bigl[ \bigl( \partial_\mu \Delta x_{\ss AB}^2 \bigr)\delta_\nu^0
	       	\!+\!\delta_\mu^0 \bigl( \partial_\nu' \Delta x_{\ss AB}^2 \bigr) \Bigr] \, ,
\label{flat tensor basis 3} 
 \\
\bigl( \partial_\mu u \bigr) \bigl( \partial_\nu' u \bigr) 
	& \;\stackrel{H_0\rightarrow0}{\sim} \;
	\bigl[ (1\!-\!\epsilon) H_0 \bigr]^2 \delta_\mu^0 \delta_\nu^0  
\, .
\label{flat tensor basis 4} 
\end{align}
and for the partial derivatives in the scalar structure functions,
\begin{equation}
\frac{\partial}{\partial y} 
	= \frac{1}{(1\!-\!\epsilon)^2\mathcal{H} \mathcal{H}' } \frac{\partial}{\partial ( \Delta x^2 )} 
	 \;\stackrel{H_0\rightarrow0}{\sim} \;
		\frac{1}{(1\!-\!\epsilon)^2H_0^2 } \frac{\partial}{\partial ( \Delta x^2 )} 
\, ,
\end{equation}
The flat space limit of the rescaled propagator function is obtained from the 
power series representation~(\ref{Fnu: Taylor series}), where the dependence 
on~$H_0$ is found in powers of~$y$ and within the gamma functions inside the index~$\nu$,
\begin{equation}
\nu = \sqrt{ \biggl( \frac{D\!-\!3\!-\!\epsilon}{2(1\!-\!\epsilon)} \biggr)^{\! \! 2}
	- \frac{m^2}{(1\!-\!\epsilon)^2 H_0^2} }
	 \;\stackrel{H_0\rightarrow0}{\sim} \;
	\frac{im}{(1\!-\!\epsilon) H_0} \, .
\end{equation}
The key is to examine the ratio of the gamma functions that appears in the power series,
and make use of 8.328.1 from~\cite{Gradshteyn:2007} to take the limit,
\begin{equation}
\frac{ \Gamma\bigl( \alpha \!+\! \nu \!+\! n \bigr) \, \Gamma\bigl( \alpha \!-\! \nu \!+\! n \bigr)}
	{\Gamma\bigl( \frac{1}{2} \!+\! \nu \bigr) \, \Gamma\bigl( \frac{1}{2} \!-\! \nu \bigr)}
	 \;\stackrel{H_0\rightarrow0}{\sim} \;
	\biggl[ \frac{m}{(1\!-\!\epsilon) H_0} \biggr]^{2n+2\alpha-1} \, ,
\end{equation}
after which the series~(\ref{Fnu: Taylor series}) reduces to
\begin{align}
\MoveEqLeft[6]
\mathcal{F}_\nu(y) 	
	\xrightarrow{H_0\to0}
	 \frac{\Gamma\bigl( \frac{D-2}{2} \bigr) \, \Gamma\bigl( \tfrac{4-D}{2} \bigr) }{(4\pi)^{D/2}} \,
\\
&
\times \sum_{n=0}^{\infty} \Biggl[
	\frac{ m^{2n} }{\Gamma\bigl( \frac{4-D}{2} \!+\! n \bigr) \, n!} \,
		\biggl( \frac{\Delta x^2}{4} \biggr)^{\! \! n - \frac{D-2}{2}} 
	- \frac{ m^{2n+D-2} }{\Gamma\bigl( \frac{D}{2} \!+\! n \bigr) \, n!} \,
		\biggl( \frac{\Delta x^2}{4} \biggr)^{\! \! n } \,
	\Biggr] \, .
\nonumber
\end{align}
This is recognized as a power series representation of the modified Bessel function of the 
second kind, according to 8.445 and 8.407.1 from~\cite{Gradshteyn:2007},
\begin{equation}
\mathcal{F}_\nu(y)
	\xrightarrow{H_0\to0}
	\frac{1}{(2\pi)^{D/2}}
		\biggl( \frac{m^2}{\Delta x^2} \biggr)^{\!\! \frac{D-2}{4}}
	K_{\frac{D-2}{2}} \Bigl(\sqrt{m^2 \Delta x^2 } \, \Bigr)
	\equiv i \Delta_m \bigl( \Delta x^2 \bigr) \, ,
\label{F flat}
\end{equation}
which corresponds to the propagator function for the vacuum state of the massive
scalar field in flat space (after inserting the appropriate~$i\varepsilon$-prescription),
\begin{equation}
\Bigl( \square - m^2 \Bigr) i \Delta_m\bigl( \Delta x_{\ss -+}^2 \bigr) = 0 \, ,
\qquad 
\Bigl( \square - m^2 \Bigr) i \Delta_m\bigl( \Delta x_{\ss ++}^2 \bigr) = i \delta^D(x\!-\!x') \, ,
\label{flat scalar propagator eom}
\end{equation}
where in flat space~$\square \!=\! \eta^{\mu\nu} \partial_\mu \partial_\nu$.

Only the first two structure functions contribute in this limit,
\begin{align}
\mathcal{C}_1(u,y)
 \; \stackrel{H_0\rightarrow0}{\sim} \;{}&
	\frac{1}{[(1\!-\!\epsilon)H_0]^2}  \!\times\! \frac{1}{m^2}
	\biggl[ - 2 \Delta x^2 \frac{\partial^2}{\partial ( \Delta x^2 )^2}
		- (D\!-\!1)  \frac{\partial}{\partial ( \Delta x^2 ) }
		\biggr] i\Delta_m \bigl( \Delta x^2 \bigr)	     \, ,
\label{C1 solution: flat}
\\
\mathcal{C}_2(u,y)  
 \;\stackrel{H_0\rightarrow0}{\sim} \;{}&
	\frac{1}{[ (1\!-\!\epsilon)H_0]^4} \!\times\!
	\frac{1}{m^2} 
		\biggl[ \frac{\partial^2}{\partial ( \Delta x^2 )^2}\biggr]
		i\Delta_m \bigl( \Delta x^2 \bigr)	     \, .
\label{C2 solution: flat}
\end{align}
Therefore, the flat space limit of the vector two-point functions is
\begin{align}
i\bigl[ \tensor*[_\mu^{\ss A \!}]{\Delta}{_\nu^{\ss \! B}} \bigr](x;x') 
	\xrightarrow{H_0\rightarrow0} \frac{2}{m^2}
&
 \Biggl\{ \eta_{\mu\nu} \biggl[ 2 (\Delta x_{\ss AB}^2 ) 
 	\frac{\partial^2}{\partial ( \Delta x_{\ss AB}^2 )^2}
	 		 + (D\!-\!1)  \frac{\partial}{\partial ( \Delta x_{\ss AB}^2 )} \biggr]
\label{photon propagator: flat}
\\
&	\hspace{2.5cm}
 	- 2(\Delta x_{\ss AB})_\mu (\Delta x_{\ss AB})_\nu 
 		\frac{\partial^2}{\partial ( \Delta x_{\ss AB}^2 )^2}
 \Biggr\}
 	i\Delta_m \bigl( \Delta x_{\ss AB}^2 \bigr)
 \, .
 \nonumber
\end{align}
The flat space limit of the Wightman function can be written in a more standard form
making use of~(\ref{flat scalar propagator eom}),
\begin{equation}
i\bigl[ \tensor*[_\mu^{\ss - \!}]{\Delta}{_\nu^{\ss \! +}} \bigr](x;x') 
	\xrightarrow{H_0\rightarrow0} 
 	\biggl( \eta_{\mu\nu} - \frac{\partial_\mu \partial_\nu}{m^2} \biggr)
 	i\Delta_m \bigl( \Delta x_{\ss -+}^2 \bigr) \, .
\end{equation}
The same can be done for the Feynman propagator, carefully accounting 
for~(\ref{flat scalar propagator eom}),
\begin{equation}
i\bigl[ \tensor*[_\mu^{\ss + \!}]{\Delta}{_\nu^{\ss \! +}} \bigr](x;x') 
	\xrightarrow{H_0\rightarrow0} 
 	\biggl( \eta_{\mu\nu} - \frac{\partial_\mu \partial_\nu}{m^2} \biggr)
 	i\Delta_m \bigl( \Delta x_{\ss ++}^2 \bigr)
 	- \frac{\delta_\mu^0 \delta_\nu^0}{m^2} \, i \delta^D(x\!-\!x') \, ,
\end{equation}
which indeed is the form the propagator is supposed to take (see {\it e.g.}
Sec. 6.5 from~\cite{Greiner:1996zu}). 
This represents
the second non-trivial check of our principal result~(\ref{covariant form})
and~(\ref{C1 solution})--(\ref{C4 solution}).
The~($++$) Green's function in flat space~(\ref{Green function}) now takes the form,
\begin{equation}
i\bigl[ \tensor*[_\mu^{\ss + }]{G}{_\nu^{\ss  +}} \bigr](x;x') 
	\xrightarrow{H_0\rightarrow0} 
 	\biggl( \eta_{\mu\nu} - \frac{\partial_\mu \partial_\nu}{m^2} \biggr)
 	i\Delta_m \bigl( \Delta x_{\ss ++}^2 \bigr) \, .
\end{equation}
%

\subsubsection{Coincidence limit}
\label{subsubsec: Coincidece limit}

The utilization of the Wightman function~(\ref{covariant form}) and the
Feynman propagator~(\ref{Feynman propagator solution}) 
in loop computations
will require their dimensionally regulated coincidence limits,
which we work out here.
The coincident limit of the variables~(\ref{y def}) and~(\ref{u and v def}) is,
\begin{equation}
y_{\ss AB} \xrightarrow{x' \! \rightarrow x} 0 \, ,
\qquad \qquad
u \xrightarrow{x' \! \rightarrow x} 2 (1\!-\!\epsilon) \ln(a) \, ,
\qquad \qquad
v \xrightarrow{x' \! \rightarrow x} 0 \, ,
\end{equation}
and of the basis tensors,
\begin{align}
&
\overline{\bigl( \partial_\mu \partial_\nu' y_{\ss -+} \bigr)} ,
\bigl( \partial_\mu \partial_\nu' y_{\ss -+} \bigr)
	\!\xrightarrow{x' \!\rightarrow x} \!
	- 2 \bigl[ (1\!-\!\epsilon) H \bigr]^2 a^2\eta_{\mu\nu} \, ,
\quad
\bigl( \partial_\mu y_{\ss AB} \bigr) \bigl( \partial_\nu' y_{\ss AB} \bigr)
	\xrightarrow{x' \! \rightarrow x} 0 \, ,
\label{coincident two point function}
\\
&
\Bigl[ \bigl( \partial_\mu y_{\ss AB} \bigr) \bigl( \partial_\nu' u \bigr) 
		\!+\! \bigl( \partial_\mu u \bigr) \bigl( \partial_\nu' y_{\ss AB} \bigr) \Bigr]
	\xrightarrow{x' \! \rightarrow x} 0 \, ,
\qquad
\bigl( \partial_\mu u \bigr) \bigl( \partial_\nu' u \bigr) \xrightarrow{x' \! \rightarrow x}
	\bigl[ (1\!-\!\epsilon) H \bigr]^2 a^2 \delta_\mu^0 \delta_\nu^0 
\, .
\nonumber
\end{align}
It follows that only the structure functions $\mathcal{C}_1$ and  $\mathcal{C}_4$
contribute to the coincident limit,
\begin{align}
i \bigl[ \tensor*[^{\ss A \!}_\mu]{\Delta}{^{\ss \! B}_\nu} \bigr](x;x) ={}&
	g_{\mu\nu}
		\times (- 2)\bigl[ (1\!-\!\epsilon) H \bigr]^2 \mathcal{C}_1 \bigl( 0,2\ln(a) \bigr)
\nonumber \\
&	\hspace{2cm}
	+ \bigl( a^2 \delta_\mu^0 \delta_\nu^0 \bigr) 
		\times \bigl[ (1\!-\!\epsilon) H \bigr]^2 \mathcal{C}_4 \bigl( 0,2\ln(a) \bigr)
\, ,
\label{Coincident correlator: form}
\end{align}
where their dimensionally regulated coincident limits are,
\begin{align}
(- 2)\bigl[ (1\!-\!\epsilon) H \bigr]^2 \mathcal{C}_1\bigl( 0,2\ln(a) \bigr) ={}&
	\frac{ \bigl[ (1\!-\!\epsilon) H\bigr]^{D-2} }{(4\pi)^{D/2}}
	\times
	\frac{\Gamma\bigl( \frac{D-1}{2} \!+\! \nu \bigr)\, 
			\Gamma\bigl( \frac{D-1}{2} \!-\! \nu \bigr)}
		{\Gamma\bigl( \frac{1}{2} \!+\! \nu \bigr) \, 
			\Gamma\bigl( \frac{1}{2} \!-\! \nu \bigr)} 
	\times
	\Gamma\bigl(\tfrac{2-D}{2} \bigr)
\qquad\quad
\nonumber\\
& \hskip-3.5cm
  \times 
    \frac{1}{D}
    \biggl[
       (D\!-\!1)
       +
       \Bigl( D\!-\!1 + \tfrac{(D-4)\epsilon}{2(1-\epsilon)} \Bigr) 
       	\Bigl( 2 + \tfrac{(D-4)(2-\epsilon)}{2(1-\epsilon)} \Bigr) 
       	\Bigl( \tfrac{(1-\epsilon)H_0}{q\overline\phi_0} \Bigr)^{\!2\,}
   \biggr] \, ,
\label{C1 coincident limit}
\\
\bigl[ (1\!-\!\epsilon) H \bigr]^2 \mathcal{C}_4\bigl( 0,2\ln(a) \bigr) ={}&
	\frac{ \bigl[ (1\!-\!\epsilon) H \bigr]^{D-2} }{(4\pi)^{D/2}}
	\times
	\frac{\Gamma\bigl( \frac{D-1}{2} \!+\! \nu \bigr)\, 
			\Gamma\bigl( \frac{D-1}{2} \!-\! \nu \bigr)}
		{\Gamma\bigl( \frac{1}{2} \!+\! \nu \bigr) \, 
			\Gamma\bigl( \frac{1}{2} \!-\! \nu \bigr)}
	\times
	\Gamma\bigl( \tfrac{2-D}{2} \bigr)
\nonumber 
\\
& \hskip0.5cm
  \times 
	\tfrac{(D-4)\epsilon}{2(1-\epsilon)}
	\Bigl( 2 + \tfrac{(D-4)(2-\epsilon)}{2(1-\epsilon)} \Bigr) 
 	\Bigl( \tfrac{(1-\epsilon)H_0}{q\overline\phi_0} \Bigr)^{\!\!2}
\, .
\label{C4 coincident limit}
\end{align}
These are obtained by (i) acting with derivative operators on 
in~(\ref{C1 solution}) and~(\ref{C4 solution}) on the power series
representation for~$\mathcal{F}_\nu(y)$ in~(\ref{Fnu: Taylor series}),
(ii) 
setting all the $D$-dependent powers of~$y$ to zero, and
(iii) setting all the positive powers of~$y$ to zero.
The coincident structure function $\mathcal{C}_1$ in~(\ref{C1 coincident limit}) 
diverges as $\propto 1/(D\!-\!4)$, which can be seen from the factor,
\begin{equation}
\Gamma\big(\tfrac{2-D}{2}\big) = \frac{2}{D\!-\!4} - 1 + \gamma_{\ss \rm E}
  + \mathcal{O}(D\!-\!4)
\,,
\label{Divergence in C1}
\end{equation}
where $\gamma_{\ss\rm E} \!\simeq\! 0.577$ 
is the Euler-Mascheroni constant.
The other coincident structure function 
$\mathcal{C}_4$ in~(\ref{C4 coincident limit})
is finite in this limit.

\subsection{Comparison with previous results}
\label{subsec: Comparison with previous results}

The existing literature on massive vector field propagators in inflation pertains only to the
de Sitter space. There are two works~\cite{Allen:1985wd,Frob:2013qsa} that we can 
directly compare to the de Sitter limit of our propagator in Sec.~\ref{subsubsec: De Sitter limit}.
The first one is the classic work of Allen and Jacobson~\cite{Allen:1985wd} where, 
among multiple results, the two-point function for the massive vector field in de Sitter is reported
in their Sec. III. This vector field 
is described by the action,
\begin{equation}
S[A_\mu] = \Lint d^{D\!}x \, \sqrt{-g} \, 
	\biggl[ - \frac{1}{4} g^{\mu\nu} g^{\rho\sigma} F_{\mu\rho} F_{\nu\sigma}
		- \frac{m^2}{2} g^{\mu\nu} A_\mu A_\nu \biggr] \, .
\end{equation}
which is assumed to result from spontaneous symmetry breaking mechanism,
and corresponds to the unitary gauge, as detailed here in 
Sec.~\ref{sec: Abelian Higgs model in cosmological spaces}.
Comparing this action to the covariant gauge-fixed action~(\ref{dS unitary gauge})
expanded in perturbations,
we see that the de Sitter limit of our two-point functions given in 
Sec.~\ref{subsubsec: De Sitter limit} should reproduce the result from~\cite{Allen:1985wd}
upon the identification~$
m^2 \!\to\!  \bigl( q \overline{\phi}_0 \bigr)^{\! 2} \!=\! {\rm const.} $
Allen and Jacobson construct the vector field two-point function by solving the
equation of motion it satisfies off-coincidence at space-like separations, under the
warranted assumption of~$D$-dimensional de Sitter isometries. 
No~$i\varepsilon$-prescriptions are necessary in that case. In fact all three
two-point functions~(\ref{Wightman function: de Sitter})--(\ref{Green function: de Sitter})
we gave in section~\ref{subsubsec: De Sitter limit} coincide for space-like separations,
and the only thing we need to check is that the scalar structure 
functions~(\ref{de Sitter limit: unitary gauge 1})--(\ref{de Sitter limit: unitary gauge 2})
are consistent with the result from~\cite{Allen:1985wd}. When transcribed to the
bi-tensor basis used here~\footnote{
{The relations between the basis tensors 
that are used in~\cite{Allen:1985wd}
and the ones we use here are given in
Ref.~\cite{Kahya:2005kj}, and they read,}
\begin{equation*}
z = 1 \!-\! \frac{y}{4} \, ,
\quad
g_{\mu\nu'} = - \frac{1}{2H^2} \bigl( \partial_\mu \partial_\nu' y \bigr)
	- \frac{1}{2H^2(4\!-\!y)} \bigl( \partial_\mu y \bigr) \bigl( \partial_\nu' y \bigr) \, ,
\quad
n_\mu n_{\nu'} = \frac{1}{H^2 (4y\!-\!y^2)} \bigl( \partial_\mu y \bigr) \bigl( \partial_\nu' y \bigr) \, .
\end{equation*}
}
the two-point function from~\cite{Allen:1985wd} reads,
\begin{equation}
i \bigl[ \tensor*[_\mu]{\Delta}{_\nu} \bigr]^{\ss \rm AJ}(x;x')
	= \bigl( \partial_\mu \partial_\nu' y \bigr) \times \mathcal{C}_1^{\ss \rm AJ}(y)
		+ \bigl( \partial_\mu y \bigr) \bigl( \partial_\nu' y \bigr)
				\times \mathcal{C}_1^{\ss \rm AJ}(y) \, ,
\end{equation}
where the two scalar structure functions,
\begin{align}
\mathcal{C}_1^{\rm \ss AJ}(y) ={}&
	\frac{1}{4(D\!-\!1) H_0^2} \biggl[ 
		- \bigl( 4y \!-\! y^2 \bigr) \frac{\partial}{\partial y}
			- (D\!-\!1) (2\!-\!y) \biggr] \gamma_{\rm \ss AJ} (y) \, ,
\label{AJ C1}
\\
\mathcal{C}_2^{\rm \ss AJ}(y) ={}&
	\frac{1}{4(D\!-\!1) H_0^2} \biggl[ 
		(2\!-\!y) \frac{\partial}{\partial y}
			- (D\!-\!1) \biggr] \gamma_{\rm \ss AJ} (y) \, ,
\label{AJ C2}
\end{align}
are expressed in terms of a single scalar function,
\begin{align}
&
\gamma_{\rm \ss AJ} (y) 
	= \frac{(1\!-\!D) \, \Gamma\bigl( \tfrac{D+1}{2} \!+\! \nu_{\rm \ss AJ} \bigr) 
	\, \Gamma\bigl( \tfrac{D+1}{2} \!-\! \nu_{\rm \ss AJ} \bigr) }{2^{D+1} \, \pi^{D/2} \, 
	\Gamma\bigl( \frac{D+2}{2} \bigr)}
\nonumber \\
&	\hspace{3cm}
	\times
	\frac{H_0^D}{m^2} \, 
	{}_2F_1\Bigl( \bigl\{ \tfrac{D+1}{2} \!+\! \nu_{\rm \ss AJ} , \tfrac{D+1}{2} \!-\! \nu_{\rm \ss AJ} \bigr\} , 
		\bigl\{ \tfrac{D+2}{2} \bigr\} , 1\!-\!\tfrac{y}{4} \Bigr) \, ,
\label{AJ gamma}
\end{align}
and
\begin{equation}
\nu_{\ss \rm AJ} = \sqrt{\Bigl( \frac{D \!-\! 3}{2} \Bigr)^{\!2} 
	- \frac{m^2}{H_0^2} } \, .
\end{equation}
Using the fact that derivatives of the hypergeometric function raises its parameters by one,
we can recognize that function in~(\ref{AJ gamma}) is
\begin{equation}
\gamma_{\rm \ss AJ}(y) = 2 (D\!-\!1) \frac{H_0^2}{m^2} 
	\frac{\partial}{\partial y} \mathcal{F}_{\nu_{\ss \rm AJ}}(y) \, .
\end{equation}
Now, upon identifying
\begin{equation}
\frac{ m^2 }{ H_0^2 } \to 
 \biggl( \frac{q \overline{\phi}_0}{H_0} \biggr)^{\!\! 2}
	= - D(D\!-\!1) \frac{q^2 \xi}{\lambda} \, ,
\end{equation}
the structure functions~(\ref{AJ C1}) and~(\ref{AJ C2}) match the de Sitter limit of 
our structure functions in~(\ref{de Sitter limit: unitary gauge 1}) 
and~(\ref{de Sitter limit: unitary gauge 2}), as they should. 

It is necessary to restore the~$i\varepsilon$-prescription for time-like separations and
on the light cone, to distinguish between different two-point functions. 
In~\cite{Allen:1985wd} the Feynman propagator is defined 
by setting~ $y\!\to\! y \!+\! i\varepsilon$ (the sign error in the original work was 
corrected in Ref.~\cite{Frob:2013qsa}) in the scalar structure functions. 
The~$i\varepsilon$-prescription in the tensor structures is, however, not discussed,
and neither are the sourced equations of motion satisfied by the Feynman propagator.
As pointed out in Sec.~\ref{subsec: Equations of motion for two-point functions} here, 
this is an important detail, which distinguishes between
the Feynman propagator and the Green's function. This distinction is important
in loop computations regarding renormalization, 
and the careful account of the~$i\varepsilon$-prescription in the
tensor structures from this section supplements the
classic result by Allen and Jacobson in this regard. 
The Allen-Jacobson
massive vector propagator in de Sitter has also been reproduced using canonical 
methods in~\cite{Cotaescu:2009yc}, and using Euclidean methods in~\cite{Narain:2014oja}.

\medskip

The second work that we can compare to is due to Fr{\"o}b and Higuchi~\cite{Frob:2013qsa} who 
report a vector field propagator for the St{ue}ckelberg model in de Sitter in 
the adapted version of the covariant~$R_\xi$ gauge (defined in~(\ref{St gf}) below). 
They report to reproduce the result by
Allen and Jacobson in the unitary gauge limit~$\alpha\!\to\!\infty$, so there is no 
need to explicitly check for that again. However, we comment on how the 
Stueckelberg model relates to the Abelian Higgs model  in 
the de Sitter limit.
The Stueckelberg model in curved space-time in given by the action
\begin{equation}
S_{\rm St.}[A_\mu, \varphi] = \Lint d^{D\!}x \, \sqrt{-g} \, \biggl[
	- \frac{1}{4} g^{\mu\rho} g^{\nu\sigma} F_{\mu\nu} F_{\rho\sigma}
	- \frac{1}{2} g^{\mu\nu} \bigl( \partial_\mu \varphi - m A_\mu \bigr)
		\bigl( \partial_\nu \varphi - m A_\nu \bigr)
	\biggr] \, ,
\label{St action}
\end{equation}
where~$m$ is the mass parameter. It is invariant under gauge transformations,
\begin{equation}
A_\mu \to A_\mu + \partial_\mu \Lambda \, ,
\qquad \qquad
\varphi \to \varphi + m \Lambda \, ,
\end{equation}
where~$\Lambda$ is an arbitrary function. 
This model in de Sitter space is equivalent to the linear level of our Abelian Higgs model.
The connection is established by fixing the de Sitter background in the action~(\ref{SQED action})
for the Abelian Higgs model, writing the complex scalar  in the polar basis,
\begin{equation}
\Phi = \frac{1}{\sqrt{2}} \bigl( \overline{\phi} + \psi \bigr) e^{-i\varphi/\overline{\phi}}
 \, ,
\end{equation}
where~$\overline{\phi}={\rm const.}$ is the 
condensate,
and expanding the action to quadratic order 
in the fluctuations $\psi$ and $\varphi$.
The action for the modulus fluctuation $\psi$ 
decouples, and takes the form 
of~(\ref{equivalent scalar action}) with $\epsilon\!=\!0$, 
while the action for the phase of the scalar, and
the vector field take the form of the Stueckelberg action~(\ref{St action}),
provided we identify~$m\!=\!q\overline{\phi}$.

The Stueckelberg model is a gauge theory, and requires gauge-fixing.
The most convenient way of imposing the gauge in de Sitter seems to be 
a version of the~$R_\xi$ gauge~\cite{tHooft:1971qjg,Fujikawa:1972fe}, whereby the 
action~(\ref{St action}) is substituted by the gauge-fixed action~$S_*\!=\!S_{\rm St.} \!+\! S_{\rm gf}$,
wit the gauge-fixing term,
\begin{equation}
S_{\rm gf}[A_\mu, \varphi] = \Lint d^{D\!}x \,
	\biggl[
	- \frac{1}{2\alpha} \Bigl( g^{\mu\nu} \nabla_\mu A_\nu - m \alpha \varphi \Bigr)^2
	\biggr] \, ,
\label{St gf}
\end{equation}
with~$\alpha$ being an arbitrary gauge-fixing parameter.
The gauge-fixed Stueckelberg action then splits into two sectors, the scalar one,
\begin{equation}
S_*^\varphi [\varphi] = \Lint d^{D\!}x \, \sqrt{-g} \,
	\biggl[
	- \frac{1}{2} g^{\mu\nu} (\partial_\mu \varphi) (\partial_\nu \varphi)
	- \frac{\alpha m^2}{2} \varphi^2
	\biggr] \, ,
\label{gf St scalar}
\end{equation}
and the vector one,
\begin{equation}
S_*^A [A_\mu] = \Lint d^{D\!}x \, \sqrt{-g} \,
	\biggl[
	- \frac{1}{4} g^{\mu\rho} g^{\nu\sigma} F_{\mu\nu} F_{\rho\sigma}
	- \frac{m^2}{2} g^{\mu\nu} A_\mu A_\nu
	- \frac{1}{2\alpha} \bigl( g^{\mu\nu} \nabla_\mu A_\nu \bigr)^{\!2}
	\biggr] \, .
\label{gf St vector}
\end{equation}
It is this vector sector gauge-fixed action that was 
taken as a starting point in~\cite{Frob:2013qsa}. 
The gauge-fixing procedure above explains the origin of this action, and why 
it can contain both the mass term and the gauge-fixing term for the vector field.
The Stueckelberg theory in de Sitter is adequate to work out the propagators.
However, one must bear in mind that it is a linear theory, and that if the propagators
are to be used in loop computations, the vertices of the interacting gauge theory
have to be worked out carefully, as we do for the Abelian Higgs model in the unitary gauge is
Sec.~\ref{sec: Abelian Higgs model in cosmological spaces}.

Fixing the unitary gauge in the Stueckelberg model via the Dirac-Bergmann algorithm of
Sec.~\ref{subsec: Hamiltonian formulation} would amount to fixing~$\varphi\!=\!0$.
However, it is a common lore that the limit~$\alpha\!\to\!\infty$ of the~$R_\xi$
gauge reproduces the unitary gauge by making the scalar in~(\ref{gf St scalar})
infinitely massive. What is in fact shown in~\cite{Frob:2013qsa} is that this limit reproduces
only the covariant Green's function, but not the Feynman propagator.

\medskip

Another noteworthy result for the vector field propagator was reported by Tsamis 
and Woodard~\cite{Tsamis:2006gj}, and is the only one out of the
three commented on in this section which has been successfully employed 
in loop computations in de Sitter 
space~\cite{Prokopec:2006ue,Prokopec:2007ak,Prokopec:2008gw,Gonzalez:2016jrn,Glavan:2015ura,Glavan:2016bvp}. 
Ref.~\cite{Tsamis:2006gj}
reported the propagator in the covariant
Landau gauge (exact covariant gauge), which corresponds to 
the~$\alpha\!\to\!0$ of the~$R_\xi$
gauge-fixing functional, and is equally well adapted to studying both massless and massive
vector fields. We cannot compare our propagator to theirs, 
as the Landau and unitary gauges 
correspond to the opposite limits of the~$R_\xi$ gauge.

\section{Field strength correlator}
\label{sec: Field strength correlator}

In this section we make use of the vector field two-point function to compute
a simple observable -- 
the off-coincident correlator of the vector field strength tensor~$F_{\mu\nu}$,
that is of central importance in primordial magnetogenesis scenarios (see {\it e.g.}
\cite{Maleknejad:2012fw,Durrer:2013pga,Subramanian:2015lua} for recent reviews).
It can be expressed in terms of derivative operators acting on the two-point function,
\begin{equation}
\bigl\langle \Omega \bigr| \hat{F}_{\mu\nu}(x) \,
	\hat{F}_{\rho\sigma}(x') \bigl| \Omega \bigr\rangle
	= 4 \bigl( \partial_{[\mu} \delta_{\nu]}^\alpha \bigr) 
		\bigl( \partial_{[\rho}' \delta_{\sigma]}^\beta \bigr) \,
		i \bigl[ \tensor*[^{\ss - \!}_\alpha]{\Delta}{^{\ss \! +}_\beta} \bigr](x;x')
\, ,
\label{F correlator}
\end{equation}
where $i \bigl[ \tensor*[^{\ss - \!}_\alpha]{\Delta}{^{\ss \! +}_\beta} \bigr](x;x')$
is the Wightman function for the vector potential 
given by~(\ref{covariant form}), with scalar structure functions
in~(\ref{C1 solution})--(\ref{C4 solution}). This correlator is gauge invariant, 
due to the vector field strength tensor being gauge invariant itself,
and it takes the same form in all gauges.
It can be expanded in the appropriate tensor basis, taking into account
the anti-symmetrization of indices,
\begin{align}
\MoveEqLeft[3]
\bigl\langle \Omega \bigr| \hat{F}_{\mu\nu}(x) \,
	\hat{F}_{\rho\sigma}(x') \bigl| \Omega \bigr\rangle =
	\bigl( \partial_{\mu} \partial'_{[\rho} y_{\ss -+} \bigr) 
			\bigl( \partial'_{\sigma]} \partial_\nu y_{\ss -+} \bigr)
		\times \mathcal{K}_1(y_{\ss -+},u)
\nonumber \\
&
	+ \bigl( \partial_{[\mu} y_{\ss -+} \bigr) \bigl( \partial_{\nu]} \partial'_{[\sigma} y_{\ss -+} \bigr) 
			\bigl( \partial'_{\rho]} y_{\ss -+} \bigr) \times \mathcal{K}_2(y_{\ss -+},u)
\nonumber \\
&
+ \Bigl[ \bigl( \partial_{[\mu} y_{\ss -+} \bigr) \bigl( \partial_{\nu]} \partial'_{[\sigma} y_{\ss -+} \bigr) 
			\bigl( \partial'_{\rho]} u \bigr)
		+ \bigl( \partial_{[\mu} u \bigr) \bigl( \partial_{\nu]} \partial'_{[\sigma} y_{\ss -+} \bigr) 
			\bigl( \partial'_{\rho]} y_{\ss -+} \bigr) \Bigr] \times \mathcal{K}_3(y_{\ss -+},u)
\nonumber \\
&
+ \bigl( \partial_{[\mu} u \bigr) \bigl( \partial_{\nu]} \partial'_{[\sigma} y_{\ss -+} \bigr)
		\bigl( \partial_{\rho]}' u \bigr) \times \mathcal{K}_4(y_{\ss -+},u)
\nonumber \\
&
+ \bigl( \partial_{[\mu} y_{\ss -+} \bigr) \bigl( \partial_{\nu]} u \bigr)
		\bigl( \partial'_{[\rho} y_{\ss -+} \bigr) \bigl( \partial'_{\sigma]} u \bigr)
		\times \mathcal{K}_5(y_{\ss -+},u) \, ,
\label{F basis expansion}
\end{align}
where the scalar structure functions~$\mathcal{K}_i$'s of this expansion
are expressed in terms of the scalar structure functions~$\mathcal{C}_i$'s
of the vector potential two-point functions,
\begin{align}
\mathcal{K}_1(y,u) ={}& 
	4 \biggl( \frac{\partial \mathcal{C}_1}{\partial y} 
	- \mathcal{C}_2 \biggr) \, ,
\label{K1 form}
\\
\mathcal{K}_2(y,u) ={}& 
	4 \frac{\partial}{\partial y} \biggl( \frac{\partial \mathcal{C}_1}{\partial y}  
	- \mathcal{C}_2 \biggr)
	= \frac{\partial \mathcal{K}_1}{\partial y} \, ,
\label{K2 form}
\\
\mathcal{K}_3(y,u) ={}& 
	4 \frac{\partial}{\partial u} \biggl( \frac{\partial \mathcal{C}_1}{\partial y}  
	- \mathcal{C}_2 \biggr)
	= \frac{\partial \mathcal{K}_1}{\partial u} \, ,
\label{K3 form}
\\
\mathcal{K}_4(y,u) ={}& 
	4 \biggl( \frac{\partial^2 \mathcal{C}_1}{\partial u^2} 
	- 2 \frac{\partial \mathcal{C}_3}{\partial u}
	+ \frac{\partial \mathcal{C}_4}{\partial y} \biggr) \, ,
\label{K4 form}
\\
\mathcal{K}_5(y,u) ={}& 
	4 \biggl( \frac{\partial^2 \mathcal{C}_2}{\partial u^2}
	- 2 \frac{\partial^2 \mathcal{C}_3}{\partial y \partial u}
	+ \frac{\partial^2 \mathcal{C}_4}{\partial y^2} \biggr)
	= - \frac{\partial^2 \mathcal{K}_1}{\partial u^2}
		+ \frac{\partial \mathcal{K}_4}{\partial y} \, .
\label{K5 form}
\end{align}
The relations above are simply derived by acting the derivatives in~(\ref{F correlator})
onto the covariantized Wightman function in~(\ref{covariant form}), and exploiting
the anti-symmetrization of derivatives.
We then plug in the~$\mathcal{C}_i$'s from~(\ref{C1 solution})--(\ref{C4 solution})
into the structure functions above, and use the hypergeometric  equation~(\ref{F equation})
to express them in terms of the rescaled propagator function,
\begin{align}
\mathcal{K}_1(y,u) ={}& 
	\frac{2 \,e^{- \frac{(D-4)\epsilon}{2(1-\epsilon)} u }}{(1\!-\!\epsilon)^2 H_0^2}
		\times \biggl[ - \frac{\partial}{\partial y} \mathcal{F}_\nu(y) \biggr] \, ,
\label{K1}
\\
\mathcal{K}_2(y,u) ={}& 
	\frac{2 \,e^{- \frac{(D-4)\epsilon}{2(1-\epsilon)} u }}{(1\!-\!\epsilon)^2 H_0^2}
		\times \biggl[ - \frac{\partial^2}{\partial y^2} \mathcal{F}_\nu(y) \biggr] \, ,
\label{K2}
\\
\mathcal{K}_3(y,u) ={}& 
	\frac{2 \,e^{- \frac{(D-4)\epsilon}{2(1-\epsilon)} u }}{(1\!-\!\epsilon)^2 H_0^2}
		\times \biggl[ \tfrac{(D-4)\epsilon}{2(1-\epsilon)} \,
		\frac{\partial}{\partial y} \mathcal{F}_\nu(y) \biggr] \, ,
\label{K3}
\\
\mathcal{K}_4(y,u) ={}& 
	\frac{2 \,e^{- \frac{(D-4)\epsilon}{2(1-\epsilon)} u }}{(1\!-\!\epsilon)^2 H_0^2}
		\times \biggl[ \tfrac{(D-4)\epsilon}{2(1 - \epsilon)} 
			\Bigl( 1 \!-\! \tfrac{(D-4)\epsilon}{2(1 - \epsilon)} \Bigr)
			\mathcal{F}_\nu(y)  \biggr] \, ,
\label{K4}
\\
\mathcal{K}_5(y,u) ={}& 
	\frac{2 \,e^{- \frac{(D-4)\epsilon}{2(1-\epsilon)} u }}{(1\!-\!\epsilon)^2 H_0^2}
		\times \biggl[ \tfrac{(D-4)\epsilon}{2(1 - \epsilon)} \,
			\frac{\partial}{\partial y} \mathcal{F}_\nu(y)  \biggr] \, .
\label{K5}
\end{align}
These structure functions, together with the tensor basis expansion~(\ref{F basis expansion})
represent our result for the vector field strength correlator in the Abelian Higgs
model in $D$-dimensional power-law inflation.

\medskip
It is worth noting that all the scalar structure functions~(\ref{K1})--(\ref{K5})
of the vector field strength correlator are regular in the limit of vanishing condensate,
despite the fact that this is a singular point of the unitary gauge.
They only depend on the condensate via the index of the rescaled propagator 
functions, which simply reduces 
to~$\nu \! \xrightarrow{ \overline{\phi}_0 \to 0 } \! (D\!-\!3\!-\!\epsilon)/[2(1\!-\!\epsilon)]$, 
introducing no singularities into the correlator.
The fact that the correlator in~(\ref{F basis expansion}) and~(\ref{K1})--(\ref{K5})
 is regular for~$\overline{\phi}\!\to\!0$ is not obvious.
The issue is clarified by writing out the components of the
correlator in terms of the transverse and longitudinal two-point functions computed in 
section~\ref{subsec: Two-point function as mode sums},
\begin{align}
&
\bigl\langle \Omega \bigr| \hat{F}_{0i}(x) \, \hat{F}_{0j}(x') \bigl| \Omega \bigr\rangle
	= \partial_0 \partial_0' 
		\bigl\langle \Omega \bigr| \hat{A}_i^{\ss \rm \, T}(x) \, 
			\hat{A}_j^{\ss \rm \, T}(x') \bigl| \Omega \bigr\rangle
\nonumber \\
&	\hspace{6cm}
			+ (aa')^{4-D} \bigl\langle \Omega \bigr| \hat{\Pi}_i^{\ss \rm \, L}(x) \, 
			\hat{\Pi}_j^{\ss \rm \, L}(x') \bigl| \Omega \bigr\rangle \, ,
\\
&
\bigl\langle \Omega \bigr| \hat{F}_{0i}(x) \, \hat{F}_{kl}(x') \bigl| \Omega \bigr\rangle
	= 2 \, \partial'_{k]}\partial_0 
		\bigl\langle \Omega \bigr| \hat{A}_i^{\ss \rm \, T}(x) \, 
			\hat{A}_{[l}^{\ss \rm \, T}(x') \bigl| \Omega \bigr\rangle \, ,
\\
&
\bigl\langle \Omega \bigr| \hat{F}_{ij}(x) \, \hat{F}_{kl}(x') \bigl| \Omega \bigr\rangle
	= 4 \, \partial'_{k]}\partial_{[i}
		\bigl\langle \Omega \bigr| \hat{A}_{j]}^{\ss \rm \, T}(x) \, 
			\hat{A}_{[l}^{\ss \rm \, T}(x') \bigl| \Omega \bigr\rangle \, .
\end{align}
It is clear that both the longitudinal and the transverse components contribute to 
the correlators in general.
It is also clear that the contribution of the longitudinal component drops out in the 
limit of vanishing condensate, as the correlator of longitudinal momenta
is suppressed by the condensate parameter~$q \overline{\phi}_0/H_0$
as seen from~(\ref{Pi_L Pi_L}). 
Thus the massless
vector field limit is recovered, where only the spatially transverse vector potential contributes
to the vector field strength.
In fact, in the limit of vanishing condensate in~$D\!=\!4$ space-time dimensions our 
correlator reduces to the flat space result for the massless~$U(1)$ vector field, 
\begin{align}
\bigl\langle \Omega \bigr| \hat{F}_{\mu\nu}(x) \, \hat{F}_{\rho\sigma}(x') \bigl| \Omega \bigr\rangle
	\xrightarrow[D\to4]{ \overline{\phi}_0 \to 0 }
	\frac{2}{\pi^2 ( \Delta x^2 )^2} \Biggl[
		\eta_{\mu [\rho} \eta_{\sigma] \nu} 
	- 4 \eta_{\alpha[\mu} \eta_{\nu] [\sigma} \eta_{\rho]\beta} 
		\frac{ \Delta x^\alpha \Delta x^\beta}{\Delta x^2 } \Biggr]
		 \, .
\end{align}
This is so because the massless~$U(1)$ vector field is coupled conformally to gravity 
in~$D\!=\!4$.
This property of regularity in the limit of vanishing condensate
can be understood by 
recalling that the field strength correlator is 
a gauge invariant quantity,
and hence it must be identical in all gauges, regardless
whether they are singular or not.

\subsection{Equal time E\&M correlators}

In the context of primordial magnetogenesis (and electrogenesis), the relevant 
vector field strength correlators are the equal-time electric and magnetic (E\&M)
field correlators in~$D\!=\!4$ space-time dimensions. 
The electric and magnetic field operators 
are given as the following components of the vector field strength operators,
\begin{equation}
\hat{E}_i = \hat{F}_{0i} \, ,
\qquad \qquad
\hat{B}_i = \tfrac{1}{2} \varepsilon_{ijk} \hat{F}_{jk} \, ,
\end{equation}
where~$\varepsilon_{ijk}$ is the 3-dimensional Levi-Civita symbol,
so that their equal-time correlators can be expressed in terms
of the equal-time correlators of the vector field strength~(\ref{F basis expansion}),
\begin{align}
&
\bigl\langle \Omega \bigr| \hat{E}_i(\eta,\vec{x} \hspace{0.3mm} ) \, 
	\hat{E}_j(\eta,\vec{x}^{\,\prime}) \bigl|\Omega\bigr\rangle =
	\bigl\langle \Omega \bigr| \hat{F}_{0i}(\eta,\vec{x} \hspace{0.3mm} ) \, 
	\hat{F}_{0j}(\eta,\vec{x}^{\,\prime}) \bigl|\Omega\bigr\rangle \, ,
\\
&
\bigl\langle \Omega \bigr| \hat{E}_i(\eta,\vec{x} \hspace{0.3mm} ) \, 
	\hat{B}_j(\eta,\vec{x}^{\,\prime}) \bigl|\Omega\bigr\rangle =
	\frac{1}{2} \varepsilon_{jkl}\bigl\langle \Omega \bigr| \hat{F}_{0i}(\eta,\vec{x} \hspace{0.3mm} ) \, 
	\hat{F}_{kl}(\eta,\vec{x}^{\,\prime}) \bigl|\Omega\bigr\rangle \, ,
\\
&
\bigl\langle \Omega \bigr| \hat{B}_i(\eta,\vec{x} \hspace{0.3mm} ) \, 
	\hat{B}_j(\eta,\vec{x}^{\,\prime}) \bigl|\Omega\bigr\rangle =
	\frac{1}{4} \varepsilon_{ikl} \, \varepsilon_{jmn}
	\bigl\langle \Omega \bigr| \hat{F}_{kl}(\eta,\vec{x} \hspace{0.3mm} ) \, 
	\hat{F}_{mn}(\eta,\vec{x}^{\,\prime}) \bigl|\Omega\bigr\rangle \, .
\end{align}
The last three scalar structure functions of the vector field strength 
correlator~(\ref{F basis expansion}) drop out in~$D\!=\!4$.
Therefore, we only need the equal-time limit of the first two tensor structures,
computed using expressions~(\ref{der1})--(\ref{der7}) for derivatives of the distance function,
\begingroup
\allowdisplaybreaks
\begin{align}
\bigl( \partial_{0} \partial'_{[0} y \bigr) \bigl( \partial'_{j]} \partial_i y \bigr)
	\xrightarrow{\eta' \rightarrow \eta}{}&
	\bigl[ (1\!-\!\epsilon) \mathcal{H} \bigr]^4
	\biggl[
		- \bigl( \mathcal{Y}^2 \!+\! 2 \bigr) \, \delta_{ij}
	+ 2 \mathcal{Y}^2
		\, \frac{\Delta x_i \Delta x_j }{ \| \Delta\vec{x} \|^2 }
	\biggr]
	 \, ,
\\
\frac{1}{2} \varepsilon_{jkl}
\bigl( \partial_{0} \partial'_{[k} y \bigr) \bigl( \partial'_{l]} \partial_i y \bigr)
	\xrightarrow{\eta' \rightarrow \eta}{}&
	\bigl[ (1\!-\!\epsilon) \mathcal{H} \bigr]^4 \biggl[
	2 \mathcal{Y} \,
		\varepsilon_{ijk} \frac{\Delta x_{k}}{\| \Delta\vec{x} \|}
	\biggr]
	\, ,
\\
\frac{1}{4} \varepsilon_{ikl} \varepsilon_{jmn}
\bigl( \partial_{k} \partial'_{[m} y \bigr) \bigl( \partial'_{n]} \partial_l y \bigr)
	\xrightarrow{\eta' \rightarrow \eta}{}&
	\bigl[ (1\!-\!\epsilon) \mathcal{H} \bigr]^4 \biggl[ 2 \, \delta_{ij} \biggr] \, ,
\\
\bigl( \partial_{[0} y \bigr) \bigl( \partial_{i]} \partial'_{[j} y \bigr) \bigl( \partial'_{0]} y \bigr)
	\xrightarrow{\eta' \rightarrow \eta}{}&
	\bigl[ (1\!-\!\epsilon) \mathcal{H} \bigr]^4 \biggl[
	- \frac{\mathcal{Y}^4}{2} \, \delta_{ij}
	+ \mathcal{Y}^2 \bigl( \mathcal{Y}^2 \!-\! 2 \bigr) \,
			 \frac{\Delta x_i \Delta x_j}{\| \Delta\vec{x} \|^2}
	\biggr] \, ,
\\
\frac{1}{2} \varepsilon_{jkl}
\bigl( \partial_{[0} y \bigr) \bigl( \partial_{i]} \partial'_{[l} y \bigr) \bigl( \partial'_{k]} y \bigr)
	\xrightarrow{\eta' \rightarrow \eta}{}&
	\bigl[ (1\!-\!\epsilon) \mathcal{H} \bigr]^4 \biggl[
	\mathcal{Y}^3 \,
		\varepsilon_{ijk} \frac{ \Delta x_{k} }{\| \Delta\vec{x} \|}
	\biggr] \, ,
\\
\frac{1}{4} \varepsilon_{ikl} \varepsilon_{jmn}
\bigl( \partial_{[k} y \bigr) \bigl( \partial_{l]} \partial'_{[n} y \bigr) \bigl( \partial'_{m]} y \bigr)
	\xrightarrow{\eta' \rightarrow \eta}{}&
	\bigl[ (1\!-\!\epsilon) \mathcal{H} \bigr]^4 \biggl[
	2 \mathcal{Y}^2 \delta_{ij}
	- 2 \mathcal{Y}^2 \, \frac{\Delta x_{i} \Delta x_{j} }{\| \Delta \vec{x} \|^2}
	\biggr] \, ,
\end{align}
\endgroup
where for convenience we define
 the quantity~$\mathcal{Y}$
as a square root of the equal-time limit of the distance function,
\begin{equation}
y \xrightarrow{\eta'\to\eta}
	\Bigl[ (1\!-\!\epsilon) \mathcal{H} \| \Delta\vec{x} \| \Bigr]^2
	\equiv \mathcal{Y}^2 \, ,
\end{equation}
and the equal-time limit of the rescaled propagator function,
\begin{equation}
\frac{ \mathcal{F}_\nu(y) }{ \bigl[ (1\!-\!\epsilon) H_0 \bigr]^2 }
\xrightarrow{\eta'\rightarrow\eta}
		\frac{ \Gamma\bigl( \tfrac{3}{2} \!+\! \nu \bigr) \,
			\Gamma\bigl( \tfrac{3}{2} \!-\! \nu \bigr) }{ (4\pi)^2 }
		{}_2F_1 \Bigl( \bigl\{ \tfrac{3}{2} \!+\! \nu , \tfrac{3}{2} \!-\! \nu \bigr\} ,
			\bigl\{ 2 \bigr\} , 1 \!-\! \tfrac{1}{4} \mathcal{Y}^2 \Bigr) 
	\equiv
	 \mathscr{F}_\nu(\mathcal{Y}) \, ,
\label{F(Y) def}
\end{equation}
where we have defined a dimensionless function~$\mathscr{F}_\nu$,
and where the index~$\nu$ is evaluated in~$D\!=\!4$,
\begin{equation}
\nu \xrightarrow{D\rightarrow 4}
	\sqrt{\frac{1}{4} - \left(\frac{q\overline\phi_0}{(1\!-\!\epsilon)H_0}\right)^{\!2}} \, \,  ,
\qquad \quad
\left(\frac{q\overline\phi_0}{H_0}\right)^{\!2}
= \frac{q^2}{\lambda} \Bigl[ \epsilon (3\!-\!2\epsilon) - 6\xi(2\!-\!\epsilon) \Bigr] > 0 \, .
\label{nu corr}
\end{equation}
Furthermore, the partial derivatives appearing in the structure functions are expressed as,
\begin{equation}
\frac{\partial}{\partial y} 
	\xrightarrow{\eta'\rightarrow\eta} \frac{1}{2(1\!-\!\epsilon)^2 \mathcal{H}^2 \| \Delta\vec{x} \|} 
	\frac{\partial}{\partial \| \Delta\vec{x} \|}
	= \frac{1}{2\mathcal{Y}} \frac{\partial}{\partial \mathcal{Y}} \, .
\end{equation}
This allows us to write the equal-time electric and magnetic field correlators as,
\begin{align}
&
\bigl\langle \Omega \bigr| \hat{E}_i(\eta,\vec{x}) \,
	\hat{E}_j(\eta,\vec{x}^{\,\prime}) \bigl| \Omega \bigr\rangle
\nonumber  \\
&	\hspace{2cm}
	= 
	\frac{\mathcal{Y}^4}{\| \Delta\vec{x} \|^4}
	\Biggl\{ \delta_{ij}
	\biggl[  \frac{2}{\mathcal{Y}} + \frac{3\mathcal{Y} }{4} 
			+ \frac{\mathcal{Y}^2}{4} \frac{\partial}{\partial \mathcal{Y} }
	\biggr]
\label{EE}
\\
&	\hspace{4cm}
+
	\biggl( \frac{\Delta x_i \, \Delta x_j}{\| \Delta\vec{x} \|^2} \biggr)
	\biggl[
		- \frac{1}{ \mathcal{Y} } - \frac{3\mathcal{Y} }{2} 
			+ \biggl( 1 - \frac{\mathcal{Y}^2}{2} \biggr) \frac{\partial}{\partial \mathcal{Y} }
	\biggr]
	\Biggr\}
	\frac{\partial}{\partial \mathcal{Y} }
	\mathscr{F}_\nu(\mathcal{Y}) \, ,
\nonumber \\
&
\bigl\langle \Omega \bigr| \hat{E}_i(\eta,\vec{x}) \,
			\hat{B}_j(\eta,\vec{x}^{\,\prime}) \bigl| \Omega \bigr\rangle
\nonumber \\
&	\hspace{2cm}
	= 
	\frac{\mathcal{Y}^4}{\| \Delta\vec{x} \|^4}
	\Biggl\{ 
	\biggl( \varepsilon_{ijk} \frac{\Delta x_k}{\| \Delta \vec{x} \|} \biggr) 
	\biggl[ - \frac{3}{2} - \frac{\mathcal{Y} }{2} \frac{\partial}{\partial \mathcal{Y} } 
		\biggr] 
	\Biggr\}
	\frac{\partial}{\partial \mathcal{Y} } \mathscr{F}_\nu(\mathcal{Y}) \, ,
\label{EB}
\\
&
\bigl\langle \Omega \bigr| \hat{B}_i(\eta,\vec{x}) \,
	\hat{B}_j(\eta,\vec{x}^{\,\prime}) \bigl| \Omega \bigr\rangle
\nonumber \\
&	\hspace{2cm}
	= 
	\frac{\mathcal{Y}^4}{\| \Delta\vec{x} \|^4}
	\Biggl\{ \delta_{ij} 
		\biggl[ - \frac{1}{\mathcal{Y}} - \frac{\partial}{\partial \mathcal{Y} } 
			\biggr] 
	+ \biggl( \frac{ \Delta x_{i} \Delta x_{j} }{\| \Delta\vec{x} \|^2} \biggr)
		\biggl[ - \frac{1}{\mathcal{Y}} + \frac{\partial}{\partial \mathcal{Y}} 
			\biggr] \Biggr\} \frac{\partial}{\partial \mathcal{Y}}
				\mathscr{F}_\nu(\mathcal{Y}) \, ,
\label{BB}
\end{align}
where we rewrote the overall factor 
as~$\bigl[ (1\!-\!\epsilon) \mathcal{H} \bigr]^4\!=\!\mathcal{Y}^4/\| \Delta\vec{x} \|^4$ 
for later convenience.

\subsubsection{Sub-Hubble limit}

The sub-Hubble limit is defined 
by~$\mathcal{Y} \!=\! (1\!-\!\epsilon)\mathcal{H}\| \Delta\vec{x} \| \! \ll \! 1$,
which we denote by~$\stackrel{{\rm sH} }{\sim}$.
In this limit the function~$\mathscr{F}_\nu(\mathcal{Y})$ from~(\ref{F(Y) def}) reduces to
({\it cf.} Eq.~(\ref{Fnu: Taylor series})),
\begin{align}
\mathscr{F}_\nu(\mathcal{Y})
	\stackrel{{\rm sH} }{\sim}
	\frac{1}{\pi^2 \mathcal{Y}^2 } \, ,
\label{F sH}
\end{align}
so that the electric and magnetic correlators~(\ref{EE})--(\ref{BB})
at leading order reduce to
\begin{align}
\bigl\langle \Omega \bigr| \hat{E}_i(\eta,\vec{x}) \,
	\hat{E}_j(\eta,\vec{x}^{\,\prime}) \bigl| \Omega \bigr\rangle
	\stackrel{{\rm sH} }{\sim}{}&
	\frac{1}{\pi^2 \| \Delta\vec{x} \|^4}
	\biggl[
	- \delta_{ij} 
	+ 2 \biggl( \frac{ \Delta x_{i} \Delta x_{j} }{\| \Delta\vec{x} \|^2} \biggr)
	+ \mathcal{O}(\mathcal{Y}^2) \biggr] \, ,
\label{sH EE}
\\
\bigl\langle \Omega \bigr| \hat{E}_i(\eta,\vec{x}) \,
	\hat{B}_j(\eta,\vec{x}^{\,\prime}) \bigl| \Omega \bigr\rangle
	\stackrel{{\rm sH} }{\sim}{}&
	\frac{1}{\pi^2 \| \Delta\vec{x} \|^4}
	\biggl[
	0
	+ \mathcal{O}(\mathcal{Y}^2) \biggr] \, ,
\label{sH EB}
\\
\bigl\langle \Omega \bigr| \hat{B}_i(\eta,\vec{x}) \,
	\hat{B}_j(\eta,\vec{x}^{\,\prime}) \bigl| \Omega \bigr\rangle
	\stackrel{{\rm sH} }{\sim}{}&
	\frac{1}{\pi^2 \| \Delta\vec{x} \|^4}
	\biggl[
	- \delta_{ij} 
	+ 2 \biggl(\frac{ \Delta x_{i} \Delta x_{j} }{\| \Delta\vec{x} \|^2}\biggr)
	+ \mathcal{O}(\mathcal{Y}^2) \biggr] \, .
\label{sH BB}
\end{align}
Interestingly, these match the E\&M correlators of 
a massless vector field.~\footnote{
The limiting form in~(\ref{F sH}) is correct as long as~$\mathcal{Y}\!<\!1/|\nu|$
in addition to~$\mathcal{Y}\!\ll\!1$. If that is not satisfied, the proper sub-Hubble
asymptotic behaviour is captured by the~$| \nu | \!\gg\! 1$ expansion 
of~$\mathscr{F}_\nu(\mathcal{Y})$, which corresponds to~(\ref{F flat}) with the
substitution~$m^2\!\to\!(q\overline{\phi}_0)^2$. 
This limit requires~$(q\overline{\phi}_0)^2 \!\gg\! H_0^2$, which is possible 
for very large and negative non-minimal coupling~$\xi$, or for very small
quartic self-coupling~$\lambda$.
We do not consider these cases here.
}

\subsubsection{Super-Hubble limit}

The interesting limit of the E\&M correlators is the super-Hubble limit
defined by~$\mathcal{Y} \!=\! (1\!-\!\epsilon)\mathcal{H}\| \Delta\vec{x} \| \! \gg \! 1$,
which we denote by~$\stackrel{{\rm SH} }{\sim}$.
In this limit the asymptotic behaviour of the function~$\mathscr{F}_\nu$ 
is given by the equal-time limit and the~$D\!=\!4$ limit of the asymptotic 
expansion~(\ref{F asy}) from Appendix~\ref{app: Rescaled propagator function F_nu},
\begin{equation}
\mathscr{F}_\nu(\mathcal{Y})
	\stackrel{{\rm SH} }{\sim} 
		\frac{\Gamma(\nu) \, \Gamma\bigl( \tfrac{3}{2} \!-\! \nu \bigr)}
		{ (2\pi)^2 \sqrt{\pi} } \,
		\mathcal{Y}^{ - 3 + 2\nu}
		+ 
		\frac{\Gamma(-\nu) \, \Gamma\bigl( \tfrac{3}{2} \!+\! \nu \bigr)}
		{ (2\pi)^2 \sqrt{\pi} } \,
		\mathcal{Y}^{ - 3 - 2\nu} \, ,
\label{scrF asymptotic}
\end{equation}
where we have used the Legendre duplication formula to simplify the coefficients.
We remind the reader that the index~$\nu$ given in~(\ref{nu corr}) either takes
real values~$0 \! \le \! \nu \!<\! 1/2 $, or is imaginary, depending on the value of the
condensate parameter~$q \overline{\phi}_0/H_0$.
 In case of~$\nu\!>\!0$
the first term in the asymptotic expansion~(\ref{scrF asymptotic}) 
dominates over the second one. However,
as~$\nu$ approaches 0, the two become of the same order. For imaginary~$\nu$
the two terms are exactly of the same order, both scaling as~$\mathcal{Y}^{-3}$, 
and they pick up an overall oscillating factor 
with a logarithmically growing frequency,
\begin{equation}
\mathscr{F}_{i|\nu|}(\mathcal{Y})
	\stackrel{{\rm SH} }{\sim} 
		2 \, {\rm Re} \biggl[\frac{\Gamma(i |\nu|) \, \Gamma\bigl( \tfrac{3}{2} \!-\! i |\nu| \bigr)}
		{(2\pi)^2 \sqrt{\pi} } 
		e^{ 2i|\nu| \ln(\mathcal{Y}) } \biggr] \, \mathcal{Y}^{-3} \, .
\end{equation}
Thus, the asymptotic form in~(\ref{scrF asymptotic}) captures the leading order super-Hubble 
behaviour in all cases. Using it in~(\ref{EE})--(\ref{BB}) captures the leading order super-Hubble
behaviour of E\&M correlators,
\begin{align}
\MoveEqLeft[2]
\bigl\langle \Omega \bigr| \hat{E}_i(\eta,\vec{x}) \,
	\hat{E}_j(\eta,\vec{x}^{\,\prime}) \bigl| \Omega \bigr\rangle
\nonumber \\
&	
	\stackrel{{\rm SH} }{\sim}
	\frac{ \mathcal{Y} }{ \pi^2 \, \| \Delta\vec{x} \|^4 }
	\biggl[ \frac{ \bigl( \frac{1}{2} \!-\! \nu \bigr) \, \Gamma(\nu) \, 
		\Gamma\bigl( \frac{5}{2} \!-\! \nu \bigr) }{ 4 \sqrt{\pi} } \, \mathcal{Y}^{2\nu}
\nonumber \\
&	\hspace{4cm}
		+
		\frac{ \bigl( \frac{1}{2} \!+\! \nu \bigr) \, \Gamma(-\nu) \, 
			\Gamma\bigl( \frac{5}{2} \!+\! \nu \bigr) }{ 4\sqrt{\pi} } \, \mathcal{Y}^{-2\nu} \biggr]
	\biggl[ \delta_{ij} - 2 \biggl( \frac{\Delta x_i \Delta x_j}{ \| \Delta\vec{x} \|^2 } \biggr) \biggr] \, ,
\label{SH EE}
\\
\MoveEqLeft[2]
\bigl\langle \Omega \bigr| \hat{E}_i(\eta,\vec{x}) \,
	\hat{B}_j(\eta,\vec{x}^{\,\prime}) \bigl| \Omega \bigr\rangle \, ,
\nonumber \\
&	
	\stackrel{{\rm SH} }{\sim}
	\frac{1}{\pi^2 \| \Delta\vec{x} \|^4} \biggl[
		- \frac{\bigl( \frac{1}{2} \!-\! \nu \bigr) \, \Gamma(\nu) \,
			\Gamma\bigl( \frac{5}{2} \!-\! \nu \bigr) }{ 2\sqrt{\pi} } \, \mathcal{Y}^{2\nu}
\nonumber \\
&	\hspace{4cm}
		- \frac{\bigl( \frac{1}{2} \!+\! \nu \bigr) \, \Gamma(-\nu) \,
			\Gamma\bigl( \frac{5}{2} \!+\! \nu \bigr) }{ 2\sqrt{\pi} } \, \mathcal{Y}^{-2\nu}
		\biggr] \biggl( \varepsilon_{ijk} \frac{\Delta x_k}{\| \Delta\vec{x} \|} \biggr) \, ,
\label{SH EB}
\\
\MoveEqLeft[2]
\bigl\langle \Omega \bigr| \hat{B}_i(\eta,\vec{x}) \,
	\hat{B}_j(\eta,\vec{x}^{\,\prime}) \bigl| \Omega \bigr\rangle 
\nonumber \\
&	
	\stackrel{{\rm SH} }{\sim}
	\frac{ \mathcal{Y}^{-1} }{ \pi^2 \, \| \Delta\vec{x} \|^4 }
	\biggl[ - \frac{ \bigl( \frac{3}{2} \!-\! \nu \bigr) \, \Gamma(\nu) \,
				\Gamma\bigl( \frac{5}{2} \!-\! \nu \bigr) }{ \sqrt{\pi} } \, \mathcal{Y}^{2\nu}
			- \frac{ \bigl( \frac{3}{2} \!+\! \nu \bigr) \, \Gamma(-\nu) \,
				\Gamma\bigl( \frac{5}{2} \!+\! \nu \bigr) }{ \sqrt{\pi} } \, \mathcal{Y}^{-2\nu} \biggr] \delta_{ij}
\nonumber
\\
&	\hspace{1.5cm}
	+ \frac{ \mathcal{Y}^{-1} }{ \pi^2 \, \| \Delta\vec{x} \|^4 }
	\biggl[ \frac{ \Gamma(\nu) \, \Gamma\bigl( \frac{7}{2} \!-\! \nu \bigr) }{ \sqrt{\pi} } \, 
				\mathcal{Y}^{2\nu}
			+ \frac{ \Gamma(-\nu) \, \Gamma\bigl( \frac{7}{2} \!+\! \nu \bigr) }{ \sqrt{\pi} } \, 
				\mathcal{Y}^{-2\nu} \biggr] 
		\biggl( \frac{\Delta x_i \Delta x_j}{ \| \Delta\vec{x} \|^2 } \biggr) \, .
\label{SH BB} 
\end{align}
It is clear from the overall factors of the expressions above that
the~$EE$ correlator at super-Hubble separations is enhanced 
compared to the conformal one, while the $BB$ correlator is suppressed.
Therefore, the effect of the scalar condensate is to enhance the electric field, 
and to screen the magnetic field, compared to the case of vanishing condensate
(the symmetric phase). Furthermore, it is also clear from the overall factors above that
the E\&M correlators satisfy a hierarchy at super-Hubble separations,
recently reported for power-law inflation in~\cite{Glavan:2019uni,Kobayashi:2019uqs},
which reads (up to numerical factors),
\begin{align}
&
\bigl\langle \hat{E}_i(\eta,\vec{x}) \, \hat{E}_j(\eta,\vec{x}^{\,\prime}) \bigr\rangle
	\Big/
\bigl\langle \hat{E}_i(\eta,\vec{x}) \, \hat{B}_j(\eta,\vec{x}^{\,\prime}) \bigr\rangle
	\Big/
\bigl\langle \hat{B}_i(\eta,\vec{x}) \, \hat{B}_j(\eta,\vec{x}^{\,\prime}) \bigr\rangle
\nonumber \\
&	\hspace{8cm}
	\! \stackrel{\rm \ss SH }{\sim} \!
\
	\bigl( \mathcal{H} \| \Delta\vec{x} \| \bigr)^2
		\Big/
	\mathcal{H} \| \Delta\vec{x} \|
		\Big/
	1 \, .
\label{EM hierarchy}
\end{align}
The scaling of the ratios above is a consequence of Faraday's law in expanding
space-times, as reported in~\cite{Kobayashi:2019uqs}, and does not depend on the
sources. The hierarchy between the correlators develops dynamically in accelerating 
space-times.

\begin{figure}[h!]
\begin{minipage}[c]{7.cm}
\vskip+2mm
\begin{center}
$\nu=1/4$
\end{center}
\raggedright
	\includegraphics[width=7.cm]{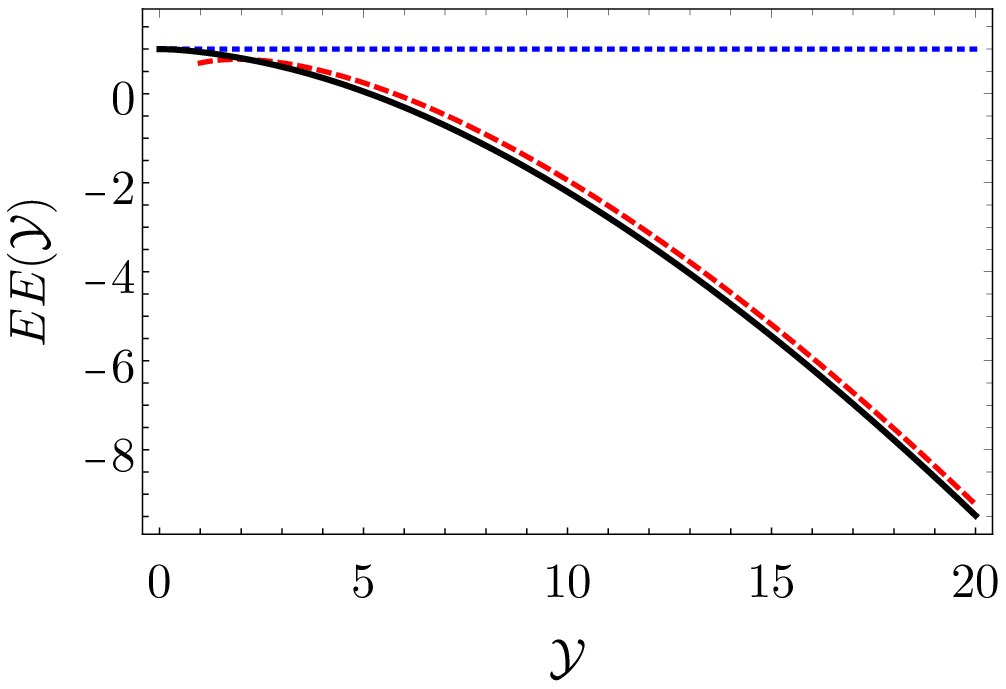}
	\vskip+6mm
	\includegraphics[width=7.cm]{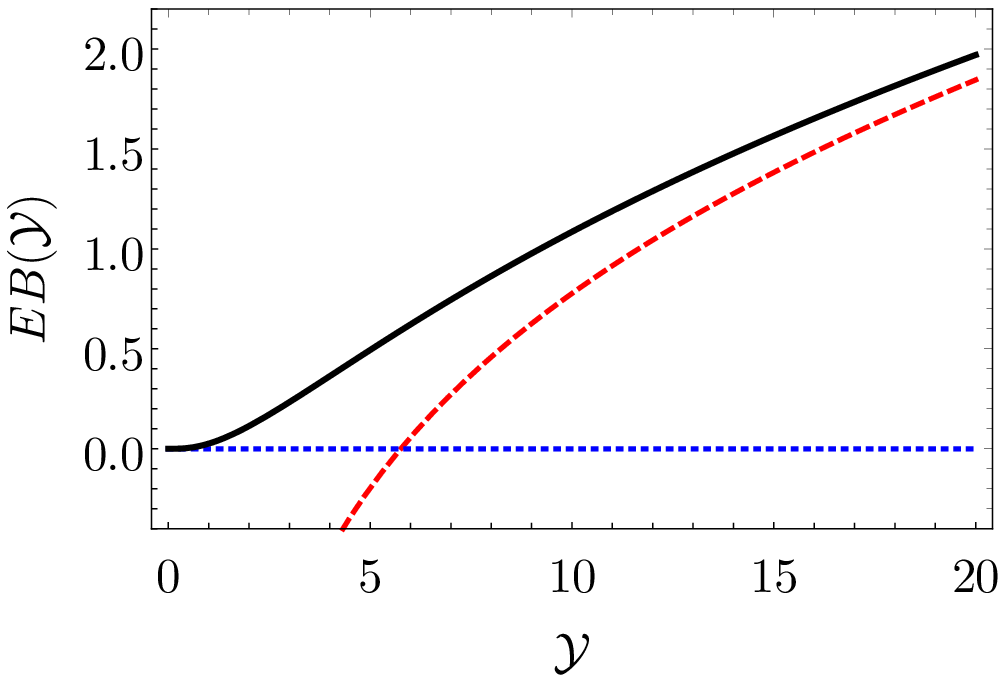}
	\vskip+6mm
	\includegraphics[width=7.cm]{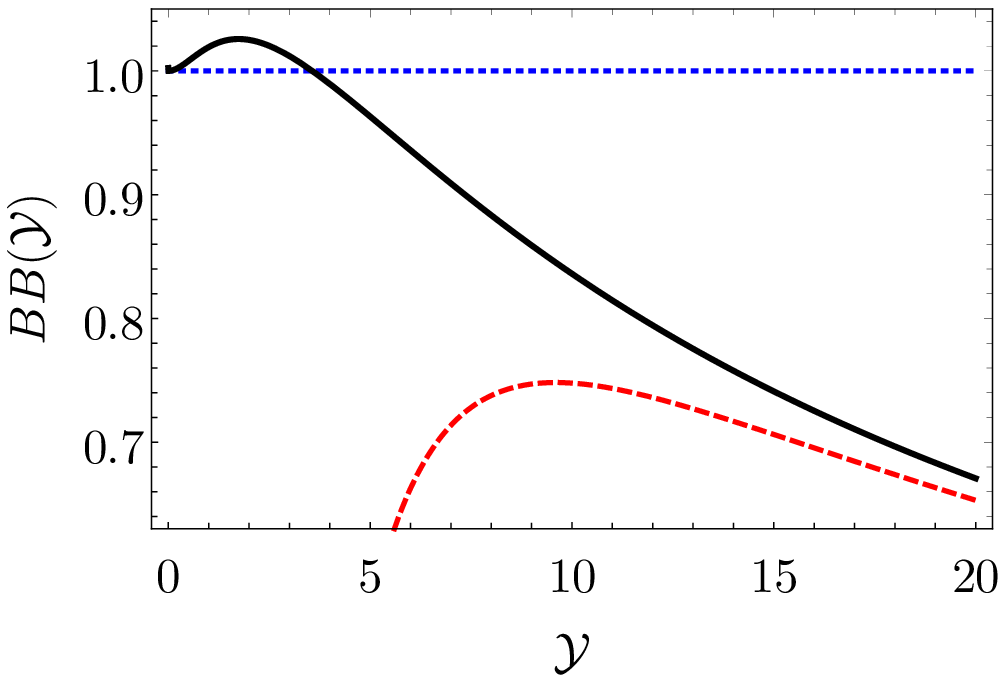}
\end{minipage}
\hfill\vline\vline\hfill
\begin{minipage}[c]{7.cm}
\vskip+2mm
\begin{center}
$\nu=3i/2$
\end{center}
\raggedleft
	\includegraphics[width=7.cm]{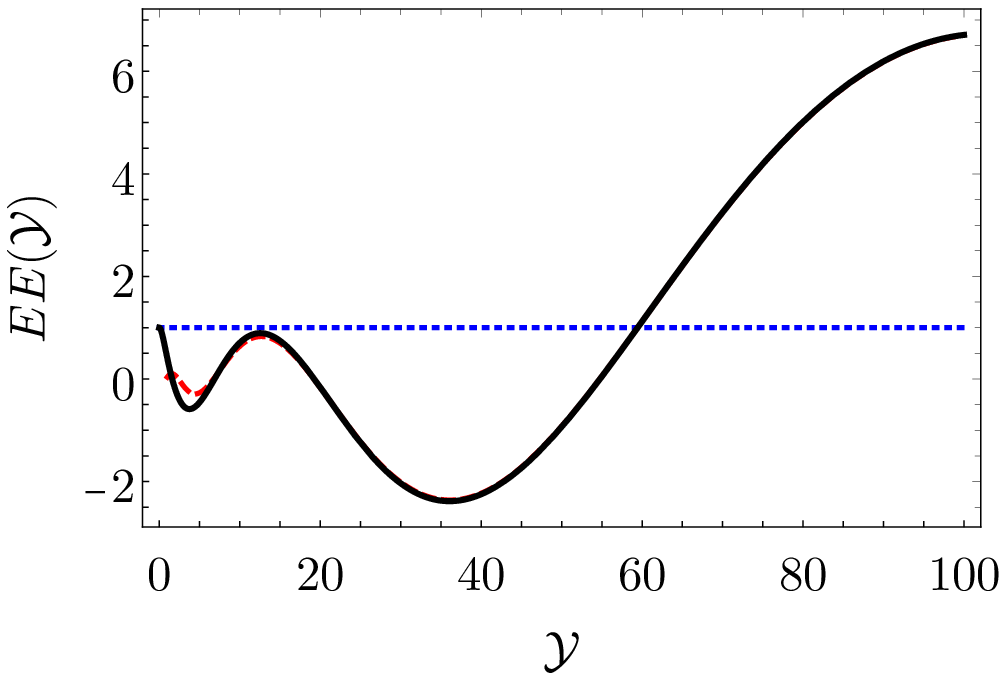}
	\vskip+6mm
	\includegraphics[width=7.cm]{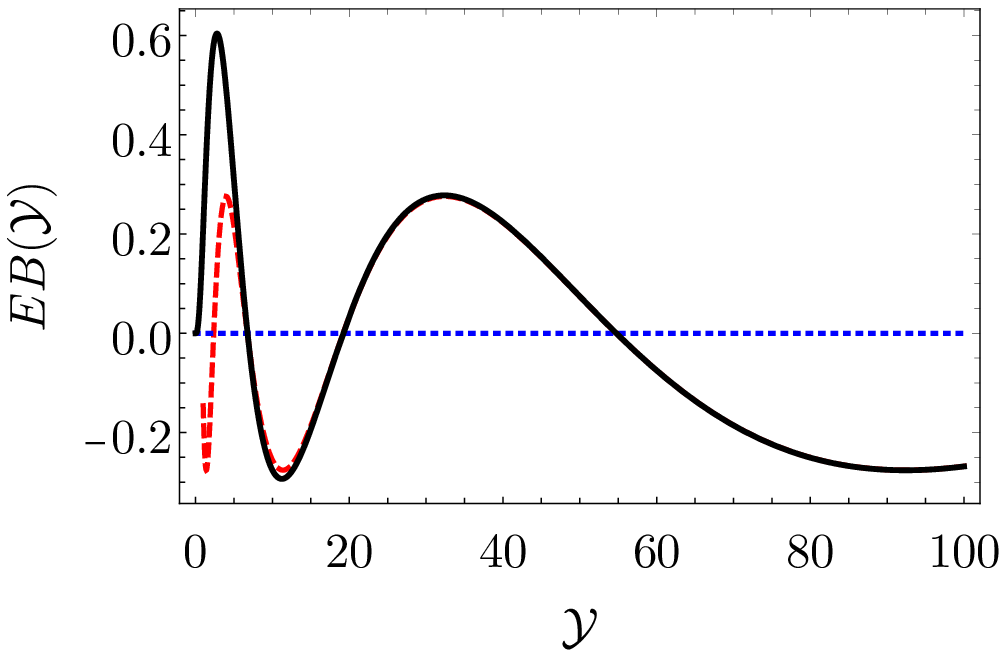}
	\vskip+7mm
	\includegraphics[width=7.cm]{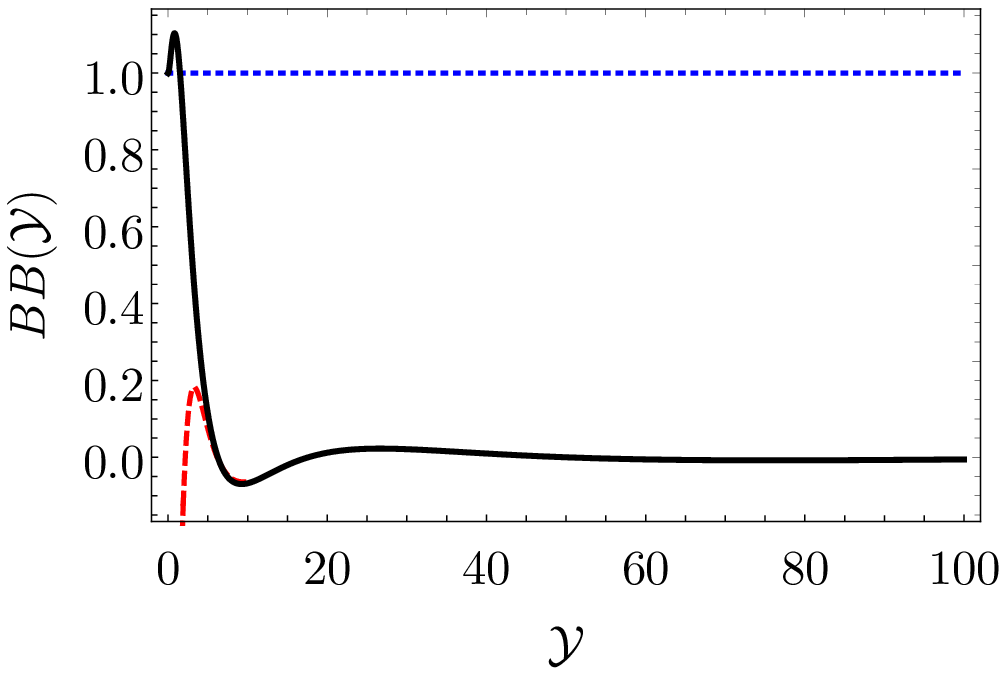}
\end{minipage}
\caption{
Plots of the contracted E\&M correlators~(\ref{EE contracted})--(\ref{BB contracted}), rescaled
by the contracted conformal correlator for two cases: $\nu\!=\!1/4$ (left), 
and~$\nu\!=\!3i/2$ (right). The solid black lines represents the full contracted 
correlators, dashed red lines are the corresponding super-Hubble asymptotic forms, and dotted blue
lined denote the conformal correlators.
}
\label{EM plots}
\end{figure}

It is instructive to examine appropriately contracted correlators, 
\begin{align}
\delta_{ij} \bigl\langle \Omega \bigr| \hat{E}_i(\eta,\vec{x}) \,
	\hat{E}_j(\eta,\vec{x}^{\,\prime}) \bigl| \Omega \bigr\rangle 
	={}&
	- \frac{1}{\pi^2 \| \Delta\vec{x} \|^4} \times EE(\mathcal{Y}) \, ,
	\qquad
\label{EE contracted}
\\
\frac{\Delta x_k}{\| \Delta\vec{x} \|}\varepsilon_{kij}\bigl\langle \Omega \bigr| \hat{B}_i(\eta,\vec{x}) \,
	\hat{E}_i(\eta,\vec{x}^{\,\prime}) \bigl| \Omega \bigr\rangle 
	={}&
	- \frac{1}{\pi^2 \| \Delta\vec{x} \|^4} \times EB(\mathcal{Y}) \, ,
\label{EB contracted}
\\
\delta_{ij} \bigl\langle \Omega \bigr| \hat{B}_i(\eta,\vec{x}) \,
	\hat{B}_j(\eta,\vec{x}^{\,\prime}) \bigl| \Omega \bigr\rangle 
	={}&
	- \frac{1}{\pi^2 \| \Delta\vec{x} \|^4} \times BB(\mathcal{Y}) \, ,
\label{BB contracted}
\end{align}
normalized to the conformal 
correlators~$\langle \hat{E}_i(\eta,\vec{x}) \, \hat{E}_i(\eta,\vec{x}^{\,\prime}) \rangle_{\rm cf}\!=\!\langle \hat{B}_i(\eta,\vec{x}) \, \hat{B}_i(\eta,\vec{x}^{\,\prime}) \rangle_{\rm cf}\!=\!-1/(\pi^2\| \Delta\vec{x} \|^4)$,
{\it i.e.} the E\&M correlators of a massless vector field.
They are plotted in Fig.~\ref{EM plots} for two cases of real and imaginary~$\nu$.
%
Note that the~$EE$ correlator is in general enhanced at super-Hubble separations
compared to the conformal one, the~$EB$ exhibits at best a mild growth, and the~$BB$
correlator is suppressed compared to the conformal value. At sub-Hubble separations
all three correlators are negative (as are the conformal ones) signifying anti-correlation of E\&M
field fluctuations on those scales.
 On super-Hubble separations, and for positive~$\nu$
only the~$EE$ correlator is positive, while the remaining two are negative, while
for imaginary~$\nu$ they all oscillate between positive and negative values
with a logarithmically growing frequency. 
The appearance of negative correlators is not unusual in quantum field theory, 
see {\it e.g.} Ref.~\cite{Ford:2004wc}.


\section{Discussion}
\label{sec: Discussion}

In this paper we consider the spectator Abelian Higgs model in {\it power-law inflation},
characterized by a constant principal slow-roll 
parameter~$\epsilon\!=\-\dot{H}/H^2 \!=\! {\rm const.}$,~$0 \!\le\! \epsilon \! < \! 1$
in~$D$ space-time dimensions.
We impose the~{\it unitary gauge} and perform canonical quantization, 
and construct the propagators for the massive scalar and the massive vector fluctuations 
of the model. 
We assume that the vector and the scalar masses are generated by 
a time-dependent condensate of the complex scalar field $\overline{\phi}(t)$.
This condensate can 
be a true scalar condensate or it can be composed of super-Hubble scalar field fluctuations,
as it is often the case in inflation~\cite{Prokopec:2002uw,Prokopec:2002jn,Prokopec:2003tm,Prokopec:2006ue,Prokopec:2008gw,Prokopec:2007ak}.
We assume that the condensate scales
as the Hubble parameter,  $\overline{\phi}(t)\!\propto\! H$, 
in which case finding the solution for the propagator dramatically simplifies.

\medskip

 For pedagogical reasons we 
fix the unitary gauge of the Abelian Higgs model by using {\it Dirac's
canonical quantization}, which is suitable for studying systems with constraints.
Even though Dirac's quantization is standard in 
some other areas of physics, it is rarely used in cosmology,
and this justifies our detailed analysis in
Sec.~\ref{sec: Abelian Higgs model in cosmological spaces}.
The symmetries of the cosmological background naturally 
select spatial momentum space, in which the analysis reduces to solving 
the equations of motion for the suitably rescaled mode functions. The details of that procedure
are given in section~\ref{subsec: Hamiltonian formulation}.
 Section~\ref{sec: Dynamics of vector perturbations}  is devoted
 to the evolution of the perturbations of the vector mode functions 
 and in section~\ref{sec: Vector field two-point functions}
 we construct the 
 {\it covariantized two-point functions} for the vector field:
 the Wightman function~(\ref{covariant form}), 
 the Feynman propagator~(\ref{Feynman propagator solution}), 
 and the Green's function~(\ref{Green function}), with the scalar
 structure functions given in~(\ref{C1 solution})--(\ref{C4 solution}).
These constitute the main results of our paper. 
The distinction between the Feynman propagator and the Green's function for the
vector propagator is a particular feature of the unitary gauge, following from the
canonical quantization, and has implications for the renormalizability of the theory,
which is not well understood in curved space-times.
Together with the interaction vertices, these propagators
represent the basic building blocks  
for dimensionally regulated perturbation theory
of the Abelian Higgs model in the unitary gauge
on accelerating cosmological spaces.~\footnote{
One can also use these propagators on decelerating cosmological backgrounds 
with~$\epsilon\!=\!{\rm const.}\!>\!1$, as the form of the propagators depends only
on the condensate evolution tracing the Hubble rate,~$\overline{\phi}\!\propto\!H$,
which is maintained here. However, one should keep in mind that in decelerating spaces 
the CTBD state is {\it not} preferred, 
as it is in accelerating spaces  based on 
simple 
physical arguments presented in 
section~\ref{subsec: Choice of the state}.
On the contrary, the quantum state depends profoundly on the evolution of
the Universe preceding the decelerating epoch. Scalar field examples 
for how to deal with such situations are given in 
Refs.~\cite{Glavan:2013mra,Glavan:2014uga,Glavan:2015cut,Glavan:2017jye}.
}

In Sec.~\ref{subsec: Various limits} 
we check the consistency of our
propagator by working out several limits, and  verifying that they reproduce the known results,
including  that it reduces to the correct massive vector propagator in the unitary 
gauge~(\ref{photon propagator: flat})
in Minkowski space-time.
We pay special attention 
to the de Sitter limit of the propagator,
which has been studied in the literature,
and verify that it 
correctly reproduces the corresponding
Allen-Jacobson massive vector propagator~\cite{Allen:1985wd}.
In addition, our propagator also provides the 
missing~$i\varepsilon$-prescriptions for the tensor structures,
responsible for the 
distinction between the Feynman propagator 
and the Green's function. 
We also comment on how the propagators of the gauge-fixed Stueckelberg theory
in de Sitter~\cite{Frob:2013qsa} are related to the Abelian Higgs model
in the broken phase.

The issue of gauge dependence is often neglected or taken lightly in cosmological studies.
The importance of understanding gauge dependence in cosmological loop calculations involving
graviton and/or gauge field loops was stressed recently in Refs.~\cite{Miao:2017feh,Miao:2018bol}. 
In order to understand gauge dependence of some (physical) quantity it is important to 
work with a proper regularization that respects all of the relevant symmetries of the problem
and have available the relevant propagators in a wide class of gauges, or at 
least in different fixed gauges. This work is a step in that direction.

The  vector propagators in the unitary gauge constructed in this work 
can be used to address various problems in cosmological spaces that involve
perturbative loop calculations, such as loop corrections induced 
by the massive vector perturbations on the cosmological perturbations,
on the effective inflaton potential,
and on the cosmological background (through quantum backreaction). 
All of these effects may leave observable imprints not only during inflation, 
but also in the subsequent radiation and matter eras, thus warranting a further study.
Here we utilized the massive vector propagator in Sec.~\ref{sec: Field strength correlator} 
to study the tree-level off-coincident electric 
and magnetic field correlators at super-Hubble separations during power-law inflation.
We find the~$EE$ correlators of the massive vector field to be larger than in the massless
vector case, while the converse is true for the~$BB$ correlators. Furthermore, the E\&M
correlators satisfy a hierarchy at super-Hubble separations, 
consistent with the results obtained from Faraday's law in~\cite{Glavan:2019uni,Kobayashi:2019uqs}. 
When evolved to late times, these correlators may be important for our understanding 
of the origin of the observed large scale magnetic fields both at the galactic and cluster scales.

\medskip

The propagators constructed here are, strictly speaking, applicable to the
Abelian Higgs model which is~(i) a spectator in power-law 
inflation,~$\epsilon\!=\!{\rm const.}$, and 
(ii) for which the scalar condensate traces the evolution of the Hubble 
rate~$\overline{\phi}\!\propto\!H\!=\!H_0  \, a^{-\epsilon}$.
Nevertheless, the assumption of the exact power-law inflationary background
can be relaxed under certain conditions, and the validity of the propagators
extended to the more general backgrounds 
of {\it slow-roll inflation}, where the 
scalar condensate is the inflaton driving a slow-roll inflation.

In slow-roll inflation the principal slow-roll 
parameter~$\epsilon\!=\! -\dot{H}/H^2$ is time-dependent,~$\dot{\epsilon}\!\neq\!0$, 
but varies slowly such that its time derivative can be considered a small parameter,
\begin{equation}
\epsilon_2 \equiv \frac{\dot{\epsilon}}{ \epsilon \,  H} \, ,
\qquad \qquad
|\epsilon_2 |\ll 1
\, ,
\label{epsilon 2}
\end{equation}
where $\epsilon_2$ is the second slow-roll parameter.
From the cosmological microwave background 
measurements~\cite{Akrami:2018vks} 
we know the scalar spectral index, 
$n_s \!\simeq -\! 2\epsilon \!-\! \epsilon_2 \!\simeq\! 0.965$,
and the tensor-to-scalar ratio, $r \!\simeq\! 16\epsilon \!<\! 0.065$,
which imply the following hierarchy, 
\begin{equation}
\epsilon \ll \epsilon_2 \ll 1, 
\label{hierarchy of slow roll}
\end{equation}
where $\epsilon_2 \!\simeq \!1-n_s \!\simeq\! 0.03$.
The Hubble rate in slow-roll inflation scales with time as, 
\begin{equation}
H(t) = H_0 \, \exp \biggl[ - \int_{t_0}^t \! dN(t') \, \epsilon(t') \biggr]
\,,
\label{Hubble rate general}
\end{equation}
where $N(t)\!=\!\ln[a(t)]$ ($N(t_0)=0$)
is the number of $e$-foldings. 
Our propagators can be seen to apply to slow-roll inflation, under certain conditions, 
in cases when the scalar condensate is a spectator, and when it is the inflaton itself driving 
the expansion. We comment below on the restrictions in these two cases.

\begin{itemize}
\item
{\bf Spectator scalar condensate.}
When slow-roll inflation is driven by some other inflaton field, 
and the Abelian Higgs model is a spectator, its scalar condensate 
is attracted towards the so-called {\it scaling regime}, where it scales as the 
Hubble rate,
\begin{equation}
\overline\phi(t)
	= \overline\phi_0 \, \exp \biggl[ - \int_{t_0}^t \! dN(t') \, \epsilon(t') \biggr]
	= \frac{\overline{\phi}_0}{H_0} H(t) \, .
\end{equation}
This is the first condition necessary for our propagators to be valid.
The second condition concerns the time-dependence of the Hubble rate,
which needs to take the same form as in power-law inflation with
the constant~$\epsilon$ promoted to a slowly varying 
one,~$H(t) \!\approx\! H_0 \, \bigl[a(t)\bigr]^{-\epsilon(t)}$.
To see under which conditions this holds, we can make a derivative 
expansion~\footnote{
The derivative expansion is generated by iterating partial integrations of the time derivatives,
\begin{align*}
&
\int_{t_0}^{t} \! dN(t') \, \epsilon(t')
  \!=\! \int_0^{N(t)} \! \! \! dN'\,\epsilon(N')
	= \epsilon(t) N(t) \biggl[ 1 - \frac{1}{2} N(t) \epsilon_2(t) \biggr]
		+ \frac{1}{2} \int_{0}^{N(t)} \!\!\! dN' {N'}^2 
	  \frac{d}{dN'} \Bigl[ \epsilon(N') \epsilon_2(N'') \Bigr]
\, ,
\end{align*}
where the derivative inside the last 
integral can be expressed in 
terms of higher slow-roll parameters defined 
as~$(d/dN)\epsilon_n \!=\! \epsilon_{n+1}$.
The last term above is the remainder integral,
which can be used to estimate the error
of the derivative expansion.
This approximation scheme is clearly 
superior to the one obtained 
by promoting the Hubble rate into a local function of time
in the de Sitter two-point functions.
}
in the exponent of~(\ref{Hubble rate general}), 
\begin{equation}
H(t) \approx
	H_0 \, \exp \biggl\{ - \epsilon(t)  N(t) 
	\biggl[ 1 - \frac{1}{2}\epsilon_2(t) N(t)\biggr] 
	\biggr\}
\,.
\end{equation}
The first term in the exponent corresponds precisely to the power-law inflation scaling 
with an adiabatically evolving principal slow-roll parameter. Therefore,
our propagators will be applicable
as long as the second term in the exponent can be neglected, implying,
\begin{equation}
\Delta N\ll \frac{2}{\epsilon_2}\simeq 60
\,,
\label{estimate accuracy}
\end{equation}
Given that currently only~$\Delta N\!\sim\!10$ $e$-foldings of inflation are observable,
our propagators can safely be used to study the spectator Abelian Higgs model in slow-roll 
inflation.

\item
{\bf Inflaton scalar condensate.} When slow-roll inflation is driven by a condensate
of the Abelian Higgs model, such as Higgs inflation (see~\cite{Bezrukov:2013fka}
and references therein), the condensate scales
in time as
\begin{equation}
\phi(t) = \overline{\phi}_0 - M_{\ss \rm P} \int_{t_0}^{t} \! dN(t') \, \sqrt{ 2 \epsilon(t') } \, ,
\label{inflaton scaling}
\end{equation}
where~$M_{\ss \rm P} \!=\! 1/\sqrt{8\pi G_{\! \ss \rm N}}$.
This is clearly not the same scaling as that of the Hubble 
rate~(\ref{Hubble rate general}). This inflaton condensate is what generates
the mass of the vector field, and our propagators apply only when it approximately
traces the evolution of the Hubble rate. This can be seen to hold for a certain number 
of~$e$-foldings. Making use of the derivative expansion, the condensate evolution 
can be written as
\begin{equation}
\overline{\phi}(t) \approx \overline{\phi}_0 \biggl\{ 1
	- N(t) \frac{\sqrt{2 \epsilon(t)} \, M_{\ss \rm P}}{\overline{\phi}_0} 
\biggl[ 1 - \frac{1}{4}N(t) \epsilon_2(t)
\biggr]	\biggr\} \, .
\end{equation}
This will correspond to the scaling of the Hubble rate provided that,
in addition to~(\ref{estimate accuracy}), 
the following two conditions hold,
\begin{equation}
\Delta N\ll \frac{1}{\epsilon} \, ,
\qquad \qquad
\overline{\phi}_0\approx \frac{2M_{\rm \ss P}}{\sqrt{2\epsilon}}
\,.
\label{conditions on applicability in slow roll}
\end{equation}
The hierarchy~(\ref{hierarchy of slow roll}) 
implies that  this first condition 
is weaker than~(\ref{estimate accuracy})
and thus not an additional constraint. 
The second condition 
in~(\ref{conditions on applicability in slow roll})
can be put in perspective if one recalls
that, for a simple class of potentials,
$V \!\propto\! \phi^n$, the inflaton in the slow roll regime
develops field values
$\overline{\phi} \!\simeq \! n M_{\rm P}/\sqrt{2\epsilon}$.
The case of the quadratic potential~$n\!=\!2$ then satisfies 
the condition~(\ref{conditions on applicability in slow roll}).
For general inflationary potentials, the condition on $\overline{\phi}_0$ 
in~(\ref{conditions on applicability in slow roll})
singles out the value of the inflaton condensate
around which our approximation applies for the range of 
$e$-foldings given in~(\ref{estimate accuracy}).

\end{itemize}

Finally, it is worth noting that our propagators may be
applicable to vector field inflation driven by a condensate
of a nonminimally coupled vector 
field~\cite{Golovnev:2008cf,Golovnev:2008hv,Dimopoulos:2006ms,Dimopoulos:2008yv,Himmetoglu:2008zp,Himmetoglu:2008hx,EspositoFarese:2009aj,Karciauskas:2010as}.
However, since in these models the vector field develops 
a condensate and thus couples to gravitational perturbations, 
a more elaborate analysis is needed to 
establish the precise conditions of applicability of 
our propagator in this class of models. The canonical quantization methods utilized here
should provide useful in such studies.

\acknowledgments

We are thankful to Gerard `t Hooft for the discussion on the unitary gauge.
D.~G. is grateful to the Division of Theoretical Physics of Rudjer Bo\v{s}kovi\'{c} 
Institute in Zagreb, and to the Institute for Theoretical Physics of Utrecht
University for their hospitality during the first course of this project. 
This work was partially supported by the Fonds de la Recherche 
Scientifique -- FNRS under Grant IISN 4.4517.08 -- Theory of fundamental
interactions, and by the D-ITP consortium, a program of 
the Netherlands Organization for Scientific Research (NWO) that is 
funded by the Dutch Ministry of Education, Culture and Science 
(OCW).

\appendix

\section{Scalar mode functions in power-law inflation}
\label{app: Mode functions}

The mode function~$\mathscr{U}_\lambda$ of a non-minimally 
coupled massless scalar field satisfies the following equation of motion in power-law inflation,
\begin{equation}
\Bigl[ \partial_0^2 + k^2 
	- \bigl( \lambda^2 \!-\! \tfrac{1}{4} \bigr) (1 \!-\! \epsilon)^2 \mathcal{H}^2 
	\Bigr] \mathscr{U}_\lambda(\eta,\vec{k} \hspace{0.3mm} ) = 0 \, ,
\label{U equation}
\end{equation}
where~$\lambda$ is either purely real, or purely imaginary, and 
depends on the scalar's non-minimal coupling, and on the
slow-roll parameter. This equation is ubiquitous to all massless fields in
power-law inflation, including the case of the vector field from 
Sec.~\ref{subsec: Dynamics of the transverse sector} 
and~\ref{subsec: Dynamics of the longitudinal sector}. 
It reduces to the Bessel equation upon a simple rescaling, and its general solution,
\begin{equation}
\mathscr{U}_\lambda(\eta, \vec{k} \hspace{0.3mm} ) = 
	A(\vec{k} \hspace{0.3mm} ) \times U_\lambda(\eta,k )
	+ B(\vec{k} \hspace{0.3mm} ) \times U_\lambda^*(\eta,k)
\, ,
\label{U general}
\end{equation}
is conveniently written in terms of suitably rescaled Hankel function of the first kind,
\begin{equation}
U_\lambda(\eta,k) = e^{\frac{i\pi}{4}(2\lambda+1)}
	e^{\frac{-ik}{(1-\epsilon)\mathcal{H}_0}}
	\sqrt{\frac{\pi}{4(1\!-\! \epsilon)\mathcal{H}}} \
	H^{\ss (1)}_{\lambda} \biggl( \frac{k}{(1\!-\!\epsilon)\mathcal{H}} \biggr) \, ,
	\label{U def}
\end{equation}
where~$A$ and~$B$ are 
constants of integration, and~$k\!=\!\|\vec k\,\|$.
The time-independent normalization of the function $U_\lambda$ in~(\ref{U def}) 
is chosen for convenience, such that its Wronskian normalization,
\begin{equation}
U_\lambda(\eta,k) \, \partial_0 U_\lambda^*(\eta,k)
	- U_\lambda^*(\eta,k) \, \partial_0 U_\lambda(\eta,k) = i \, ,
\label{Wronskian}
\end{equation}
is valid for both cases of real and imaginary~$\lambda$, and that its flat space 
limit is simple,
\begin{equation}
U_\lambda(\eta,k) \xrightarrow{H_0\to0}
	\frac{e^{-ik\eta}}{\sqrt{2k}} \, .
\end{equation}
For spatially homogeneous states the complex 
integration constants
$A$ and $B$ depend on the magnitude of the momentum
$k \!=\! \|\vec k\|$ only, and furthermore the state normalization
condition
(given by the Klein-Gordon product) imposes, 
$|A|^2 \!-\! |B|^2=1$. The choice $A\!=\!1, B\!=\!0$ ($\forall \vec k$)
corresponds to the CTBD vacuum 
state~\cite{Chernikov:1968zm,Bunch:1978yq}.
In general FLRW space-times formal solutions for the scalar mode functions 
were given in~\cite{Tsamis:2003zs}. However, explicit solutions are available
only in a handful of cases, power-law inflation we consider here being one.

\section{Rescaled propagator function~$\mathcal{F}_\nu$}
\label{app: Rescaled propagator function F_nu}

In~(\ref{F def}) we have introduced the rescaled propagator function,
\begin{equation}
\mathcal{F}_\lambda(y) \!=\!
	\frac{ \bigl[ (1\!-\!\epsilon) H_0 \bigr]^{D-2} }{(4\pi)^{D/2}} 
	\frac{\Gamma\bigl( \frac{D-1}{2} \!+\! \lambda \bigr) \,
			\Gamma\bigl( \frac{D-1}{2} \!-\! \lambda \bigr)}
		{ \Gamma\bigl( \frac{D}{2} \bigr)} \,
	{}_2F_1 \Bigl( \bigl\{ \tfrac{D-1}{2} \!+\! \lambda , \tfrac{D-1}{2} \!-\! \lambda 
			\bigr\} , \bigl\{ \tfrac{D}{2} \bigr\} , 1\!-\! \tfrac{y}{4} \Bigr) 
\, .
\label{app F function}
\end{equation}
where~${}_2F_1$ is the Gauss' hypergeometric function.
It appears in the  solution for the two-point functions of a massless non-minimally coupled
scalar in power-law inflation~(\ref{scalar 2-pt})
(see Refs.~\cite{Janssen:2008dw,Janssen:2008px}),
\begin{equation}
i \bigl[ \tensor*[^{\ss \! A \!}]{\Delta}{^{\ss \! B }}\bigr]_{\nu_\xi}(x;x')
	= (aa')^{-\frac{(D-2)\epsilon}{2}} \times \mathcal{F}_{\nu_\xi}(y_{\ss AB}) \, ,
\end{equation}
where $A,B$ are the Schwinger-Keldysh polarities of the distance function
defined in~(\ref{y AB}), and parameter~$\nu_\xi$ is related
to the non-minimal coupling of the scalar as,
\begin{equation}
\nu_\xi^2 = \biggl( \frac{D\!-\!1\!-\!\epsilon}{2(1\!-\!\epsilon)} \biggr)^{\!2}
	- \frac{(D\!-\!1) (D\!-\!2\epsilon)}{(1\!-\!\epsilon)^2} \xi \, .
\end{equation}
The rescaled propagator function satisfies the hypergeometric equation in disguise,
\begin{equation}
\biggl[ (4y \!-\! y^2) \frac{\partial^2}{\partial y^2}
	+ D(2\!-\!y) \frac{\partial}{\partial y}
	+ \lambda^2 - \bigl( \tfrac{D-1}{2} \bigr)^2 \biggr]
	\mathcal{F}_\lambda(y) = 0 \, .
\label{F equation}
\end{equation}
At multiple instances we need the power series expansion of the rescaled propagator function
around~$y=0$. To derive it
we first use the identity (15.8.4) from~\cite{Olver:2010,Olver:2010b} 
(or identity~(9.131.2) from~\cite{Gradshteyn:2007}) 
for the transformation of argument 
of the hypergeometric function, and write~(\ref{app F function}) as
\begin{align}
\mathcal{F}_\lambda(y) ={}&
	\frac{\bigl[ (1\!-\!\epsilon) H_0 \bigr]^{D-2}}{(4\pi)^{D/2}} 
		\Gamma\bigl( \tfrac{D-2}{2} \bigr) \Biggl\{
	\Bigl( \frac{y}{4} \Bigr)^{\! - \frac{D-2}{2}} \times
	{}_2F_1 \Bigl( \bigl\{ \tfrac{1}{2} \!+\! \lambda , \tfrac{1}{2}\!-\!\lambda \bigr\} ,
		\bigl\{ \tfrac{4-D}{2} \bigr\} , \tfrac{y}{4} \Bigr)
\label{Fnu: 2 hypergeometric}
 \\
&	\hspace{0.5cm}
	+ \frac{\Gamma\bigl( \frac{2-D}{2} \bigr) \, 
			\Gamma\bigl( \frac{D-1}{2}\!+\! \lambda \bigr) \,
			\Gamma\bigl( \frac{D-1}{2} \!-\! \lambda \bigr)}
		{\Gamma\bigl( \frac{D-2}{2} \bigr) \,
			\Gamma\bigl( \frac{1}{2}\!+\!\lambda \bigr) \,
			\Gamma\bigl( \frac{1}{2} \!-\! \lambda \bigr)}
	\times 
	{}_2F_1 \Bigl( \bigl\{ \tfrac{D-1}{2} \!+\! \lambda , 
		\tfrac{D-1}{2} \!-\! \lambda \bigr\} , 
		\bigl\{ \tfrac{D}{2} \bigr\} , \tfrac{y}{4} \Bigr)
	\Biggr\} \, ,
\nonumber
\end{align}
which is then straightforward to expand in the desired power series,
\begin{align}
&
\mathcal{F}_\lambda(y) = \bigl[ (1\!-\!\epsilon) H_0 \bigr]^{D-2} \times
	\frac{\Gamma\bigl( \frac{D-2}{2} \bigr)}{(4\pi)^{D/2}} \times \Biggl\{
	\Bigl( \frac{y}{4} \Bigr)^{\! - \frac{D-2}{2}}
	+ \frac{\Gamma\bigl( \frac{4-D}{2} \bigr)}
		{\Gamma\bigl( \frac{1}{2} \!+\! \lambda \bigr) \,
			\Gamma\bigl( \frac{1}{2} \!-\! \lambda \bigr)}
	\times \sum_{n=0}^{\infty}
\label{app power series}
\\
&	\hspace{0.5cm}
	\times \biggl[
	\frac{\Gamma\bigl( \frac{3}{2} \!+\! \lambda \!+\! n \bigr) \,
			\Gamma\bigl( \frac{3}{2} \!-\! \lambda \!+\! n \bigr) }
		{\Gamma\bigl( \frac{6-D}{2} \!+\! n \bigr) \, \Gamma(n\!+\!2)}
		\Bigl( \frac{y}{4} \Bigr)^{\! n - \frac{D-4}{2}}
	- \frac{\Gamma\bigl( \frac{D-1}{2} \!+\! \lambda \!+\! n \bigr) \,
			\Gamma\bigl( \frac{D-1}{2} \!-\! \lambda \!+\! n \bigr) }
		{\Gamma\bigl( \frac{D}{2} \!+\! n \bigr) \, \Gamma(n\!+\!1)}
		\Bigl( \frac{y}{4} \Bigr)^{\! n}
	\biggr]
	\Biggr\} \, .
\nonumber 
\end{align}
We also need an asymptotic expansion for large~$y$ in Sec.~\ref{sec: Field strength correlator}. 
To obtain it we first apply the identity 9.132.1 from~\cite{Gradshteyn:2007} to transform the argument
of the hypergeometric function,
\begin{align}
\mathcal{F}(y) ={}& 
	\frac{\bigl[ (1\!-\!\epsilon) H_0 \bigr]^{D-2}}{(4\pi)^{D/2}} 
\\
&	\times \Biggl\{
	\frac{\Gamma(2\lambda) \, \Gamma\bigl( \frac{D-1}{2} \!-\! \lambda \bigr) }
			{ \Gamma\bigl( \frac{1}{2} \!+\! \lambda \bigr) }
		\Bigl( \frac{y}{4} \Bigr)^{\! - \frac{D-1}{2} + \lambda}
		{}_2F_1 \Bigl( \bigl\{ \tfrac{D-1}{2} \!-\! \lambda , \tfrac{1}{2} \!-\! \lambda \bigr\} 
				, \bigl\{ 1 \!-\! 2\lambda \bigr\} , \tfrac{4}{y} \Bigr)
\nonumber \\
&	\hspace{1cm}
	+ \frac{\Gamma(-2\lambda) \, \Gamma\bigl( \frac{D-1}{2} \!+\! \lambda \bigr) }
			{\Gamma\bigl( \frac{1}{2} \!+\! \lambda \bigr)}
		\Bigl( \frac{y}{4} \Bigr)^{\! - \frac{D-1}{2} - \lambda}
		{}_2F_1 \Bigl( \bigl\{ \tfrac{D-1}{2} \!+\! \lambda , \tfrac{1}{2} \!+\! \lambda \bigr\} 
				, \bigl\{ 1 \!+\! 2\lambda \bigr\} , \tfrac{4}{y} \Bigr)
	\Biggr\} \, ,
\nonumber 
\end{align}
after which the expansion for large~$y$ is straightforward,
\begin{align}
\mathcal{F}_\lambda(y)
	\stackrel{y\rightarrow\infty}{\sim}{}&
	\frac{\bigl[ (1\!-\!\epsilon) H_0 \bigr]^{D-2}}{(4\pi)^{D/2}}
	\Biggl\{
	\frac{\Gamma(2\lambda) \, \Gamma\bigl( \tfrac{D-1}{2} \!-\! \lambda \bigr)}
		{\Gamma\bigl( \tfrac{1}{2} \!+\! \lambda \bigr)} 
			\Bigl( \frac{y}{4} \Bigr)^{\! - \frac{D-1}{2}+\lambda}
		\biggl[ 1 + \frac{(D\!-\!1\!-\!2\lambda)}{y} + ... \, \biggr]
\nonumber \\
&	\hspace{0.5cm}
+ \frac{\Gamma(-2\lambda) \, \Gamma\bigl( \tfrac{D-1}{2} \!+\! \lambda \bigr)}
		{\Gamma\bigl( \tfrac{1}{2} \!-\! \lambda \bigr)} \Bigl( \frac{y}{4} \Bigr)^{\! - \frac{D-1}{2}-\lambda}
	\biggl[ 1 + \frac{(D\!-\!1\!+\!2\lambda)}{y} + ... \, \biggr]
	\Biggr\}  .
\label{F asy}
\end{align}
%

\section{Useful identity for~$\mathcal{F}_\nu$}
\label{app: Useful identity for F_nu}

In this appendix we derive a very useful identity which allows us to 
eliminate the inverse Laplace operator from the expression for the 
vector two-point function in Sec.~\ref{subsec: Two-point function as mode sums}.
The variables used to express the two-point function introduced in~(\ref{y def})
and~(\ref{u and v def}) are,
\begin{equation}
y = (1\!-\!\epsilon)^2 \mathcal{H} \mathcal{H}'
	\Bigl[ \| \Delta\vec{x} \|^2 - (\eta\!-\!\eta')^2 \Bigr] \, ,
\qquad
u = (1\!-\!\epsilon) \ln(aa') \, ,
\qquad
v = (1\!-\!\epsilon) \ln(a/a') \, ,\quad
\end{equation}
where~$\mathcal{H} \!=\! H_0 \, a^{1-\epsilon} \!=\! H_0 \, e^{(u+v)/2}$,
$\mathcal{H}' \!=\! H_0 \, {a'}^{1-\epsilon} \!=\! H_0 \, e^{(u-v)/2}$, and
$(1\!-\!\epsilon)^2 \mathcal{H} \mathcal{H}'(\eta\!-\!\eta')^2 \!=\!2 \left[\ch(v)\!-\!1\right]$.
In this appendix all expressions are valid for~$y_{\ss -+}(x;x')$ only, but we suppress the
explicit Schwinger-Keldish polarities.
At multiple instances we make use of the first and second space or time derivatives
of $y(x;x')$,
\begin{align}
&
(\partial_0 y) = (1\!-\!\epsilon) \bigl[ \mathcal{H} y  
	- 2(\mathcal{H} \!-\! \mathcal{H}') \bigr]
	= (1\!-\!\epsilon) H_0 \, e^{\frac{u+v}{2}}
		\bigl[ 2 e^{-v} \!-\! (2\!-\!y) \bigr] \, ,
\label{der1}
\\
&
(\partial_0' y) = (1\!-\!\epsilon) \bigl[ \mathcal{H}' y  
	+ 2(\mathcal{H} \!-\! \mathcal{H}') \bigr]
	= (1\!-\!\epsilon) H_0 \, e^{\frac{u-v}{2}}
		\bigl[ 2 e^{v} \!-\! (2\!-\!y) \bigr] \, ,
\label{der2}
\\
&
(\partial_i y ) = - (\partial_i' y) =
	2 (1\!-\!\epsilon)^2 \mathcal{H} \mathcal{H}' \Delta x_i 
	= 2 (1\!-\!\epsilon)^2 H_0^2 \, e^u \Delta x_i
\\
&
( \partial_0\partial_0' y ) =
	(1\!-\!\epsilon)^2 \bigl[ 2 \mathcal{H}^2 + 2 (\mathcal{H}')^2
	- \mathcal{H} \mathcal{H}' (2\!-\!y) \bigr] 
	= (1\!-\!\epsilon)^2 H_0^2 \, e^u \bigl[
	4 \ch(v) \!-\! (2\!-\!y) \bigr] \, ,
\label{der4}
\\
&
(\partial_0 \partial_i y )
	= - (\partial_0 \partial_i' y )
	= (1\!-\!\epsilon) \mathcal{H} (\partial_i y)
	= (1\!-\!\epsilon) H_0 \, e^{\frac{u+v}{2}} (\partial_i y) \, ,
\\
&
(\partial_0' \partial_i y )
	= - (\partial_0' \partial_i' y )
	= (1\!-\!\epsilon) \mathcal{H}' (\partial_i y)
	= (1\!-\!\epsilon) H_0 \, e^{\frac{u-v}{2}} (\partial_i y) \, ,
\\
&
(\partial_i \partial_j y) = - (\partial_i \partial_j' y)
	= - (\partial_i' \partial_j y) = (\partial_i' \partial_j' y)
	=  2 \delta_{ij} (1\!-\!\epsilon)^2 \mathcal{H} \mathcal{H}'
	= 2 \delta_{ij} (1\!-\!\epsilon)^2 H_0^2 \, e^u
\,.
\label{der7}
\end{align}
Using the derivatives above we can derive a simple derivative identity
for the functions dependent on $y$ only,
\begin{equation}
\frac{1}{2} \nabla^2 I[f](y) =
	(1\!-\!\epsilon)\Bigl[ \mathcal{H}' \partial_0 + \mathcal{H} \partial_0'
		+ (D\!-\!1) (1\!-\!\epsilon) \mathcal{H} \mathcal{H}' \Bigr] f(y) 
\, ,
\label{simple useful id}
\end{equation}
where $ I[f](y)\!\equiv\!\int^y \! dy' \, f(y')$ denotes the primitive function of $f(y)$.
Proving it involves showing that
\begin{equation}
\bigl[ \mathcal{H}' \partial_0 + \mathcal{H} \partial_0' \bigr] y
	= \frac{1}{2(1\!-\!\epsilon)} (\vec{\nabla} y) \!\cdot\! (\vec{\nabla} y)
\end{equation}
by making use of~(\ref{der4}), (\ref{der1}), and (\ref{der2}), and then using this 
to compare the two sides of equality~(\ref{simple useful id}).
Using the simple identity~(\ref{simple useful id}), and the differential 
equation~(\ref{F equation}) satisfied by $\mathcal{F}_\lambda(y)$ we can
derive a more involved identity,
\begin{align}
\partial_0 \partial_0' \bigl[ (aa')^\rho \mathcal{F}_\lambda(y) \bigr] ={}&
	(aa')^\rho \times \biggl\{
	- \frac{1}{2} \nabla^2 I 
		\Bigl[ (2\!-\!y) \tfrac{\partial \mathcal{F}_\lambda}{\partial y}
		- \tfrac{\rho}{(1\!-\!\epsilon)} \mathcal{F}_\lambda \Bigr]
\nonumber \\
&	\hspace{2cm}
	- (1\!-\!\epsilon)^2 \mathcal{H} \mathcal{H}'
		\Bigl[ \lambda^2 
		- \Bigl(\tfrac{D-1}{2}\!-\!\tfrac{\rho}{1-\epsilon} 
			\Bigr)^{\!\!2\,} \Bigr] \mathcal{F}_\lambda
	\biggr\} 
\, ,
\label{important identity}
\end{align}
which we use in Sec.~\ref{subsec: Two-point function as mode sums}.
This identity can be proven in several steps outlined below:
\begin{itemize}
\item[(i)]
Show the two relations,
\begin{align}
(\partial_0 y) (\partial_0' y) ={}&
	- (1\!-\!\epsilon) (2\!-\!y) \bigl[ \mathcal{H} \partial_0'
		+ \mathcal{H}' \partial_0 \bigr] y
	+ (1\!-\!\epsilon)^2 \mathcal{H} \mathcal{H}' (4y \!-\! y^2) \, ,
\label{rel 1}
\\
(\partial_0 \partial_0' y) ={}&
	(1\!-\!\epsilon) \bigl[ \mathcal{H} \partial_0' + \mathcal{H}' \partial_0 \bigr] y
	+ (1\!-\!\epsilon)^2 \mathcal{H} \mathcal{H}' (2\!-\!y) \, .
\label{rel 2}
\end{align}\
that follow from the basic derivative relations~(\ref{der1}--\ref{der7}).

\item[(ii)]
Use the two relations~(\ref{rel 1}) and~(\ref{rel 2}) to write out two derivatives
action on some function dependent only on the length function $y$,
\begin{align}
\partial_0 \partial_0' f(y) ={}&
	- (1\!-\!\epsilon) (2\!-\!y) \frac{\partial^2 f}{\partial y^2}
		\bigl[ \mathcal{H} \partial_0' + \mathcal{H}' \partial_0 \bigr] y
	+ (1\!-\!\epsilon) \frac{\partial f}{\partial y} 
		\bigl[ \mathcal{H} \partial_0' + \mathcal{H}' \partial_0 \bigr] y
\nonumber \\
&	+ (1\!-\!\epsilon)^2 \mathcal{H} \mathcal{H}' 
		(4y \!-\! y^2) \frac{\partial^2 f}{\partial y^2}
	+ (1\!-\!\epsilon)^2 \mathcal{H} \mathcal{H}' 
		(2\!-\!y) \frac{\partial f}{\partial y}  \, .
\end{align}
Recognizing that $[\partial f/\partial y] (\partial_0 y) \!=\! \partial_0 f$
and applying the Leibnitz rule write the above expression as,
\begin{equation}
\partial_0 \partial_0' f(y) =
	- (1\!-\!\epsilon) \bigl[ \mathcal{H} \partial_0' + \mathcal{H}' \partial_0 \bigr] 
		\biggl[ (2\!-\!y) \frac{\partial f}{\partial y} \biggr]
	+ (1\!-\!\epsilon)^2 \mathcal{H} \mathcal{H}' 
		\biggl[ (4y \!-\! y^2) \frac{\partial^2 f}{\partial y^2}
			+ (2\!-\!y) \frac{\partial f}{\partial y} \biggr]  \, .
\label{rel 3}
\end{equation}

\item[(iii)]
Specialize to $f(y) \!=\! \mathcal{F}_\lambda(y)$ given in~(\ref{app F function}), 
and use the differential equation~(\ref{F equation}) to get rid of the second 
derivative in the last bracket of~(\ref{rel 3}),
\begin{align}
\partial_0 \partial_0' \mathcal{F}_\lambda(y) ={}&
	- (1\!-\!\epsilon) \Bigl[ \mathcal{H} \partial_0' + \mathcal{H}' \partial_0
		+ (D\!-\!1) (1\!-\!\epsilon) \mathcal{H} \mathcal{H}' \Bigr] 
		\biggl[ (2\!-\!y) \frac{\partial \mathcal{F}_\lambda}{\partial y} \biggr]
\nonumber \\
&	- (1\!-\!\epsilon)^2 \mathcal{H} \mathcal{H}' 
		\Bigl[ \lambda^2 - \bigl( \tfrac{D-1}{2} \bigr)^2 \Bigr] \mathcal{F}_\lambda \, .
\label{rel 4}
\end{align}

\item[(iv)]
Apply the expression from the previous step to the left hand side
of~(\ref{important identity}),
\begin{align}
\MoveEqLeft[3]
\partial_0 \partial_0' \bigl[ (aa')^{\rho} \mathcal{F}_\lambda(y) \bigr] =
	(aa')^\rho \times \biggl\{
	- (1\!-\!\epsilon)^2 \mathcal{H} \mathcal{H}'
		\Bigl[ \lambda^2 
			- \Bigl(\tfrac{D-1}{2}\!-\!\tfrac{\rho}{1-\epsilon} 
			\Bigr)^{\!\!2\,}  \Bigr] \mathcal{F}_\lambda
\label{rel 5}
 \\
&	
	- (1\!-\!\epsilon) \Bigl[ \mathcal{H} \partial_0' + \mathcal{H}' \partial_0
		+ (D\!-\!1) (1\!-\!\epsilon) \mathcal{H} \mathcal{H}' \Bigr] 
		\biggl[ (2\!-\!y) \frac{\partial \mathcal{F}_\lambda}{\partial y}
		- \frac{\rho}{(1\!-\!\epsilon)} \mathcal{F}_\lambda \biggr]
	\biggr\} \, .
\nonumber 
\end{align}

\item[(v)]
Apply the simple identity~(\ref{simple useful id}) to the right hand side
of~(\ref{rel 5}), after which the identity~(\ref{important identity}) follows.

\end{itemize}

\section{Various derivative identities}
\label{app: Various derivative identities}

The covariantized solutions for the vector two-point functions given 
in~(\ref{covariant form}) and (\ref{Feynman propagator solution}), with 
the scalar structure functions given in~(\ref{C1 solution})--(\ref{C4 solution})
have to respect properties dictated by canonical quantization. Namely, they have to
satisfy transversality relations~(\ref{Wightman transversality})--(\ref{non-transverse 2}),
and the equations of motion~(\ref{EOM: Wightman functions})--(\ref{EOM: Feynman propagator 2})
Demonstrating that these hold involves taking multiple derivatives of the distance
function~$y_{\ss AB}\!=\!(1\!-\!\epsilon)^2 \mathcal{H} \mathcal{H}'\Delta x^2_{\ss AB}$, 
with either the~$(-+)$ or the~$(++)$~$i\varepsilon$-prescription,
\begin{equation}
\Delta x^2_{\ss -+} = 
	\| \vec{x} \!-\! \vec{x}^{\,\prime} \|^2 - \bigl(\eta\!-\! \eta'  \!-\! i \varepsilon \bigr)^2 \, ,
\qquad \quad
\Delta x^2_{\ss ++} = 
	\| \vec{x} \!-\! \vec{x}^{\,\prime} \|^2 - \bigl( |\eta\!-\! \eta' | \!-\! i \varepsilon \bigr)^2 \, .
\end{equation}
In this appendix we collect the necessary derivative identities.

When taking derivatives of the~$y_{\ss -+}$ appearing in the Wightman function,
in practice we may just neglect the $i\varepsilon$-prescription, and restore it to each~$y$ at the end.
However, when the~$i\varepsilon$-prescription corresponds to the Feynman or Dyson prescription, taking derivatives is particularly tricky, since local
terms are generated. It is two or more time derivatives acting on~$y_{\ss ++}$
that generate~$\delta$-functions in time,
\begin{align}
&
\bigl( \partial_0 \partial_0 y_{\ss ++} \bigr)
	= \bigl( \partial_0 \partial_0 y \bigr)
	+ 4 (1\!-\!\epsilon)^2 \mathcal{H}^2 
	\times i \varepsilon \delta(\eta\!-\!\eta') \, ,
\\
&
\bigl( \partial_0 \partial_0' y_{\ss ++} \bigr)
	= \bigl( \partial_0 \partial_0' y \bigr)
	- 4 (1\!-\!\epsilon)^2 \mathcal{H}^2 
	\times i \varepsilon \delta(\eta\!-\!\eta') \, ,
\\
&
\bigl( \partial_0 \partial_0 \partial_0' y_{\ss ++} \bigr)
	= \bigl( \partial_0 \partial_0 \partial_0' y \bigr)
	- 4 (1\!-\!\epsilon)^2 \mathcal{H}^2 \Bigl[ 
		\partial_0 + 2 (1\!-\!\epsilon) \mathcal{H} \Bigr] i \varepsilon \delta(\eta\!-\!\eta') \, .
\end{align}
Whenever an ordinary spatial derivative acts on~$y_{\ss ++}$ no~$\delta$-function
in time can be generated from it.
Identities with different combinations of primed and unprimed derivatives
are derived from the above ones by simply switching the time arguments.
This will be the case for all the expressions in this appendix, including spatial
derivatives. 

The tensor structures that harbour local terms in time multiply 
the scalar structure functions~(\ref{C1 solution})--(\ref{C4 solution}),
which can be seen as power series in~$y_{\ss++}$, and this can result in a space-time local
term that survives the limit~$\varepsilon\!\to\!0$. The space-time local terms are identified 
by making use of a representation for the covariant delta-function,
\begin{equation}
i \varepsilon \delta(\eta\!-\!\eta') \times \Bigl( \frac{y_{\ss ++}}{4} \Bigr)^{\!\! - \frac{D}{2}}
	= \bigl[ (1\!-\!\epsilon) H \bigr]^{-D} \, \frac{(4\pi)^{D/2}}{\Gamma\bigl( \frac{D}{2} \bigr)}
		\times \frac{i \delta^D(x\!-\!x')}{\sqrt{-g}} \, .
\label{fundamental delta}
\end{equation}
Different combinations of derivatives acting on the identity above 
generate further identities,
\begingroup
\allowdisplaybreaks
\begin{align}
&
i \varepsilon \delta(\eta\!-\!\eta') 
	(\partial_i y) \Bigl( \frac{y_{\ss ++}}{4} \Bigr)^{\!\! - \frac{D+2}{2}}
	= \bigl[ (1\!-\!\epsilon) H \bigr]^{-D} \, \frac{(4\pi)^{D/2}}{\Gamma\bigl( \frac{D}{2} \bigr)}
		\times \bigl( - \tfrac{8}{D} \bigr) \partial_i \frac{i \delta^D(x\!-\!x')}{\sqrt{-g}} \, ,
\label{covariant delta 1}
\\
&
i \varepsilon \delta(\eta\!-\!\eta') \,
	\partial_i \biggl[ (\partial_j y) \Bigl( \frac{y_{\ss ++}}{4} \Bigr)^{\!\! - \frac{D+2}{2}} \biggr]
	= \bigl[ (1\!-\!\epsilon) H \bigr]^{-D} \, \frac{(4\pi)^{D/2}}{\Gamma\bigl( \frac{D}{2} \bigr)}
		\times \bigl( - \tfrac{8}{D} \bigr) \partial_i \partial_j \frac{i \delta^D(x\!-\!x')}{\sqrt{-g}} \, ,
\label{covariant delta 2}
\\
&
 i \varepsilon \delta(\eta\!-\!\eta') 
	\Bigl( \frac{y_{\ss ++}}{4} \Bigr)^{- \frac{D+2}{2}}
	= \bigl[ (1\!-\!\epsilon) H \bigr]^{-D-2} \, \frac{(4\pi)^{D/2}}{\Gamma\bigl( \frac{D}{2} \bigr)}
		\times \bigl( \tfrac{4}{3D} \bigr) a^{-2} \nabla^2 \frac{i \delta^D(x\!-\!x')}{\sqrt{-g}} \, ,
\label{covariant delta 3}
\\
&
i \varepsilon \delta(\eta\!-\!\eta') 
	\bigl( \partial_0 y_{\ss ++} \bigr) \Bigl( \frac{y_{\ss ++}}{4} \Bigr)^{\!\! - \frac{D}{2}}
	= \bigl[ (1\!-\!\epsilon) H \bigr]^{-D} \, \frac{(4\pi)^{D/2}}{\Gamma\bigl( \frac{D}{2} \bigr)}
		\times 4 (1\!-\!\epsilon) \mathcal{H} \frac{i \delta^D(x\!-\!x')}{\sqrt{-g}} \, ,
\label{covariant delta 4}
\\
&
i \varepsilon \partial_0 \delta(\eta\!-\!\eta') \Bigl( \frac{y_{\ss ++}}{4} \Bigr)^{\!\! - \frac{D}{2}}
	= \bigl[ (1\!-\!\epsilon) H \bigr]^{-D} \, \frac{(4\pi)^{D/2}}{\Gamma\bigl( \frac{D}{2} \bigr)}
		\times
		\Bigl[ \partial_0 + \tfrac{1}{2} D(1\!+\!\epsilon) \mathcal{H} \Bigr] 
			\frac{i \delta^D(x\!-\!x')}{\sqrt{-g}} \, ,
\label{covariant delta 5}
\\
&
i \varepsilon \partial_0 \delta(\eta\!-\!\eta') (\partial_i y)
	\Bigl( \frac{y_{\ss ++}}{4} \Bigr)^{\!\! - \frac{D+2}{2}} 
\nonumber \\
&	\hspace{1.5cm}
	= \bigl[ (1\!-\!\epsilon) H \bigr]^{-D} \, \frac{(4\pi)^{D/2}}{\Gamma\bigl( \frac{D}{2} \bigr)}
		\!\times\!
		\bigl( - \tfrac{8}{D} \bigr)
		\Bigl[ \partial_0 + \tfrac{1}{2} D(1\!+\!\epsilon) \mathcal{H} \Bigr] 
			\partial_i \frac{i \delta^D(x\!-\!x')}{\sqrt{-g}} \, .
\label{covariant delta 6}
\end{align}
\endgroup
%
\begin{table}[h!]
\renewcommand{\arraystretch}{1.3}
\centering
\begin{tabular}{l r}
\hline
$\quad
	g^{\mu\nu} \bigl( \partial_{\mu} y \bigr) \bigl( \partial_{\nu} y \bigr)$
	&
	$(1\!-\!\epsilon)^2 H^2 \bigl( 4y \!-\! y^2 \bigr) \quad$
\\
\hline
$\quad
	g'^{\rho\sigma} \bigl( \partial_{\rho}' y \bigr) \bigl( \partial_{\sigma}' y \bigr)$
	&
	$(1\!-\!\epsilon)^2 H'^2 \bigl( 4y \!-\! y^2 \bigr) \quad$
\\
\hline
$\quad
	g^{\mu\nu} \bigl( \partial_{\mu} y \bigr) \bigl( \partial_{\nu} u \bigr) $
	&
	$(1\!-\!\epsilon)^2 H^2 \bigl( 2 \!-\! y \!-\! 2 e^{-v} \bigr) \quad$
\\
\hline
$\quad
	g'^{\rho\sigma} \bigl( \partial_{\rho}' y \bigr) \bigl( \partial_{\sigma}' u \bigr) $
	&
	$(1\!-\!\epsilon)^2 H'^2 \bigl( 2 \!-\! y \!-\! 2 e^{v} \bigr) \quad$
\\
\hline
$\quad
	g^{\mu\nu} \bigl( \partial_{\mu} u \bigr) \bigl( \partial_{\nu} u \bigr) $
	&
	$- (1\!-\!\epsilon)^2 H^2 \quad$
\\
\hline
$\quad
	g'^{\rho\sigma} \bigl( \partial_{\rho}' u \bigr) \bigl( \partial_{\sigma}' u \bigr) $
	&
	$ - (1\!-\!\epsilon)^2 H'^2 \quad$
\\
\hline
$\quad
	g^{\mu\nu} \bigl( \partial_{\mu} y \bigr) \bigl( \partial_{\nu} \partial_{\!\rho}' y \bigr) $
	&
	$ (1\!-\!\epsilon)^2 H^2 (2\!-\!y) \bigl( \partial_{\rho}' y \bigr) \quad$
\\
\hline
$\quad
	g'^{\rho\sigma} \bigl( \partial_{\mu} \partial_{\rho}' y \bigr) \bigl( \partial_{\sigma}' y \bigr)$
	&
	$ (1\!-\!\epsilon)^2 H'^2 (2\!-\!y) \bigl( \partial_{\mu} y \bigr) \quad$
\\
\hline
$\quad
	g^{\mu\nu} \bigl( \partial_{\mu} u \bigr) \bigl( \partial_{\nu} \partial_{\rho}' y \bigr) $
	&
	$- (1\!-\! \epsilon)^2 H^2 
		\bigl[ \bigl(\partial'_{\rho}y \bigr) + 2 e^{-v} \bigl( \partial'_{ \rho} u \bigr) \bigr]
		\quad$
\\
\hline
$\quad
	g'^{\rho\sigma} \bigl( \partial_{\mu} \partial_{\rho}' y \bigr) \bigl( \partial'_{\sigma} u \bigr) $
	&
	$ - (1\!-\! \epsilon)^2 H'^2 
		\bigl[ \bigl(\partial_{\mu}y \bigr) + 2 e^v \bigl( \partial_{ \mu} u \bigr) \bigr]
		\quad$
\\
\hline
$\quad
	g^{\mu\nu} \bigl( \partial_{ \mu} \partial_{\rho}' y \bigr) 
	\bigl( \partial_{\nu} \partial_{\sigma}' y \bigr) $
	&
	$\qquad (1\!-\!\epsilon)^2 H^2 \bigl[ 4 (1\!-\!\epsilon)^2 H'^2 g'_{\rho\sigma}
			- \bigl( \partial'_\rho y \bigr) \bigl( \partial'_\sigma y \bigr) \bigr] \quad$
\\
\hline
$\quad
	g'^{\rho\sigma} \bigl( \partial_{\mu} \partial_{\rho}' y \bigr) 
	\bigl( \partial_{\nu} \partial_{\sigma}' y \bigr) \qquad$
	&
	$\qquad (1\!-\!\epsilon)^2 H'^2 \bigl[ 4 (1\!-\!\epsilon)^2 H^2 g_{\mu\nu}
			- \bigl( \partial_\mu y \bigr) \bigl( \partial_\nu y \bigr) \bigr] \quad$
\\
\hline
\end{tabular}
\caption{
Contractions of tensor structures. The expressions  in the table are strictly speaking valid for 
distance functions~$y$ without the~$i\varepsilon$-prescription. In case the distance functions
have~$i\varepsilon$-prescriptions~$y_{\ss -+}$ or~$y_{\ss ++}$ the expressions above
are valid up to an innocuous term linear in~$\varepsilon$, which can be neglected in all the
expressions where contractions are used in this paper. In the case of double derivatives 
and the~$(++)$~$i\varepsilon$-prescription,
the modified derivative~$\overline{\bigl( \partial_\mu \partial'_\rho y_{\ss ++} \bigr)}$ 
defined in Eq.~(\ref{modified tensor}) is assumed. }
\label{tensor contractions}
\end{table}

It is useful
to be able to reduce the derivative order, which is accomplished using the
following two expressions,
\begin{align}
&
\bigl( \nabla_{\!\mu} \nabla_{\!\nu} y \bigr) \!= 
	g_{\mu\nu} (1\!-\!\epsilon) H^2 \bigl( 2 \!-\! y \!-\! 2\epsilon \, e^{-v} \bigr)
	\!- \frac{\epsilon}{(1\!-\!\epsilon)}
		\Bigl[ \bigl( \partial_\mu y \bigr) \bigl( \partial_\nu u \bigr)
				\!+\! \bigl( \partial_\mu u \bigr) \bigl( \partial_\nu y \bigr) \Bigr] \, ,
\label{derivative reduction 1}
\\
&
\bigl( \nabla_{\!\mu} \nabla_{\!\nu} u \bigr) \!= 
	- g_{\mu\nu} (1\!-\!\epsilon) H^2
	- \Bigl( \frac{1\!+\!\epsilon}{1\!-\!\epsilon} \Bigr) 
		\bigl( \partial_\mu u \bigr) \bigl( \partial_\nu u \bigr) \, ,
\label{derivative reduction 2}
\end{align}
where for higher derivatives we must keep in mind that the Hubble rate 
is not a constant, but~$H^2 \!=\! H_0^2 \, e^{-\frac{(u+v)\epsilon}{(1-\epsilon)}}$.
Furthermore, contractions of different tensor structures appear in the
transversality relations, and equations of motion, and it is useful to
have contraction identities given in Table~\ref{tensor contractions}.
The expressions below are ready made to be used in checking that the
Wightman function and the Feynman propagator solutions from 
Sec.~\ref{subsec: Covariantizing two-point functions}
satisfy the correct transversality relations, and the correct equations of motion 
from Sec.~\ref{subsec: Equations of motion for two-point functions},
%
%
\begin{align}
&
\bigl( \nabla_\mu \nabla_\nu y_{\ss ++} \bigr)
	=
	g_{\mu\nu} (1\!-\!\epsilon) H^2 \Bigl[ 2 - y_{\ss ++} - 2\epsilon \, e^{-v} \Bigr]
\label{++ derivative 1} \\
&	\hspace{3cm}
	- \frac{2\epsilon}{(1\!-\!\epsilon)} \bigl( \nabla_{(\mu} u \bigr)
		\bigl( \nabla_{\nu)} y_{\ss ++} \bigr)
	+ \bigl( \nabla_\mu u \bigr) \bigl( \nabla_\nu u \bigr)
		\times 4 i \varepsilon \delta(\eta\!-\!\eta') \, ,
\nonumber 
\\
&
\bigl( \nabla_\mu \nabla'_\rho y_{\ss ++} \bigr)
	= \bigl( \overline{\nabla_\mu \nabla'_\rho y_{\ss ++}} \bigr)
		- \bigl( \nabla_\mu u \bigr) \bigl( \nabla'_\rho u \bigr) \times 
			4 i \varepsilon \delta(\eta\!-\!\eta') \, ,
\label{++ derivative 2}
\\
&
\bigl( \nabla_\mu \nabla_\nu \nabla'_\rho y_{\ss ++} \bigr)
	=
	- (1\!-\!\epsilon) H^2 g_{\mu\nu} \bigl( \nabla'_\rho y_{\ss ++} \bigr)
\label{++ derivative 3}
\\
&	\hspace{1cm}
	- 2 (1\!-\!\epsilon) H^2 g_{\mu\nu} \bigl( \nabla'_\rho u \bigr) 
		\Bigl[ \epsilon \, e^{-v}
			-2 i \varepsilon \delta(\eta\!-\!\eta') \Bigr]
	- \frac{2\epsilon}{(1\!-\!\epsilon)} \bigl( \nabla_{(\mu} u \bigr)
		\bigl( \overline{ \nabla_{\nu)} \nabla'_\rho y_{\ss ++} } \bigr)
\nonumber \\
&	\hspace{2cm}
	+ \bigl( \nabla_\mu u \bigr) \bigl( \nabla_\nu u \bigr) \bigl( \nabla'_\rho u \bigr)
		\biggl[ \frac{8 \epsilon}{(1\!-\!\epsilon)} i \varepsilon \delta(\eta\!-\!\eta')
			+ \frac{4}{(1\!-\!\epsilon) \mathcal{H}' } \partial_0' i \varepsilon \delta(\eta\!-\!\eta') \biggr] \, ,
\nonumber 
\\
&
\bigl( \nabla_\mu \overline{\nabla_\nu \nabla'_\rho y_{\ss ++} } \bigr)
	=
	- (1\!-\!\epsilon) H^2 g_{\mu\nu} \bigl( \nabla'_\rho y_{\ss ++} \bigr)
\label{++ derivative 4} \\
&	\hspace{3.5cm}
	- 2 \epsilon (1\!-\!\epsilon) H^2 e^{-v} g_{\mu\nu} \bigl( \nabla'_\rho u \bigr) 
	- \frac{2\epsilon}{(1\!-\!\epsilon)} \bigl( \nabla_{(\mu} u \bigr)
		\bigl( \overline{ \nabla_{\nu)} \nabla'_\rho y_{\ss ++} } \bigr) \, ,
\nonumber\\
&
\bigl( \nabla_\alpha \nabla_\mu \overline{\nabla_\nu \nabla'_\rho y_{\ss ++} } \bigr)
	=
	- (1\!-\!\epsilon) H^2 g_{\mu\nu} \bigl( \overline{ \nabla_\alpha \nabla'_\rho y_{\ss ++} } \bigr)
\label{++ derivative 5} \\
&	\hspace{2cm}
	- 2 (1\!-\!\epsilon)  g_{\mu\nu} \bigl( \nabla'_\rho u \bigr) \nabla_\alpha 
		\Bigl[ H^2 \epsilon \, e^{-v} \Bigr]
	- \frac{2\epsilon}{(1\!-\!\epsilon)} \bigl( \nabla_\alpha \nabla_{(\mu} u \bigr)
		\bigl( \overline{ \nabla_{\nu)} \nabla'_\rho y_{\ss ++} } \bigr)
\nonumber \\
&	\hspace{4cm}
	- \frac{2\epsilon}{(1\!-\!\epsilon)} \bigl( \nabla_{(\mu} u \bigr)
		\bigl(  \nabla_\alpha \overline{ \nabla_{\nu)} \nabla'_\rho y_{\ss ++} } \bigr)
\nonumber \, .
\end{align}
%

\bibliographystyle{JHEP}
\bibliography{Abelian_Higgs_References.bib}

\end{document}